\documentclass[]{aa}

\usepackage{graphicx}
\usepackage{natbib}
\usepackage{txfonts}
\usepackage{wasysym}

\usepackage[bottom]{footmisc}
\begin{document} 

\title{Blue, white, and red ocean planets}
\subtitle{\large Simulations of orbital variations in flux and polarization colors}

\author{V.J.H.\ Trees\thanks{\emph{Present address: Atmospheric, Oceanic, and 
        Planetary Physics, Clarendon Laboratory, University of Oxford, 
        Parks Road, Oxford OX1 3PU, UK,
        e-mail: victor.trees@physics.ox.ac.uk.}}
         \and D.M.\ Stam}
        
\institute{Faculty of Aerospace Engineering, Delft University of Technology,
           Kluyverweg 1, 2629 HS Delft, The Netherlands}

\date{Received 5 March, 2019; accepted 16 April, 2019}

 
\abstract
{An exoplanet's habitability will depend strongly on the presence of
liquid water. Flux and/or polarization measurements of starlight that is 
reflected by exoplanets could help to identify exo-oceans. 
}
{We investigate which broadband
spectral features in flux and polarization phase functions of reflected
starlight uniquely identify exo-oceans.}
{With an adding-doubling algorithm, we compute 
total fluxes $F$ and polarized fluxes $Q$ of 
starlight that is reflected by cloud-free and 
(partly) cloudy exoplanets, for wavelengths from 350 to 865~nm. 
The ocean surface has waves composed of Fresnel reflecting
wave facets and whitecaps, and scattering within the water body
is included.}
{Total flux $F$, polarized flux $Q$, and degree of polarization
$P$ of ocean planets change color from blue, through white, 
to red at phase angles $\alpha$ ranging from 
$\sim 134^\circ$ to $\sim 108^\circ$ for $F$, and from 
$\sim 123^\circ$ to $\sim 157^\circ$ for $Q$,
with cloud coverage fraction $f_{\rm c}$
increasing from 0.0 (cloud-free) to 1.0 (completely cloudy) 
for $F$, and to 0.98 for $Q$.
The color change in $P$ only occurs for $f_{\rm c}$ 
ranging from 0.03 to 0.98, with the color crossing angle $\alpha$ 
ranging from $\sim 88^\circ$ to $\sim 161^\circ$. The total flux
$F$ of a cloudy, zero surface albedo planet can also
change color, and for $f_{\rm c}=0.0$, an ocean planet's $F$
will not change color for surface pressures $p_{\rm s}$ 
$\gtrapprox$ 8~bars. 
Polarized flux $Q$ of a zero surface albedo planet does not 
change color for any $f_{\rm c}$.}
{The color change of $P$ of starlight 
reflected by an exoplanet, from blue, through white, to red with increasing
$\alpha$ above 88$^\circ$, appears to identify a (partly) cloudy exo-ocean.
The color change of polarized flux $Q$ with increasing $\alpha$ 
above 123$^\circ$ appears to uniquely identify an exo-ocean, 
independent of surface pressure or cloud fraction. 
At the color changing phase angle, the angular 
distance between a star and its planet is much larger than 
at the phase angle where the glint appears in reflected light. 
The color change in polarization thus offers 
better prospects for detecting an exo-ocean.}


\titlerunning{Blue, white, and red ocean planets} 

\maketitle

\section{Introduction}

Liquid water is considered to be a prerequisite for life as we know it. 
The presence of liquid water on the surface of an exoplanet will depend 
on the planet's surface temperature and therefore on the separation between 
the planet and its host star, the stellar luminosity, the surface 
albedo, the 
composition and structure of the planetary atmosphere (which also strongly
influence the planet's albedo) and the atmospheric
surface pressure. While the amount of stellar flux that is incident
on the planet can be measured, the surface albedo, and the atmospheric
properties are more difficult to determine. Yet they are crucial
factors in determining a planet's habitability.

Some information about the composition and structure of the upper 
atmosphere of exoplanets can be derived using transit spectroscopy,
where the spectral variation of the stellar flux that is transmitted 
through the atmosphere of a planet during its transit, is measured.
Indeed, detections have been reported of water vapor in the atmospheres of 
hot Jupiter-like planets \citep[see for example][]{Tinetti2007,Deming2013,Kreidberg2014} 
and a low-mass Neptune sized planet \citep{Fraine2014}.
A large number of transiting exoplanets is already known, mostly thanks 
to the space observatories Corot ('Convection, Rotation and planetary Transits,' 
by ESA/CNES) and, of course, especially Kepler (NASA).
New and upcoming space observatories, like TESS ('Transiting Exoplanet Survey
Satellite,' NASA, launched in 2018, \citet{2015JATIS...1a4003R}) 
and PLATO ('PLAnetary Transits and 
Oscillations of stars,'
ESA, planned for launch in 2026, \citet{2014ExA....38..249R}), and ground-based telescope networks like WASP 
('Wide Angle Search for Planets')
\citep{2015JATIS...1a4003R} and MASCARA 
('Multi-site All-Sky CAmeRA') \citep{2017A&A...601A..11T}
are designed to search for transiting
exoplanets around bright, nearby stars. Such transiting 
exoplanets will be excellent
targets for follow-up transit spectroscopy measurements with, e.g.\ 
the JWST (the 'James Webb Space Telescope,' NASA) 
and ARIEL ('Atmospheric 
Remote-sensing Infrared Exoplanet Large-survey,' ESA, planned for launch
after PLATO).

Transit spectroscopy, however, does not provide information about the lower
atmospheric layers because the optical path-length through these layers
is too large to allow the transmission of starlight, in particular in the 
presence of clouds. And, rocky exoplanets are so small compared to their
star, and their atmospheres are so thin compared to the planetary radius, that obtaining a reliable signal-to-noise ratio could take several tens 
to hundreds of planetary transits. For rocky exoplanets in the habitable zones
of solar-type stars, that would thus take several tens to hundreds of
Earth-years \citep{2009ApJ...698..519K}. 

A secondary eclipse spectrum, derived from the change in the 
total flux spectrum when a planet starts to disappear behind and/or reappears from behind 
its star as seen from the observer, could provide information about the planetary albedo, and thus potentially about 
the planet's surface albedo, the atmospheric gaseous scattering, 
and the presence of clouds (not their actual composition),
although all blended together due to the lack of spatial resolution. 
Unless the atmospheric gaseous optical thickness, the cloud coverage 
fraction and the cloud physical properties
(especially cloud optical thickness) are known, a secondary eclipse 
spectrum will thus not provide information about the planet's surface 
albedo (thus, without the atmosphere) and the probability of liquid 
surface water (which is expected to have a very small albedo) will thus
remain elusive in secondary eclipse measurements.

For identifying liquid surface water, measuring the 
light of the parent star that is reflected by an orbiting exoplanet, appears
to be the most straightforward tool. 
The spectral and temporal variations of this reflected light as the planet 
orbits its parent star, holds information about the composition and structure
of the planetary atmosphere and/or the surface (if present).
This is particularly true for the state of polarization of the
reflected light \citep[see][and references therein]{2015psps.book..439W}.  
The angular distribution of the degree of polarization of singly scattered 
light is much more sensitive to the size, composition and shape of the
scattering  particles in the exoplanet's atmosphere 
than the total flux (unpolarized + polarized fluxes) is
\citep{Hansen1974}, 
and in addition, the angular distribution of the polarized signal
is preserved when multiple light is added. In particular, 
the phase angles\footnote{A planetary phase angle is the angle between the 
directions towards the parent star and the observer as measured from 
the center of the planet.} where the degree of polarization of the 
reflected starlight attains a local maximum value or where it equals 
zero (the so-called 'neutral points') can be directly translated into the 
scattering angles where the degree of polarization of the
singly scattered light attains a local maximum value or 
zero\footnote{Given a planetary phase angle $\alpha$, the single
scattering angle $\Theta$ equals 180$^\circ- \alpha$.}.
And, because the degree of polarization is a relative measure 
(i.e.\ the ratio of the polarized flux to the total flux), it is
insensitive to parameters such as distance, planetary radii, and the
incoming flux. 
The degree of polarization is also relatively insensitive to the Earth's 
atmosphere for ground-based observations, as the transmission of
light through the Earth's atmosphere is virtually independent of the state 
of polarization of the light ($F$, $Q$, and $U$ are equally affected
by scattering and/or absorption)
and it is insensitive to several instrumental parameters
(depending on the instrument design), such as the degradation of filters
or lenses in time. 

A famous application of these advantages of polarimetry 
is the characterisation of the particles composing the 
main cloud deck on Venus by \citet{1974JAtS...31.1137H}. They used
disk-integrated polarization measurements at wavelengths 
across the 0.34 - 1.00~$\mu$m region of Venus with Earth-based 
telescopes at a wide range of phase angles to determine that the clouds
particles consist of 75\% sulfuric acid solution, and have an effective radius
of 1.05~$\mu$m with an effective variance of 0.07. 
Because, as seen from Earth, Venus 
is an inner planet, it can be observed at phase angles ranging from almost 
0$^\circ$ to almost 180$^\circ$, while outer Solar System planets can only 
be observed at small phase angles (except of course from a spacecraft
in the vicinity of an outer planet). The phase angle range at which an
exoplanet can be observed depends on the orbital inclination angle $i$
(defined such that $i=0^\circ$ for a face-on orbit),
as seen from Earth: $90^\circ - i \leq \alpha \leq 90^\circ + i$.
In particular, transiting exoplanets, which 
TESS \citep{2015JATIS...1a4003R} and 
PLATO \citep{2014ExA....38..249R} are and will
be searching for, will be observable across a broad phase angle range
with direct detection techniques. 

Simulations of orbital variations of the (polarized) flux and degree
of polarization of starlight that is reflected by 
habitable exoplanets explore the potential observables of these planets,
are crucial for the design and optimization of future instruments and
telescopes aimed at the direct detection of planetary signals, and 
can be used to devise observational strategies (such as integration times 
and temporal coverage) and (future) data analysis algorithms. 
An example of a direct detection 
instrument that is designed to measure polarization of starlight 
reflected by exoplanets is the Exoplanet Imaging Camera and 
Spectrograph (EPICS) for the European Extremely Large Telescope (E-ELT)
\citep[]{2010SPIE.7735E..2EK}.
EPOL is the imaging polarimeter for EPICS \citep[]{2010SPIE.7735E..6GK}.
NASA's LUVOIR (Large UV Optical Infrared Surveyor) space telescope concept
\citep[]{2018arXiv180909668T} includes POLLUX, a polarimeter for the 
ultraviolet wavelength region \citep[]{2018SPIE10699E..3BB}. 
Another concept study by NASA is 
the Habitable Exoplanet Observatory (HabEx) \citep[]{2018arXiv180909674G},
that, as far as we know, does not (yet) include a polarimeter.

In this paper, we focus on the numerical simulations of the total and polarized 
fluxes and the degree of polarization of starlight that is reflected by 
Earth-like exoplanets that are covered by a liquid ocean with surfaces
that are rough due to the presence of waves.
Fresnel reflection of light by an ocean on an exoplanet results in 
two main observable phenomena: a glint on the ocean surface, 
whose size increases with increasing surface roughness (i.e. 
increasing wind speed), and a maximum degree of polarization at the 
Brewster angle. Indeed, as the strength of both phenomena depends on 
the local reflection angles, they are strongly phase angle dependent. 

The signatures of an ocean in the orbital variations of the total flux $F$ of starlight that is reflected by an exoplanet have been studied before.  
\citet{Robinsonetal2010} modeled the phase curves of the total flux of 
the Earth as an exoplanet, and \citet{OakleyandCash2009}, 
\citet{Cowanetal2009} and \citet{Lustig-Yaeger2018} aim to reconstruct a 
surface type map from modeled total flux phase curves at a single or 
multiple wavelength(s). Those papers, however, did not include the 
polarization of the reflected starlight.
\citet{Stam08} computed the total flux $F$ and degree of polarization $P$ as 
functions of the wavelength and planetary phase angle for Earth-like 
exoplanets some of which with a Fresnel reflecting ocean, but did not 
include wind-driven waves on the ocean surface, thus yielding an 
infinitely narrow beam of reflected light representing the stellar glint
(the spatial distribution of this glint was lost upon integration 
of the reflected starlight across the illuminated and visible part of the
planetary disk).
\citet{Williams2008} presented the phase curves of an ocean planet 
with a rough surface and found strong signatures of Fresnel reflection 
in the phase angle dependence of the degree of polarization $P$.
They, however, neglected atmospheric contributions such as Rayleigh 
scattering and scattering by clouds.  
\citet{Zugger2010,ZuggerErratum2011,Zugger2011} did include atmospheric 
Rayleigh scattering, and the reflection by clouds modeled as horizontal,
Lambertian (i.e.\ isotropic and non-polarizing) reflecting 
surfaces\footnote{Indeed, the 'rainbow peaks' in the degree of polarization 
phase curves in the papers of \citet{Zugger2010,ZuggerErratum2011,Zugger2011} are caused by their assumption of spherical maritime aerosols, not by their model clouds. 
It should be noted, however, that maritime (sea-salt) aerosols may be 
non-spherical \citep[see Chap. 18 of][]{BookMishchenko2000}.} 
(not only ignoring gas below and in the clouds, but also above the clouds), 
and scattering by maritime aerosols (modeled as spherical particles) in the 
cloud-free atmosphere above a rough ocean surface.
They found that the ocean signature in $P$ is limited by the scattering by 
maritime aerosols and clouds. 

The model planets of 
\cite{Zugger2010,ZuggerErratum2011,Zugger2011} are \textit{horizontally 
homogeneous}, that is, the atmosphere and surface are invariant with 
latitude and longitude. Therefore, they could, for example, not study 
the signatures of the glint appearing and hiding behind \textit{patchy clouds}, 
which are clouds of varying horizontal shapes distributed across the planetary
disk. \citet{Rossi2017} showed the features that could potentially be 
observed in the flux and polarization phase curves of \textit{horizontally
inhomogeneous} ocean planets using various cloud cover types. Their ocean
surface was, however, still a flat air-water interface, without waves,
and thus with an infinitely narrow glint. Also, \citet{Karalidi12} and
\citet{KaralidiStam2012} presented computed flux and polarization phase curves 
for horizontally inhomogeneous, cloudy Earth-like exoplanets, focusing
on the signatures of the clouds, and representing the ocean by a flat, zero albedo (i.e.\ black) surface (without the glint).

In this paper, we present the computed planetary phase curves of the total and 
polarized fluxes and the degree of polarization of starlight that is 
reflected by oceanic exoplanets with rough, wavy surfaces beneath 
terrestrial-type atmospheres. We fully include scattering of (unpolarized
and polarized) light within the water body (i.e. below the rough ocean surface)
and the reflection by wind-generated foam on the ocean surface. We compute 
the phase curves for various wind speeds and surface pressures, at 
wavelengths ranging from 350 to 865~nm. We use two types of model ocean 
planets: cloud-free planets, and cloudy planets. The clouds are patchy 
and embedded within the gaseous atmosphere (there is thus gas below, within
and above the clouds), just like those on Earth, and our model clouds consist
of liquid water droplets.

This paper is structured as follows. 
In Sects.~\ref{sect_definitions} and~\ref{sect_calculating_reflected_starlight}, 
we provide our definitions and describe our numerical method to compute both 
the total and polarized fluxes, and the degree of polarization $P$ of starlight 
that is reflected by our model ocean planets. 
In Sect.~\ref{sect_atmosphere_surface_models}, we describe the physical and 
optical properties of the atmosphere and the ocean surface and water body below. 
In Sect.~\ref{sect_results}, we present the computed reflected total and
polarized fluxes and $P$ for our model planets for various
atmospheric and surface parameters, and we describe the influence of these 
parameters on the planet's color change in total flux $F$, polarized flux $Q$, 
and $P$. In Sect.~\ref{sect_discussion}, we discuss the color change and with 
that the detectability of an ocean beneath clouds. 
In Sect.~\ref{sect_summary}, finally, we summarize our results. 

\section{Numerical method}
\label{sect_numerical}

\subsection{Definitions of fluxes and polarization}
\label{sect_definitions}

We describe light as the following column vector
\citep[see e.g.][]{Hansen1974,2004Hovenier}
\begin{equation}
   \mathbf{F} = \left[ \begin{array}{c}
                         F \\ Q \\ U \\ V
                         \end{array} \right],
\label{eq_stokes}
\end{equation}
where $F$ describes the total flux, $Q$ and $U$ the linearly polarized 
fluxes, and
$V$ the circularly polarized flux. The dimension of these fluxes is Wm$^{-2}$
or in Wm$^{-3}$, when accounting for the dependence on 
the wavelength $\lambda$, but because of our normalization
(see below), we do not use these dimensions explicitly. 
Linearly polarized fluxes $Q$ and $U$ are defined with respect 
to a reference plane. 
For our computations of starlight that is reflected by a planet as a whole,
we use the planetary scattering plane, i.e.\ the plane that 
contains the centers of the star, the planet, and the observer,
as the reference plane.
Fluxes $Q$ and $U$ can be redefined with respect to 
another reference plane, such as an instrument's optical plane, 
using a so-called rotation matrix \citep[see][]{1983A&A...128....1H}.

We assume that the starlight that is incident on the planet 
is unidirectional and unpolarized. The latter is based on
the very small disk--integrated polarized fluxes of solar-type stars, as measured for active and inactive FGK--stars by \citet{2017MNRAS.467..873C}
(with, respectively, 23 $\pm$ 2.2 ppm and 2.9 $\pm$ 1.9 ppm polarized flux) and 
\citet{Kemp1987} for the Sun itself, and the very 
small degree of stellar linear 
polarization (on the order of $10^{-6}$) that is expected from 
symmetry-breaking disturbances such as spots and/or flares 
\citep{2011ApJ...728L...6B,kostogryz2015polarization}.
The incident starlight is thus described as
${\bf F}_0 = F_0 \bf{1}$, with $\pi F_0$ the stellar flux measured
perpendicular to the direction of propagation, and ${\bf 1}$ the
unit column vector. 

Starlight that is reflected by a planet will become polarized
by scattering by gases and aerosol or cloud particles in the
planetary atmosphere and/or by reflection off the surface. 
We define the degree of polarization of the reflected starlight as 
\begin{equation}
    P = \frac{\sqrt{Q^2 + U^2}}{F}.
\label{eq:def-pol-tot}
\end{equation}
Unlike $Q$ and $U$, $P$ is independent of a reference plane. 
We ignore the circularly polarized flux $V$ 
of the reflected starlight. This flux is expected to be very small 
\citep[for examples for rocky exoplanets, see][and references therein]{2018A&A...616A.117R,2015IJAsB..14..379G},
and ignoring circular polarization in the radiative transfer computations 
yields only very small errors in the total and linearly polarized fluxes
\citep{stametal05}.

\subsection{Calculating reflected starlight}
\label{sect_calculating_reflected_starlight}

We compute starlight that is reflected by different types of exoplanets, both 
the total flux and polarization, and for a range of wavelengths $\lambda$.
Given a planet orbiting a star, with $\pi F_0 {\bf 1}$ the 
flux vector that is incident on the planet,
we compute the flux vector $\pi{\bf F}$ 
that is reflected by the planet as a whole
and that arrives at an observer situated at a large distance $d$ of the
planet and star, 
by integrating the locally reflected light across the illuminated 
and visible part of the planetary disk. We perform this integration 
using a grid of pixels that cover the planetary disk 
\citep[see][]{rossi2018pymiedap},
and by summing the fluxes reflected by each of the $N$ pixels that are 
both illuminated by the parent star and visible to the observer, 
as follows:
\begin{equation}
   \pi {\bf F}(\lambda,\alpha) = \frac{F_0(\lambda)}{d^2} \hspace*{0.1cm}
   \sum_{i=1}^{N} \mu_i \mu_{0i} \hspace*{0.1cm}
   {\bf L}(\beta_i)
   {\bf R}_i(\lambda,\mu_i,\mu_{0i},\phi_i-\phi_{0i}) {\bf 1} \hspace*{0.1cm}
   dO_i.
\label{eq_1}
\end{equation}
Here $\alpha$ is the planetary phase angle, which is defined as
the angle between the star and the observer measured from the 
center of the planet (at $\alpha=0^\circ$, the planet is thus behind
the star as seen from the observer, and at $\alpha=180^\circ$, the
planet is precisely in front of the star). Angle $\alpha$ and 
the location on the planetary disk determine
the local illumination and viewing angles across the planet: 
$\mu = \cos \theta$, with $\theta$ the local viewing zenith angle, 
$\mu_0 = \cos \theta_0$, with $\theta_0$ the local illumination angle,
and $\phi-\phi_0$, the local azimuthal angle 
\citep[see][for the precise definitions of these angles]{deHaan87}.

Matrix ${\bf R}_i$ in Eq.~\ref{eq_1},
is a local reflection matrix, i.e.\ it describes how the area that 
pixel $i$ covers on the planet, reflects the incident light. 
This matrix depends on the composition and structure of the local atmosphere
and the underlying surface (see Sect.~\ref{sect_atmosphere_surface_models}), 
on the wavelength, as the optical properties
of the atmosphere and surface are wavelength dependent, 
and on the local illumination and viewing angles 
(hence, on phase angle $\alpha$ and/or the location of the pixel on the 
planet). Matrix ${\bf L}$ is a rotation matrix
\citep[see][]{1983A&A...128....1H} that rotates each locally 
reflected flux vector, that is defined with respect to
the local meridian plane through the local zenith and the local 
direction towards the observer, to the planetary scattering plane.
Angle $\beta_i$ is the local rotation angle.
Furthermore, $dO_i$ is the area that grid pixel $i$ covers on the 3D-planet.
The area $dO_i$ a pixel covers on the
3D-planet depends on the local viewing angle $\theta_i$. Indeed, 
$\mu_i dO_i$ equals the pixel size, which we choose the same for all pixels. 

With Eq.~\ref{eq_1}, we can compute the flux vectors reflected by locally 
horizontally homogeneous planets, i.e. where the surface and atmosphere 
are the same for each pixel across the planetary disk,
and by horizontally inhomogeneous planets, i.e. where the
surface and atmosphere can be different for different pixels.
The error that results from using the summation of Eq.~\ref{eq_1} for 
the integration across the planetary disk depends on the size of the pixels,
on the horizontal variations of the surface and atmosphere, and on the 
phase angle $\alpha$ \citep[see][]{rossi2018pymiedap}. 
For our computations, we use a minimum of 40 pixels across the planetary 
equator for every $\alpha$, which provides a good relation between
the integration error and the computing time.

To compute $\pi {\bf F}$, we first assign
an atmosphere and surface model to each pixel on the planetary disk. 
Then, given the phase angle $\alpha$, and thus the local illumination and
viewing angles, the local matrices ${\bf R}_i$ are computed and rotated 
using the local rotation angles $\beta_i$. Then, the summation in 
Eq.~\ref{eq_1} is evaluated.
The computation of the matrices ${\bf R}_i$ is performed with an efficient 
adding--doubling algorithm \citep{deHaan87,Stam08}, extended with reflection 
by and transmission through a rough 
ocean surface and scattering in the ocean body (see 
Sect.~\ref{sect_atmosphere_surface_models}), 
fully including polarization for all orders of scattering. Our adding--doubling algorithm is a multi-stream radiative transfer code.
To accurately capture the angular variations in the reflected light signals,
we use 130 streams for our simulations and a Fourier-series representation  
for the azimuthal direction \citep[for details, see][]{deHaan87,rossi2018pymiedap}.

We normalize computed reflected flux vectors such that at $\alpha=0^\circ$,
the total reflected flux equals the planet's geometric albedo $A_{\rm G}$
\citep[see, e.g.][]{Stam2006,rossi2018pymiedap}.
The total and polarized fluxes presented in this article
can thus straightforwardly be scaled to the relevant
parameters of a given star-planet system.
The degree of polarization $P$ is a relative measure and thus
independent of the stellar flux, planetary radius, and/or the 
distance to the planet. 
\subsection{The planetary surface--atmosphere models}
\label{sect_atmosphere_surface_models}

Each pixel on the planetary disk has an atmosphere bounded below by a surface.
The local atmosphere and surface are horizontally homogeneous, and 
rotationally symmetric around the local vertical. 
In principle, every pixel can have a unique atmosphere and surface model.
In practice, we limit the number of different models across the disk to
two. Detailed descriptions of the atmosphere and surface models
are given below.

\subsubsection{The atmosphere}
\label{sect_atmosphere_model}

Our model atmosphere is Earth--like. Locally, the planetary atmosphere is composed of a stack of horizontally
homogeneous layers, each containing either only gas or gas with a cloud 
added. 
The gas molecules scatter as anisotropic Rayleigh scatterers with a 
wavelength independent depolarization factor equal to 0.03
\citep[see][for details regarding their scattering properties]{Stam08}.
We perform our computations at wavelengths where absorption by the gas is small, and can safely be ignored. The standard surface pressure is 1~bar, the acceleration of gravity
9.81 m/s$^{2}$, the mean molecular mass 29~g/mole, and we assume hydrostatic
equilibrium, which yields a total
atmospheric gas optical thickness of 0.096 at $\lambda= 550$~nm
\citep[see][and references therein for the relevant equations]{Stam08}.

\begin{table}[b]
\caption{Parameters of the atmosphere. Unless stated otherwise, the values 
         listed are used for the cloudy model planet. For the clear model planets, 
         the cloud parameters may be ignored.}   
\label{table_atmos}     
\centering 
\begin{tabular}{l l l} \hline \hline             
Atmosphere parameter & Symbol & Value \\ \hline           
Surface pressure [bar]        & $p_{\rm s}$   & 1.0   \\
Depolarization factor         & $\delta$   & 0.03  \\
Mean molecular mass [g/mole]   & $m_{\rm g}$       & 29    \\
Acceleration of gravity [m/s$^2$]  & $g$       & 9.81    \\
Gas optical thickness (at $\lambda$ = 550 nm) & $b^{\rm m}$       & 0.096 \\
Cloud particle effective radius [$\mu$m]  & $r_{\rm eff}$   & 10.0   \\
Cloud particle effective variance        & $v_{\rm eff}$   & 0.1  \\
Cloud optical thickness (at $\lambda=550$ nm)   & $b^{\rm a}$       & 4.926 \\
Cloud-top pressure (bar)                 & $p_{\rm ct}$          & 0.7 \\
Cloud-bottom pressure (bar)                 & $p_{\rm cb}$          & 0.8 \\
\hline       
\end{tabular}
\end{table}

Clouds strongly affect the signal of the glint
on a planet, because they prevent light from reaching the ocean and
they will block the light reflected by the ocean from reaching the observer.
Our model clouds are composed of spherical particles, which we choose to be
made of liquid water. The particles are distributed in size according
to a two-parameter gamma distribution 
\citep[see][]{Hansen1974,de1984expansion}, with the cloud
particle effective radius equal to 10~$\mu$m and the effective 
variance equal to 0.1.
The optical properties of the cloud particles are
computed using Mie-theory as described by \citet{de1984expansion}
and a wavelength independent refractive index of 1.33. 
Table~\ref{table_atmos} lists the parameters describing the atmosphere model
including the clouds. Note that the reflection by cloudy regions 
on the planet models used by
\citet{Zugger2010,ZuggerErratum2011,Zugger2011} is described by the 
reflection by white, Lambertian surfaces, and
scattering by gas below, in and above the clouds is ignored by 
\citet{Zugger2010,ZuggerErratum2011,Zugger2011}.
 
The influence of clouds on the planet signals will not only depend on 
the cloud physical properties, but also on the location of the cloudy pixels
on the planetary disk. We will investigate the effect of the cloud cover on the planet's flux and
polarization signals for patchy clouds, 
modelling them as horizontal slabs with optical thickness $b^{\rm a}$ that are distributed across the planetary disk. The location of the clouds is
determined by the cloud fraction $f_{\rm c}$ (defined on the full disk, 
not only on the 
illuminated part) and asymmetry parameters that describe a
zonal-oriented pattern similar to that observed on Earth 
\citep[for the algorithm, see][]{2017A&A...607A..57R}. 
Examples for different cloud coverage fractions $f_{\rm c}$ can be seen
in Fig.~\ref{fig_diskresolved}. 

\subsubsection{The surface}
\label{sect_surface_model}

The reflection by the surface is either Lambertian, i.e. isotropic and 
completely depolarizing, or oceanic, i.e. the reflection is described 
by a Fresnel reflecting and transmitting air-water interface and 
sub-interface water layers. 

The air-water interface can be rough due to wind-driven waves. The 
roughness is modeled with a Gaussian distribution for the wave facet 
inclinations, 
with a standard deviation determined by the horizontally isotropic wind speed 
\citep[][]{CoxandMunk1954}. 
Unless stated otherwise, we use a wind speed of 7 m/s which corresponds to 
a moderate breeze\footnote{See for example \url{https://www.metoffice.gov.uk/guide/weather/marine/beaufort-scale.}} 
and is a reasonable value for above the Earth's oceans where the long-term
annual mean wind speed varies approximately between 5 m/s and 9 m/s
\citep{MishchenkoandTravis1997,Hsiung1986}.
The reflection and transmission matrices for 
polarized light for a flat wave facet follow from the Fresnel equations 
and a proper treatment of the changing solid angle of the light beam when 
traveling from one medium to another  
\citep[][]{Garcia2012,Zhaietal2012,Garcia2012ResponsetoComment,Zhaietal2010}. 
We use the shadowing function of \citet{Smith1967} and \citet{Sancer1969} 
to account for the energy budget across the rough air-water interface 
at grazing  angles caused by neglecting wave shadowing 
\citep[see also][]{tsang1985theory,MishchenkoandTravis1997}. 
The interface reflection and transmission matrices are corrected for 
the energy deficiency caused by neglecting reflections of light between 
different wave facets \citep{NakajimaandTanaka1983}.
The elements of our matrix describing light reflected by the air-water
interface have been verified against the code of \citet{MishchenkoandTravis1997}, 
and the energy balances across the interface (i.e. the sum of the reflection 
and transmission, for both illumination from below and above) are compared
to Fig.~4 of \citet{NakajimaandTanaka1983}. The elements 
of our reflection and transmission matrices have been verified against those 
of \citet{Zhaietal2010}.

The sub-interface ocean is modeled as a stack of horizontally homogeneous 
layers bounded below by a black surface. The reflection matrix of the sub-interface
ocean is computed with the adding-doubling algorithm \citep{deHaan87}. 
Scattering of light in pure seawater is caused by small density fluctuations. 
However, the matrix describing the single scattering in this water may 
be approximated by that for anisotropic Rayleigh scattering 
\citep[][]{Hansen1974}, 
using a depolarization factor of 0.09 \citep[][]{morel1974optical,Chowdharyetal2006}. 
The single scattering albedo of the water is computed according to
\begin{equation}
   \omega_{\rm w}(\lambda) = \frac{b_{\rm w}(\lambda)}
                                  {b_{\rm w}(\lambda)+ a_{\rm w}(\lambda)},
\end{equation}
where $b_{\rm w}$ and $a_{\rm w}$ are the wavelength dependent scattering 
and absorption coefficients for pure seawater, respectively, which are both 
expressed in m$^{-1}$. For $b_{\rm w}$, we use the values tabulated by 
\citet{SmithandBaker1981} and for $a_{\rm w}$ the values of 
\citet{popeandfry1997}\footnote{The wavelength range is extended from 
$\lambda=~380$~nm to $\lambda=350$~nm with the values of 
\citet{SogandaresandFry1997}. 
Between $\lambda=727.5$~nm and 800~nm, we use the values of \citet{SmithandBaker1981}. 
For $\lambda > 800$~nm, the sub-interface ocean albedo is set equal to zero.}. 
The ocean optical thickness is computed from the beam attenuation coefficient
$c(\lambda) = b_w(\lambda) + a_w(\lambda)$ and the ocean's 
geometric depth. Note that the ocean water does not include hydrosols 
(such as phytoplankton, detritus and bubbles). This would require a 
determination of the hydrosol scattering matrix elements 
\citep[see e.g.][]{Chowdharyetal2006}, which is beyond the scope of this study. 
We compared our wavelength dependent sub-interface ocean albedo at a local 
solar zenith angle $\theta_0$ of 30$^\circ$ with the bio-optical model for 
the so-called Case~1 Waters of \citet{MorelandMaritorena2001} and find 
our ocean albedos, at the wavelengths employed in this paper, 
correspond to those of water with phytoplankton chlorophyll 
concentrations between 0~and 0.1 mg~m$^{-3}$.

\begin{table}[t]
\caption{Parameters of the ocean. Unless stated otherwise, the values listed 
         are used for the ocean model planet.}
\label{table_ocean} 
\centering            
\begin{tabular}{l l l}  \hline\hline     
Ocean parameter & Symbol & Value \\  \hline                      
Wind speed {[}m/s{]}                                  & $v$          & 7.0     \\
Foam albedo                                  & $a_{\rm foam}$          & 0.22     \\
Depolarization factor                                 & $\delta_{\rm w}$ & 0.09  \\
Real air refractive index    & $n_1$       & 1.0   \\
Real water refractive index & $n_2$       & 1.33  \\
Chlorophyll concentration {[}mg/m$^3${]}              & {[}Chl{]}  & 0     \\
Ocean depth {[}m{]}                                   & $z$          & 100   \\ 
Ocean bottom surface albedo                           & $a_{\rm bs}$  & 0 \\ 
\hline
\end{tabular}
\end{table}

The reflection by the ocean system, i.e. by both the air-water interface 
and the sub-interface ocean, is computed with the adding equations described 
by \citet{deHaan87} and by assuming an infinitely thin interface 
\citep[see also][]{Xuetal2016}. Light traveling from the atmosphere into 
the water is refracted into a sharp cone, resulting in a lower angular
resolution in the cone compared to the resolution in the atmosphere. In 
order to accurately cover the sharp cone, we use rectangular super-matrices
\citep[see][for the use of the super-matrices]{deHaan87}
for the interface in the adding-doubling algorithm, allowing us to employ 
more and different Gaussian quadrature points inside the ocean layers 
than inside the atmosphere layers \citep[see also Section 3.2.4 by][]
{Chowdhardy1999PhDthesis}. 

Finally, we adapt the hence obtained 'clean' ocean reflection for the 
reflection by wind-generated foam, or white caps, using a weighted sum of the 
clean ocean reflection matrix and the Lambertian (i.e. isotopic and 
non-polarizing) reflection matrix by foam. As a baseline, we assume an 
effective foam albedo of 0.22 \citep{koepke1984effective} 
and the empirical relation of \citet{Monahan1980} for the wind speed 
dependent fraction of white caps. 

In App.~\ref{app_equations}, we provide the governing equations that we 
use to compute the ocean reflection of polarized light as described in 
this section. The ocean parameters are listed in Tab.~\ref{table_ocean}. 
Note that the real refractive index of air is set equal to 1.0 in the 
air-water interface computations, while in reality, it will vary slightly
with the wavelength (this variation is indeed taken into account in the 
computation of the scattering optical thickness of the atmospheric layers,
see Sect.~\ref{sect_atmosphere_model}). The refractive index of water 
is assumed to be wavelength independent and equal to 1.33 
\citep{hale1973optical}. Varying this refractive index 
between 1.34 (representative for liquid water and short wavelengths) 
and 1.31 (representative for water ice) yielded negligible differences 
in the phase angle dependent, disk-integrated, reflected fluxes and degree 
of polarization. 

\section{Results}
\label{sect_results}

Here, we present the results of our computations of the flux and
polarization phase curves of ocean planets, starting with horizontally
homogeneous, cloud-free planets (Sect.~\ref{sect_HHplanets}), followed 
by planets with different cloud coverage fractions
(Sect.~\ref{sect_HIplanets}).

\subsection{Horizontally homogeneous, cloud-free planets}
\label{sect_HHplanets}

\begin{figure*}
	\includegraphics[width=0.33\textwidth]{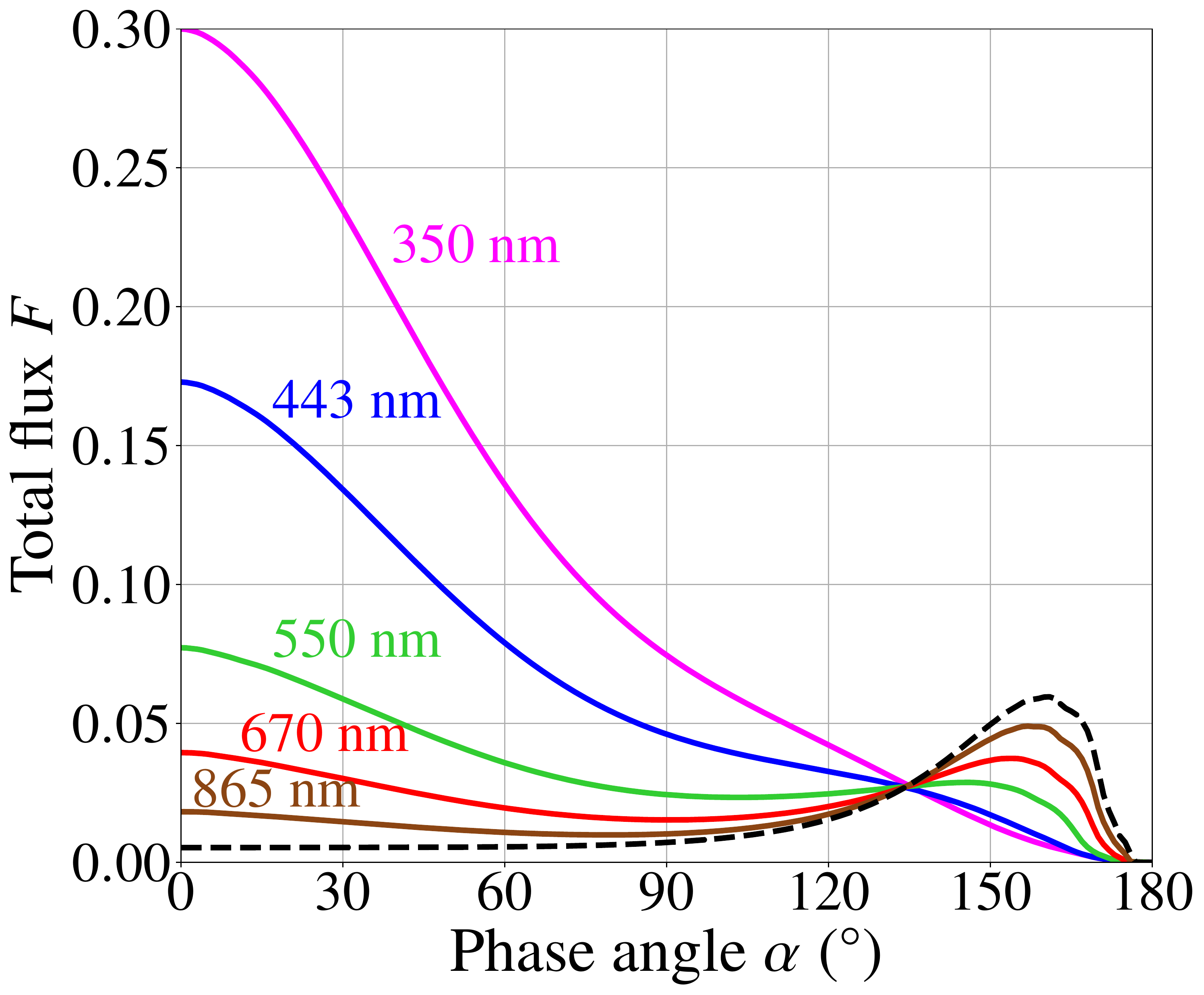}
	\includegraphics[width=0.33\textwidth]{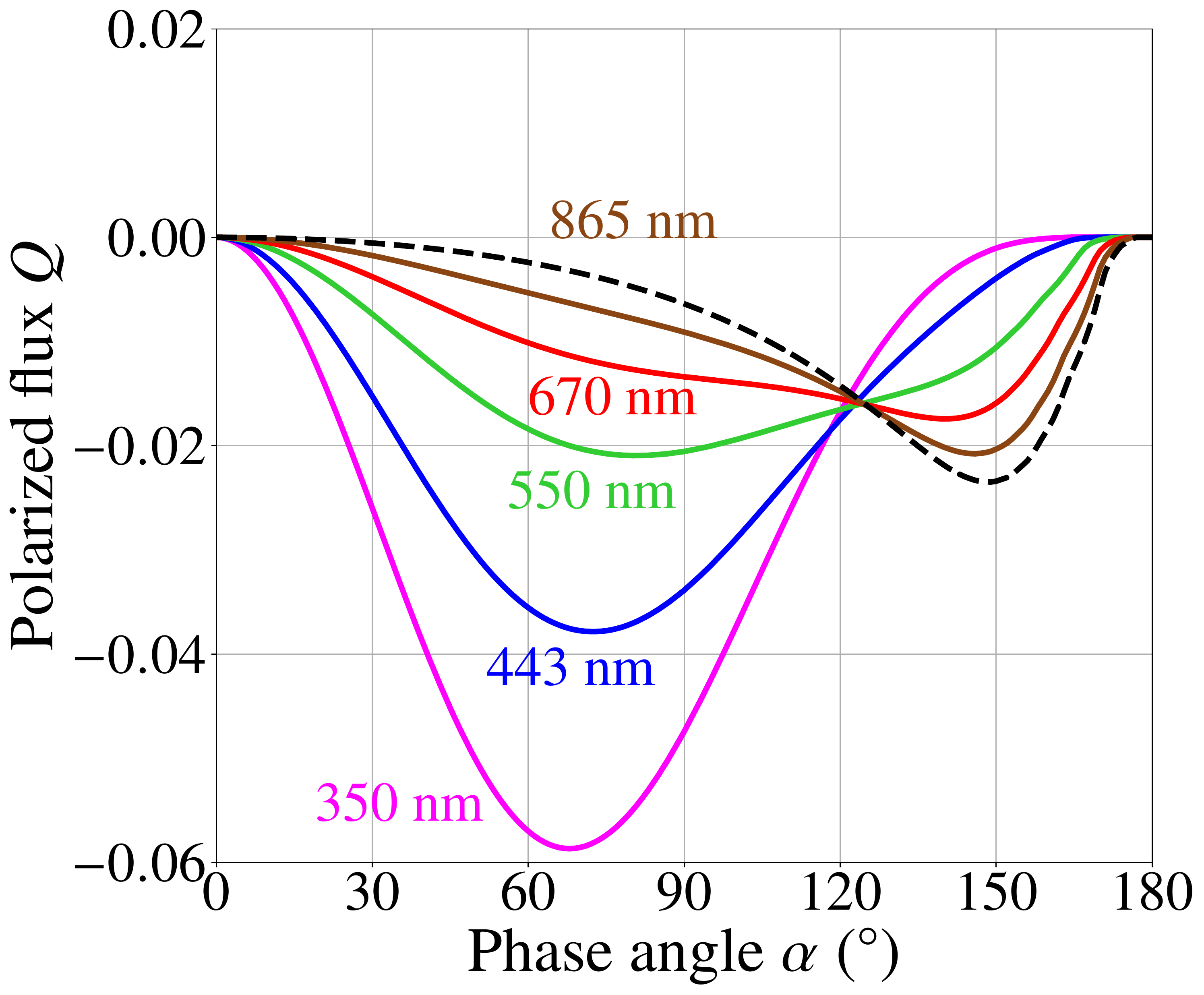}
	\includegraphics[width=0.33\textwidth]{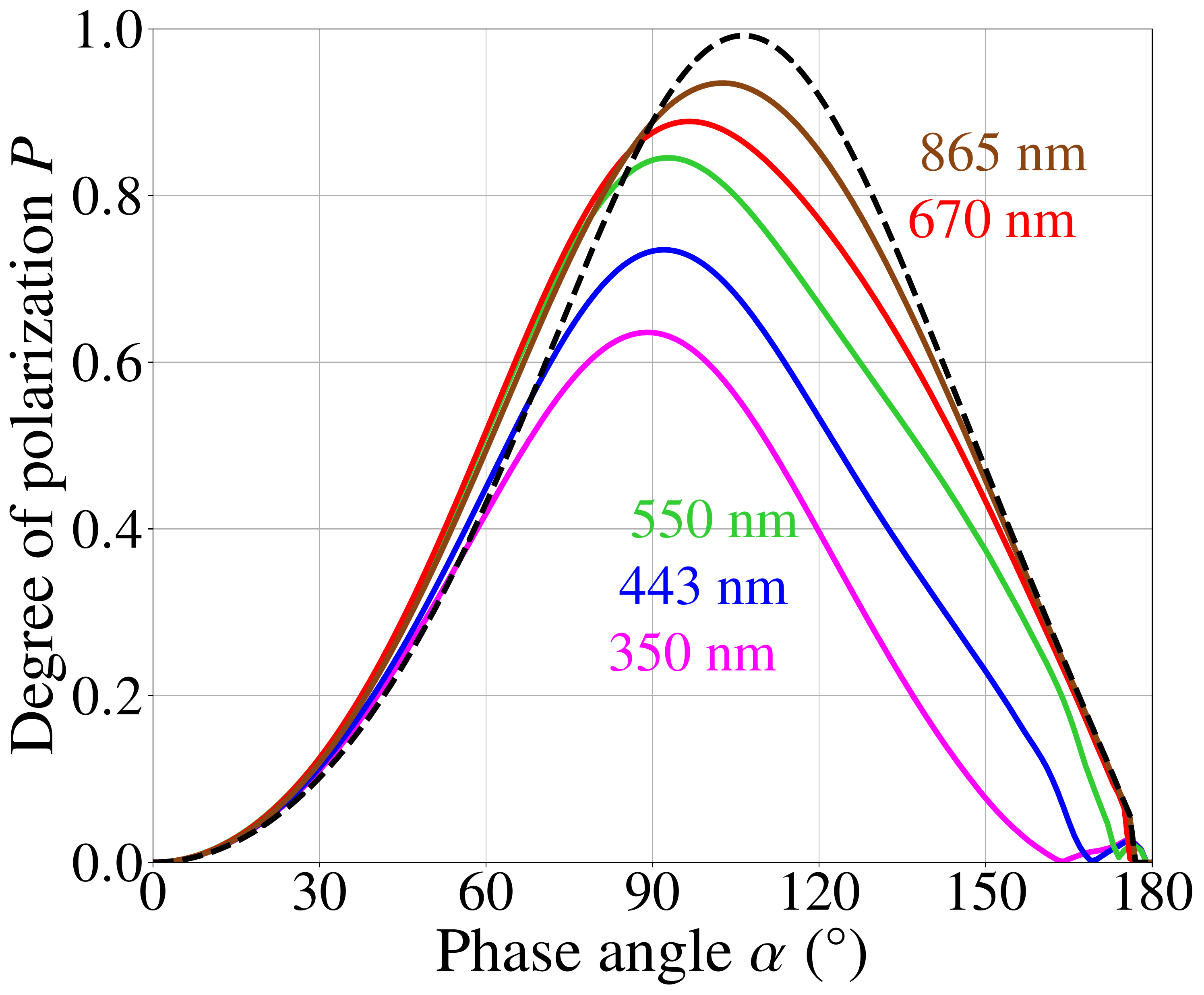} \\
	
	\includegraphics[width=0.33\textwidth]{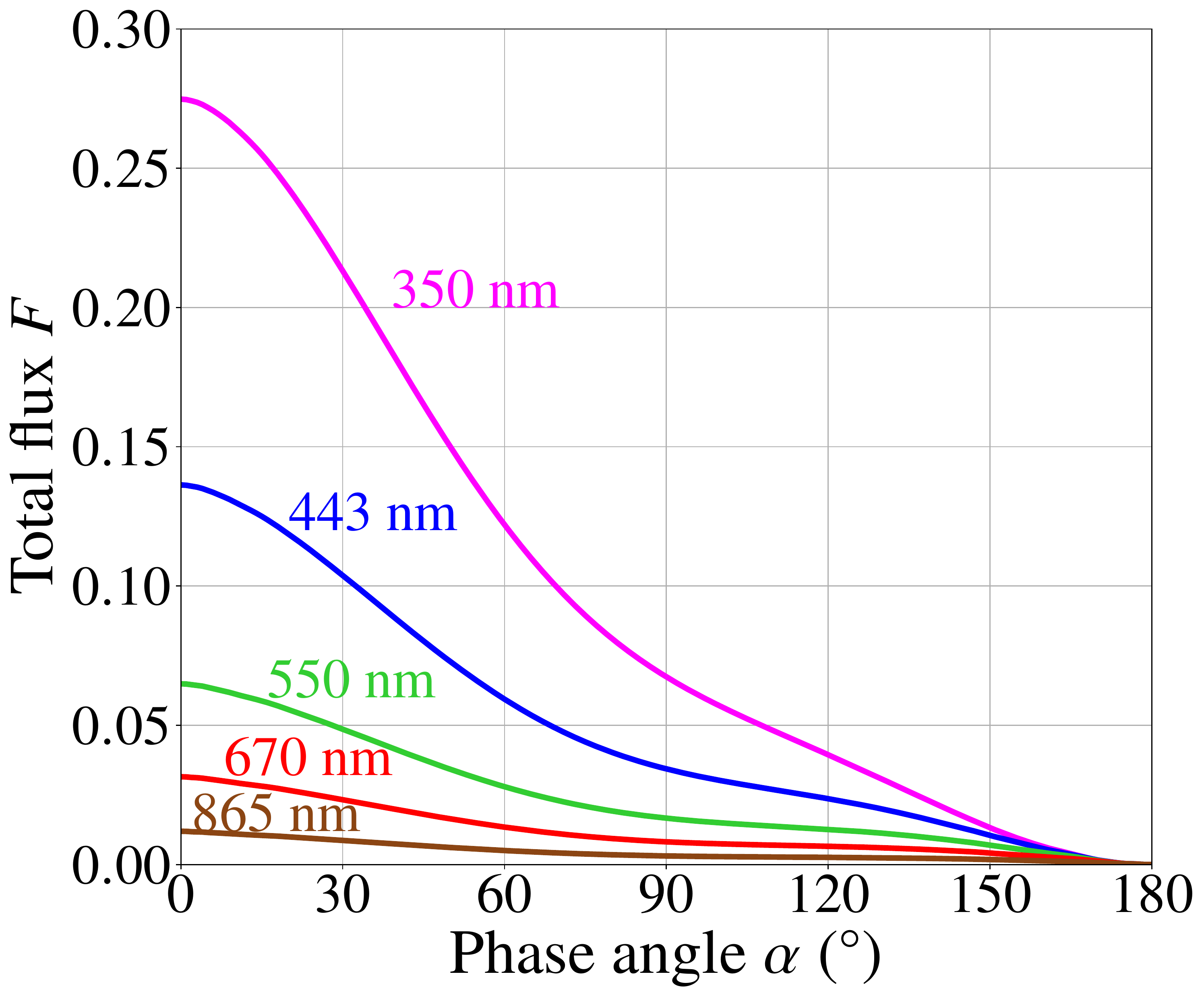}
	\includegraphics[width=0.33\textwidth]{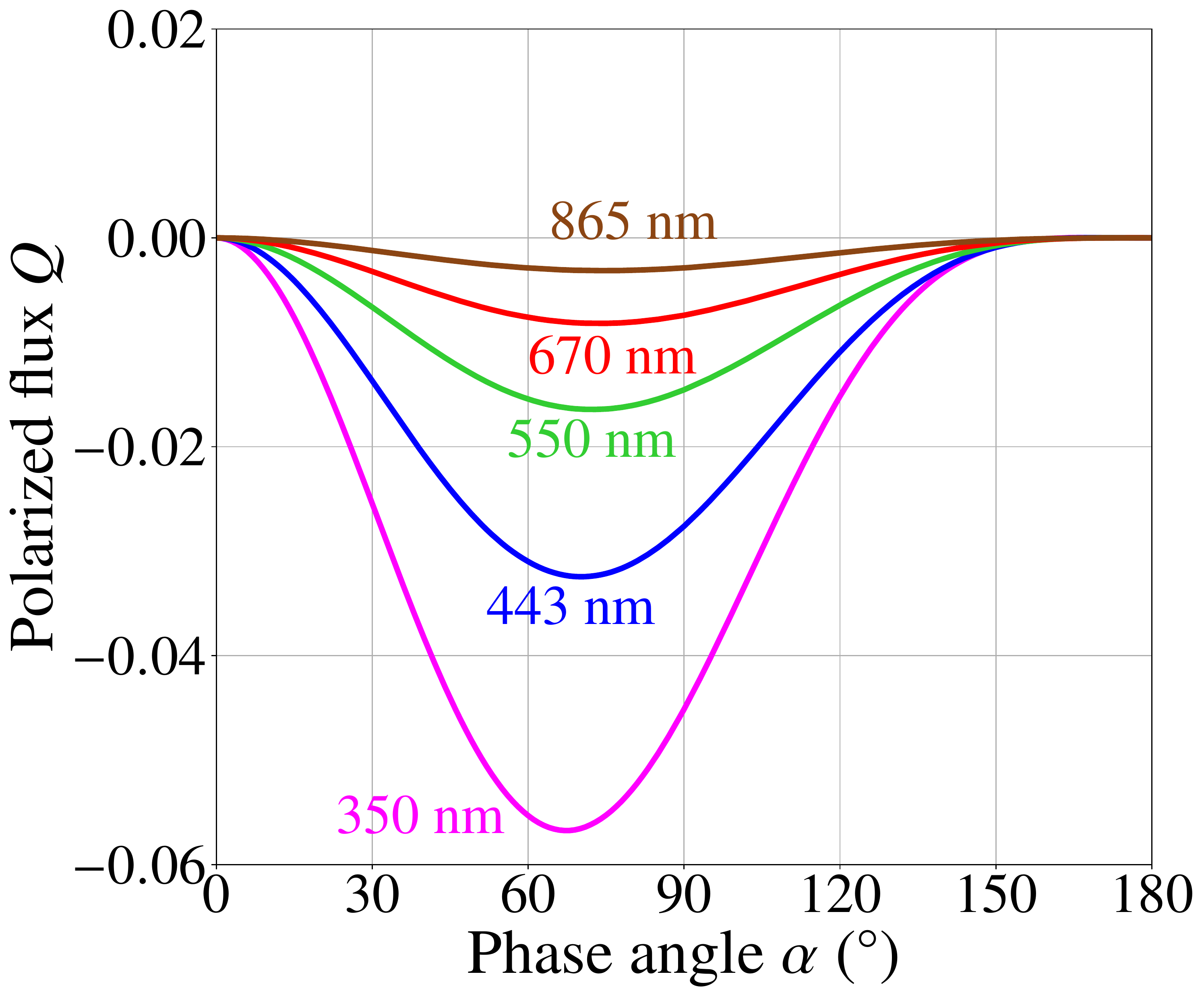}
	\includegraphics[width=0.33\textwidth]{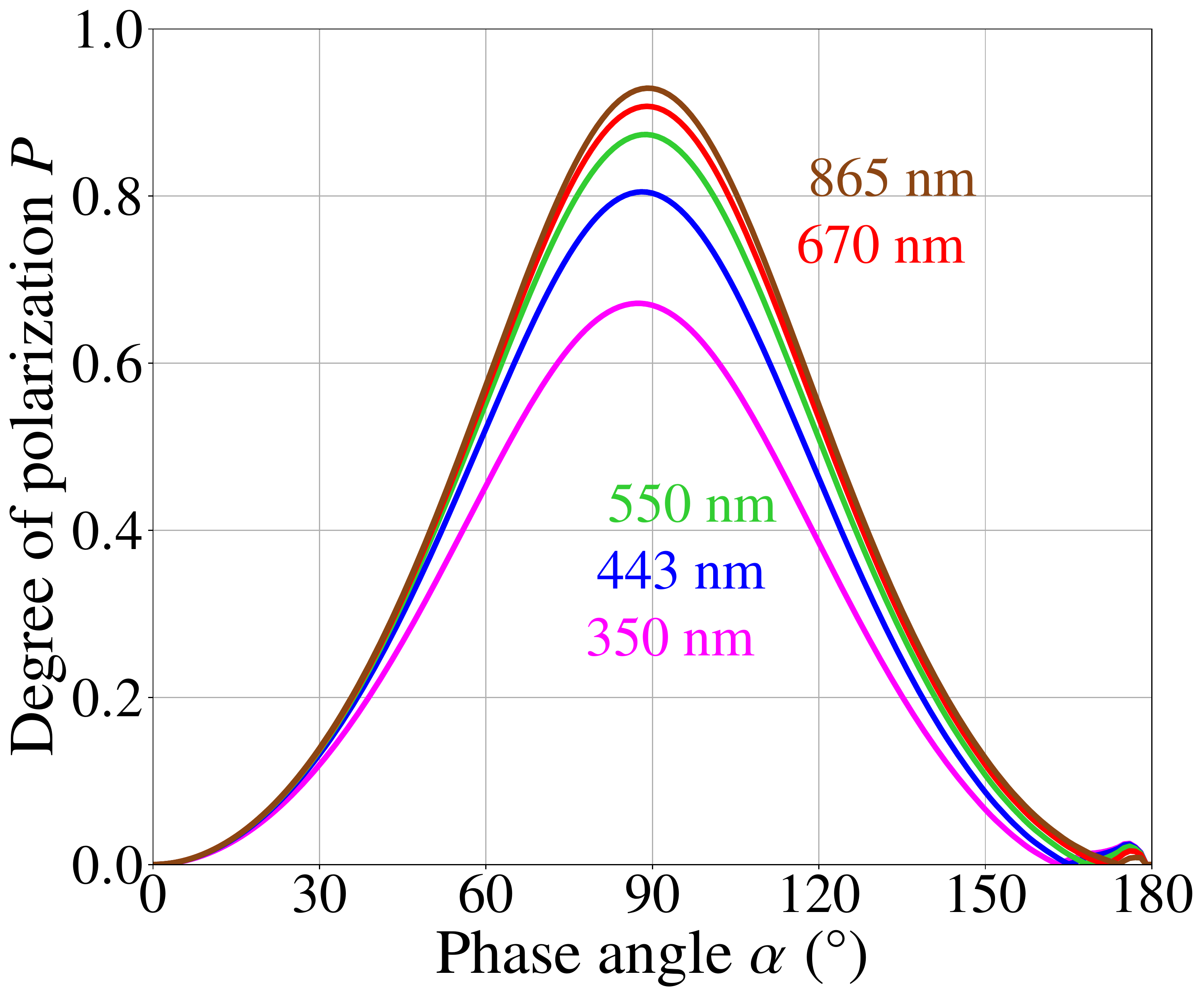}
\caption{The (disk-integrated) reflected total flux $F$ (left column), linearly polarized flux $Q$  (middle column) and degree of polarization $P$ (right column) of an ocean planet with a rough, Fresnel reflecting surface (top row) and a planet with a black surface (bottom row) as functions of the phase angle,
         for various wavelengths. Both planets have a pure gaseous
         atmosphere and are cloud-free. The wind speed above the ocean is 7~m/s, and the surface coverage of white caps 0.28$\%$.
         The black dashed line in the ocean planet plots
         (top row) pertains to a planet covered by the rough Fresnel
         surface, with a black surface below, thus without sub-surface ocean,
         and without atmosphere above.}	
\label{fig_HHplanets}         
\end{figure*}

Figure~\ref{fig_HHplanets} shows the phase curves of
an 'ocean planet', i.e.\ a planet with a rough, Fresnel reflecting 
interface on top of a 100~m deep ocean that is bounded below by a 
black surface, and a 'black surface planet', i.e.\ a planet with a black surface. The wind speed is 7~m/s, and the percentage of white caps 
thus 0.28\% \citep[see][]{Monahan1980}.
The planets have the same gaseous atmosphere, 
with a 1~bar surface pressure. The curves are shown
for $\lambda$ ranging from 350~to 865~nm.
Because these model planets are symmetric with respect 
to the reference plane, their polarized flux $U$ 
(see Eq.~\ref{eq_stokes}) is zero at all phase angles. 
Also included in Fig.~\ref{fig_HHplanets}, are the phase curves 
of a planet with the same rough Fresnel reflecting interface, 
but without an atmosphere and without an ocean (on this planet,
the Fresnel interface is thus
bounded below by a black surface). While such a planet is physically
impossible (without atmosphere, there would be no liquid water),
its curves are included for comparison.
Because we assume a wavelength independent refractive index for the
Fresnel interface and because this planet has no atmosphere, its 
phase curves are wavelength independent.

The phase curves for the black surface planet (with atmosphere) are as expected
\cite[see e.g.][]{buenzli2009grid,Stam08}:
with increasing $\lambda$, the contribution of the gaseous atmosphere 
decreases because of the $\sim\lambda^{-4}$ dependence of the 
Rayleigh scattering cross-section, hence fluxes $F$ and $|Q|$ 
decrease with $\lambda$ at all phase angles.
Degree of polarization $P$ increases with $\lambda$, because of the 
decreasing amount of multiple scattering, which usually lowers $P$,
and the non-reflecting, black surface.
The maximum $P$ occurs around $\alpha=90^\circ$ for all $\lambda$
($\alpha$ where $P$ is maximum increases slightly with $\lambda$). 
With increasing surface albedo, the location of this maximum would shift to 
much larger phase angles while the maximum $P$ would decrease
(see e.g.\ Fig.~4 of \citet{Stam08}).

Up to $\alpha \sim 80^\circ$, the flux phase curves 
for the ocean planet are very similar to those of the black surface planet,
although the ocean planet is somewhat brighter, especially at shorter 
wavelengths, where the ocean is a more efficient reflector.
In particular, at 443~nm, the ocean planet's geometric albedo is more
than 20\% larger than that of the black surface planet.
Retrieving such small geometric albedo differences from observational 
data would require knowing the planetary radius to within a few percent.

With increasing $\alpha$, the differences between the black surface and the
ocean planet increase. To start with, at the longer wavelengths,
where the atmosphere scatters less, the Fresnel reflection significantly
brightens up the planet. This can also be seen from the phase curves of
the planet with the Fresnel interface but without the atmosphere.
Also, the ocean planet's flux curves for 
the different wavelengths cross each other at $\alpha \sim 134^\circ$.
Indeed, while at smaller phase angles, the ocean planet is blue,
around 134$^\circ$ it would appear white. With further increasing
$\alpha$, the ocean planet would turn red, before darkening completely 
when $\alpha > 160^\circ$. 
The ocean planet's polarized flux $Q$ shows a similar crossing of the
phase curves, and hence a color change from blue, through white,
to red in polarization, except the crossing happens at a smaller phase 
angle: at about $123^\circ$. 

This color change of the total and polarized fluxes with phase 
angle could be used to identify an ocean on an exoplanet,
as the black surface planet does not show this color change with phase angle.
Note that a color change in the \textit{total} reflected 
flux $F$ only, thus without a detection of a color change in the polarized
flux, would be ambiguous because a color change in the total flux
can also occur in the presence of clouds (see Sect.~\ref{sect_HIplanets}) 
or for planets without an ocean and with a non-zero surface 
albedo\footnote{The total 
flux, $F$, phase curves of planets with cloud-free atmospheres and a surface 
albedo higher than 0.8 show two phase angles where the phase curves 
for different wavelengths slightly cross ('slightly' because the curves 
are almost parallel at those crossings).}. Because the black surface planet does not show a crossing in $Q$ in the presence of clouds (Fig.~\ref{fig_HH_clouds}), 
a color change in $Q$ would indicate an exo-ocean. 
\citet{Zugger2010} emphasize that it will be difficult to  
distinguish between a liquid and a frozen water surface when
measuring total flux $F$, because of the small difference in 
refractive index (see Sect.~\ref{sect_surface_model}). 
Measuring a crossing in $Q$, however, would allow such a distinction
to be made, 
because Fresnel reflection by an icy surface strong
enough to allow confusion with a liquid water surface
will only occur when the ice is clean, i.e.\ not covered by rocks, snow 
or other rough materials. 

The color change can be attributed to the wavelength dependence of the
molecular scattering cross-section: with increasing $\alpha$,
the average atmospheric gaseous optical path-length encountered by 
the light that is received by the distant observer increases. While the 
light with short wavelengths (blue) is scattered in the 
higher atmospheric layers, the light with the long wavelengths (red)
will reach the surface and is reflected by the ocean, back towards
space and the observer. Without the gaseous atmosphere, all colors of 
incident light would be reflected by the ocean at the
larger phase angles, as can be seen from the curves of the planet
without atmosphere but with a Fresnel reflecting interface.
Such an atmosphere-free planet would thus appear white at the 
largest phase angles.
The influence of the atmospheric optical thickness (i.e.\ the 
surface pressure) on the color change of a planet 
will be discussed in Sect.~\ref{sect_surfacepressure}.

We assume the incident starlight to be white: a strong
wavelength dependent stellar spectrum would influence the planet's actual
color in total and polarized flux. On the other hand,
in the case of real observations, the planet's fluxes can
be normalized to the spectrum of the incident starlight to retrieve
the normalized planetary color shown in our figures.

\begin{figure*}[ht!]
\begin{center}
\includegraphics[width=0.48\textwidth]{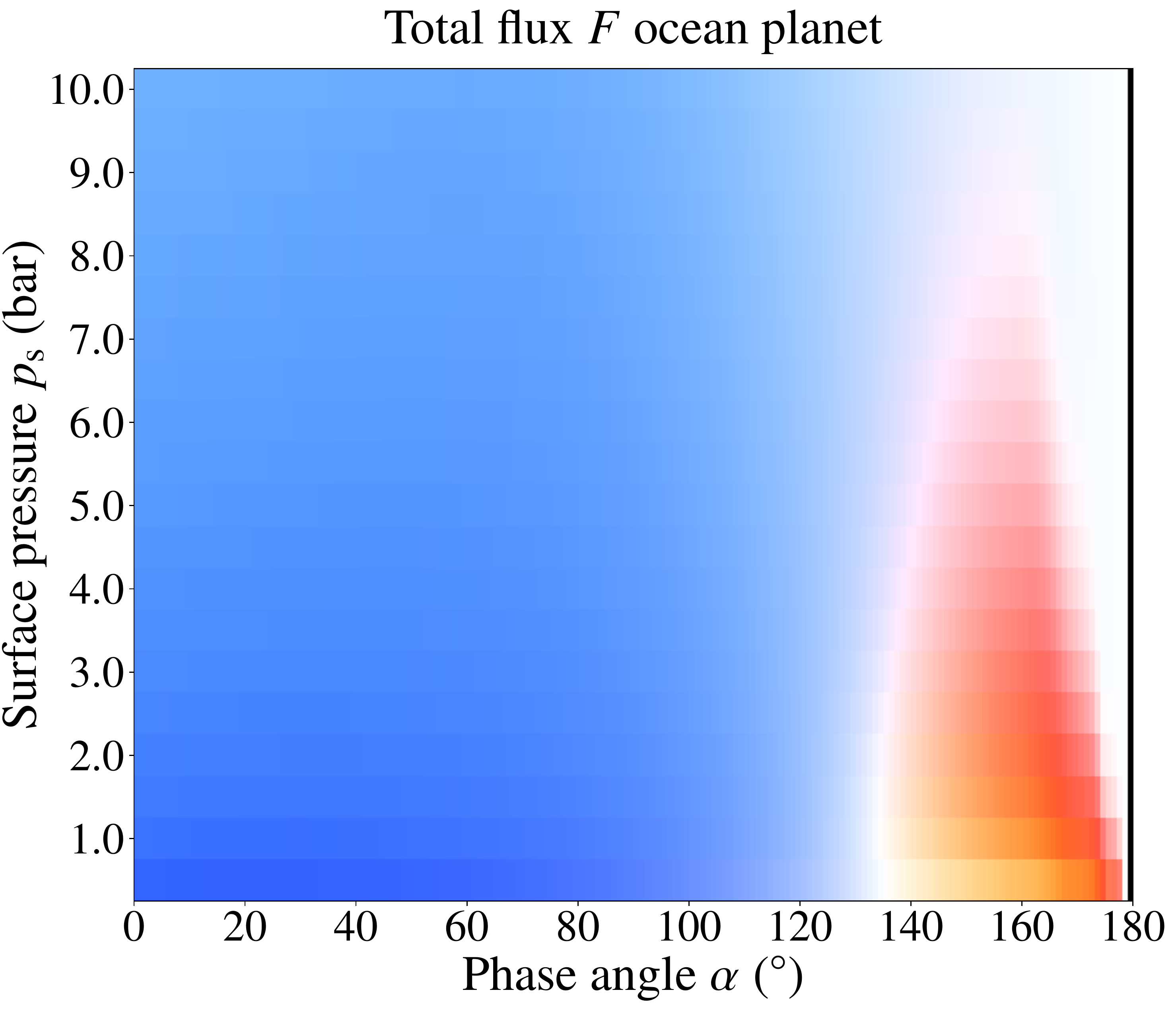}
\label{fig:colorpressuresI}
\includegraphics[width=0.48\textwidth]{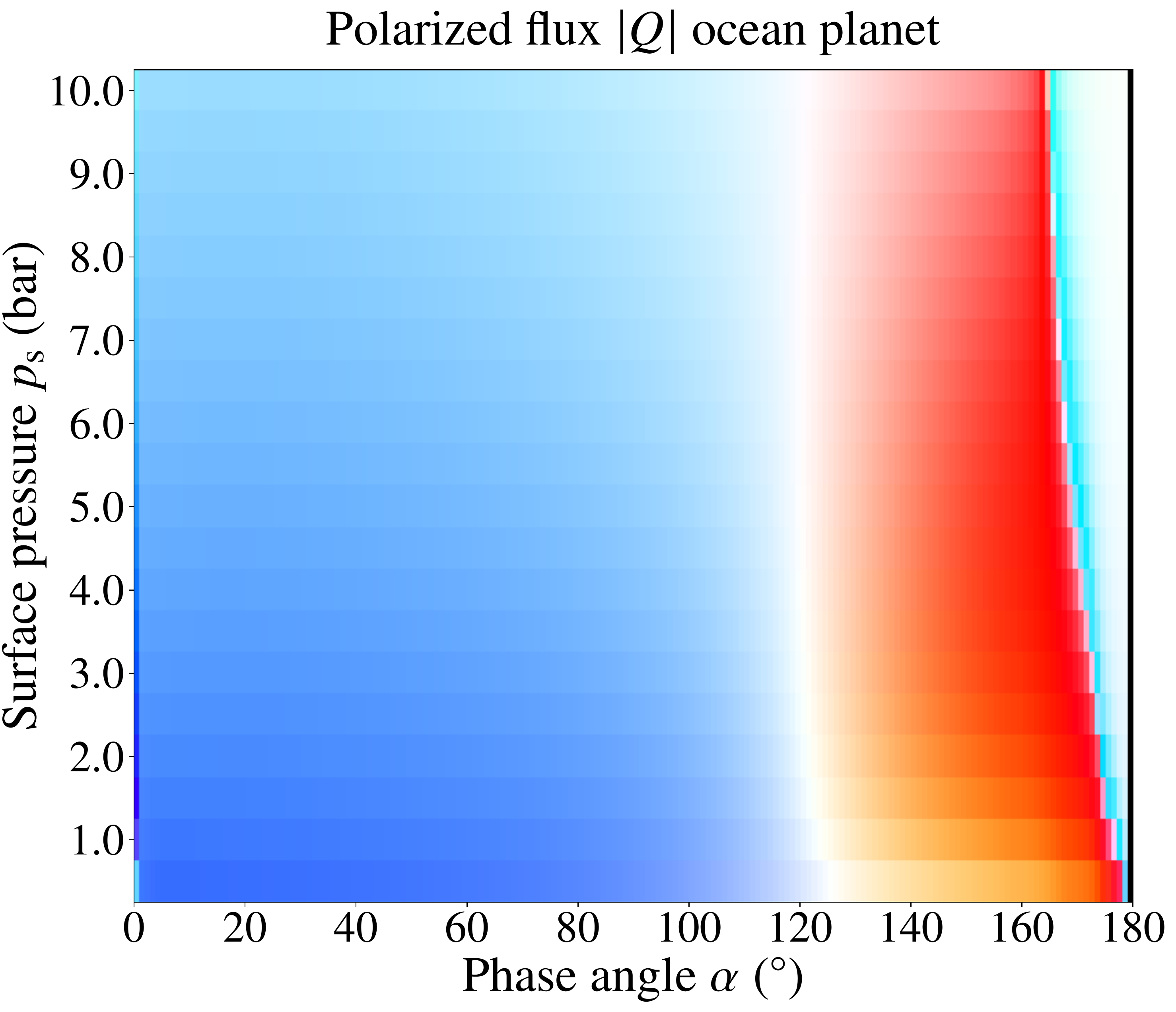}
\label{fig:colorpressuresQ}
\includegraphics[width=0.48\textwidth]{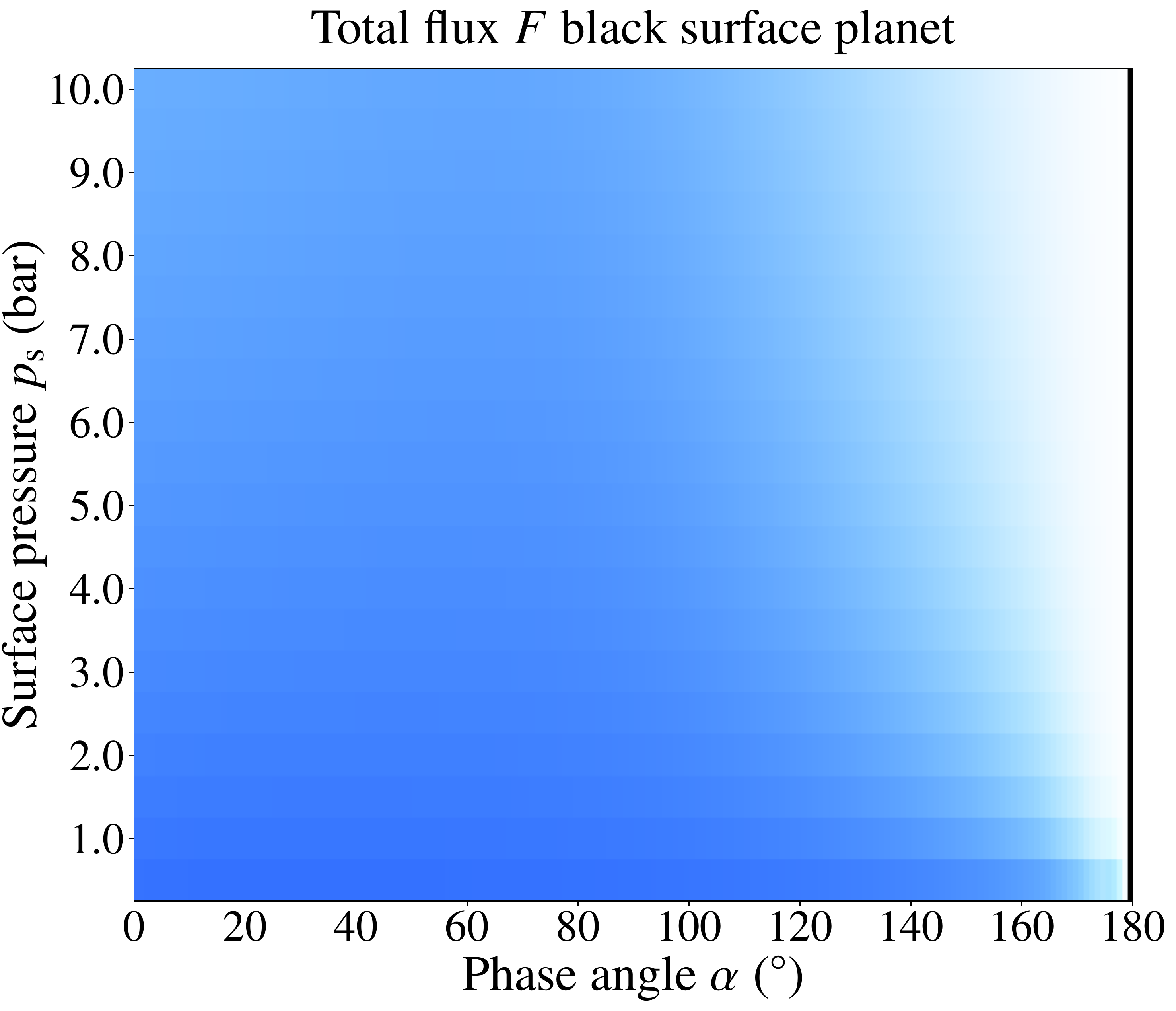}
\label{fig:colorpressuresIlamb}
\includegraphics[width=0.48\textwidth]{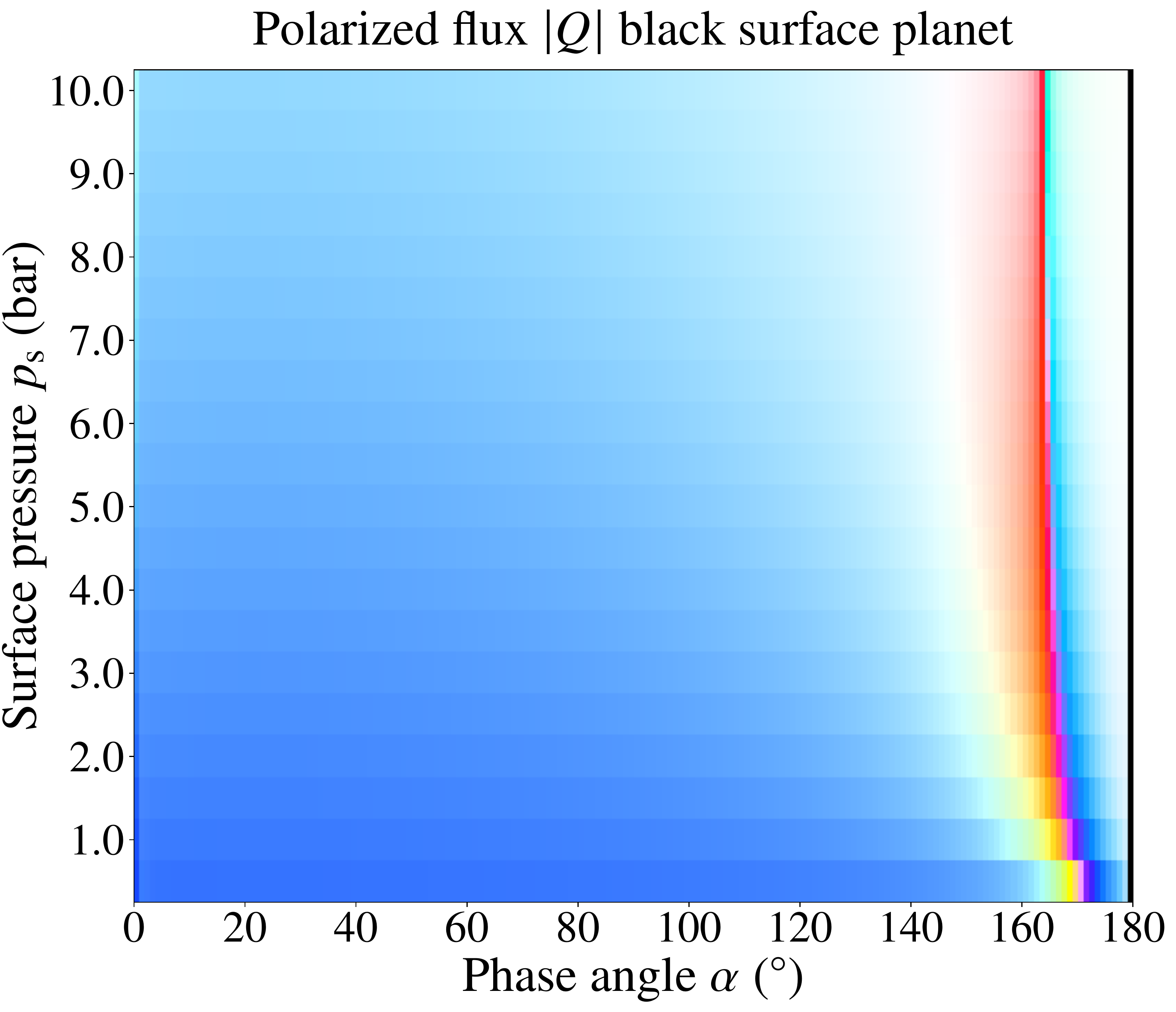}
\label{fig:colorpressuresQlamb}
\end{center}
\vspace*{-6mm}
\caption{RGB colors of the (disk-integrated) reflected $F$ 
         (left column) and $|Q|$ (right column) of the cloud-free 
         ocean planet (top row) and the cloud-free black surface planet 
         (bottom row) for 
         surface pressures ranging from 0.5 to 10~bar in steps of 0.5 bar. 
         Like in Fig.~\ref{fig_HHplanets}, the wind speed is 7~m/s.
         The RGB colors are computed using weighted additive color mixing 
         of the fluxes at $\lambda=443$~nm (blue), 550~nm (green),
         and 670~nm (red), such that when these fluxes are equal, the
         color is white.}
\label{fig_RGB_surfacepressure}         
\end{figure*}

Up to $\alpha \sim 80^\circ$, the phase curves of the degree of 
polarization $P$ for the ocean planet are very similar to those of the 
black surface planet, although the ocean planet's $P$ is slightly lower at 
the same wavelength and phase angle.
At the shorter wavelengths, the maximum $P$ of the ocean planet is 
smaller than that of the black surface planet due to the increased amount of 
multiple scattered, thus low polarized light in the ocean planet's 
reflected signal. 
With increasing $\lambda$, the influence of the atmosphere on the 
planet signal decreases and that of the Fresnel reflecting surface 
increases, shifting the maximum $P$ to larger phase angles. 
The maximum $P$ shifts towards twice the Brewster angle, 
thus towards $2 \arctan{n_2/n_1}= 106^\circ$ for our water ocean. 
This is indeed the phase angle of maximum $P$ for the ocean planets 
without an atmosphere, as can be seen in Fig.~\ref{fig_HHplanets},
and as was also shown by \citet{Zugger2010}.

The degree of polarization $P$ is of course a relative measure and usually 
not associated with colors, although the color of polarization has also been
used by, for example, \citet{2019Sterzik} in their analysis
of polarized Earth-shine observations. 
Translating the phase curves of $P$ as shown in Fig.~\ref{fig_HHplanets} 
into colors, both the black surface and the ocean planet would be white at smaller
phase angles, and red at intermediate angles. With increasing phase
angle, the ocean planet would remain red in $P$,
while the black surface planet would return to its white color.

\subsubsection{Influence of the surface pressure $p_{\rm s}$}
\label{sect_surfacepressure}

\begin{figure*}[ht!]
	\includegraphics[width=0.33\textwidth]{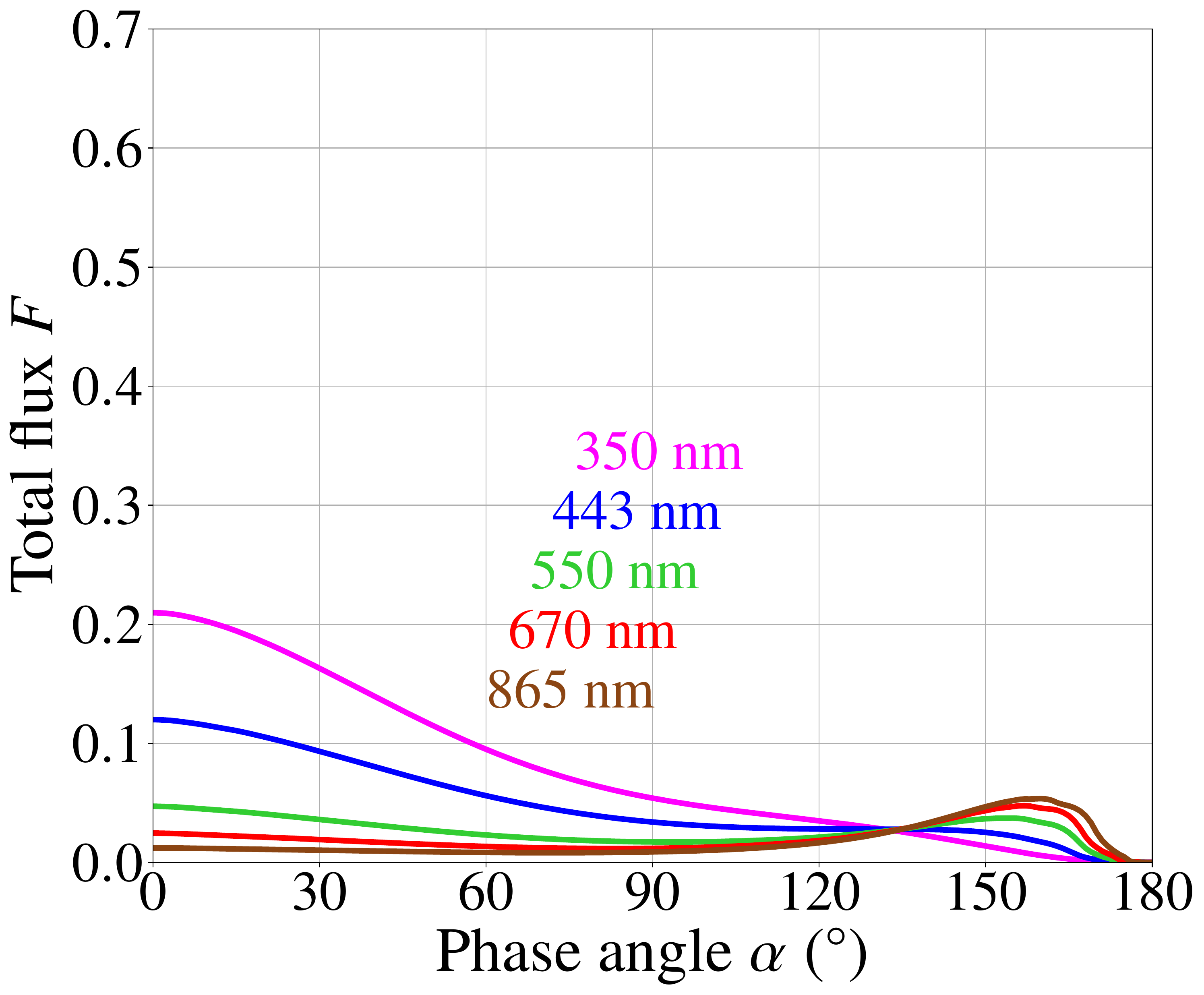}
	\includegraphics[width=0.33\textwidth]{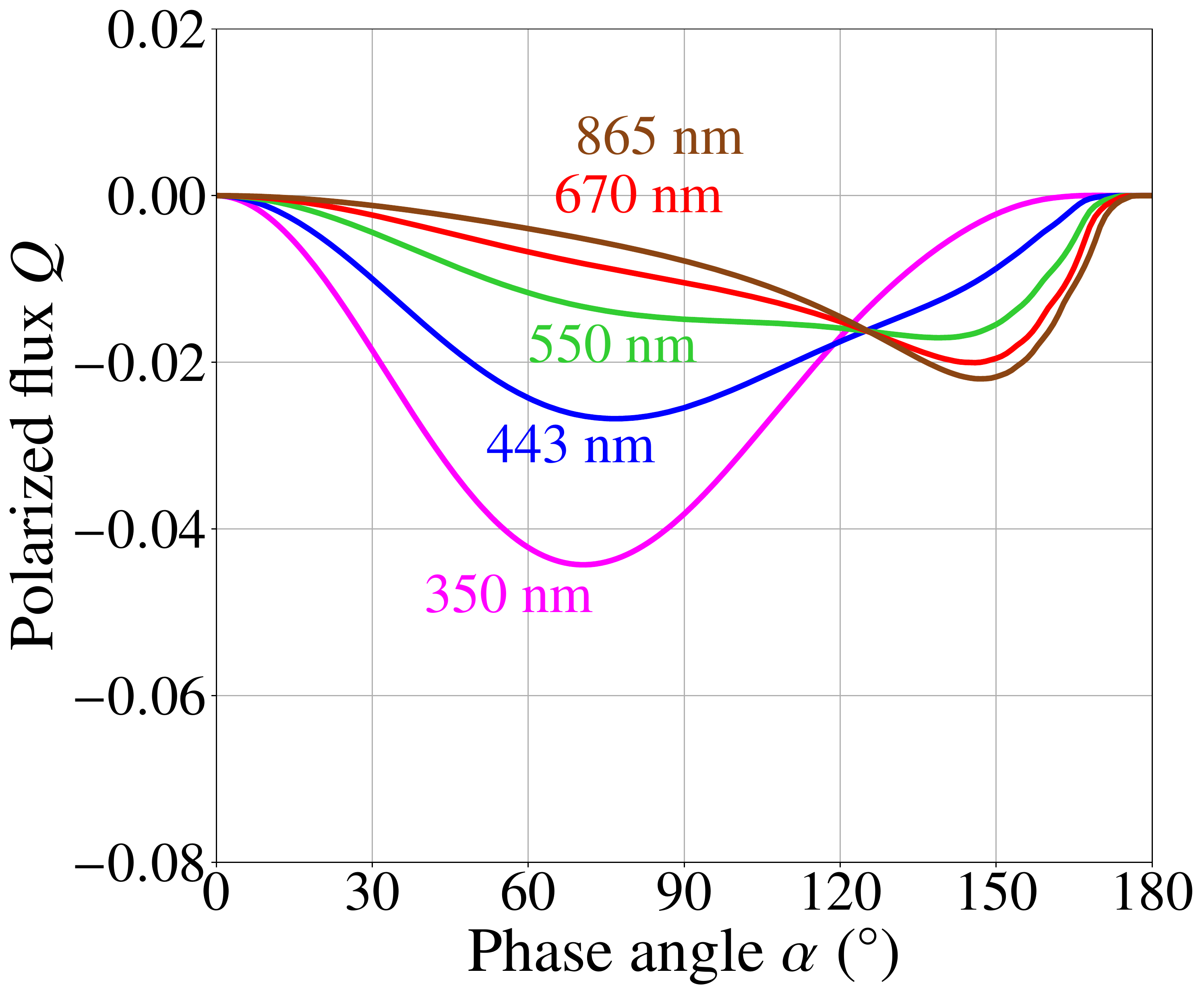}
	\includegraphics[width=0.33\textwidth]{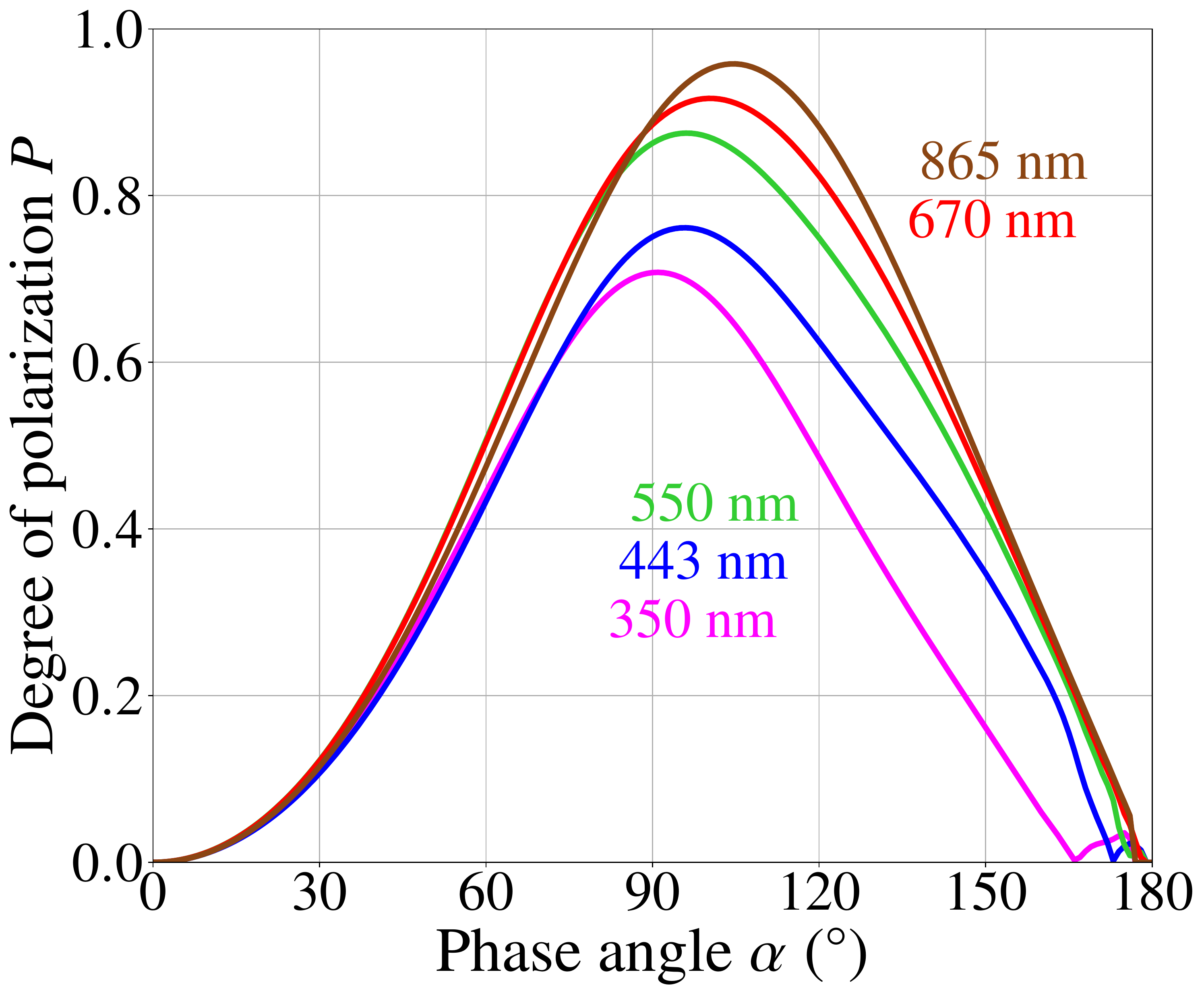}
	
	\includegraphics[width=0.33\textwidth]{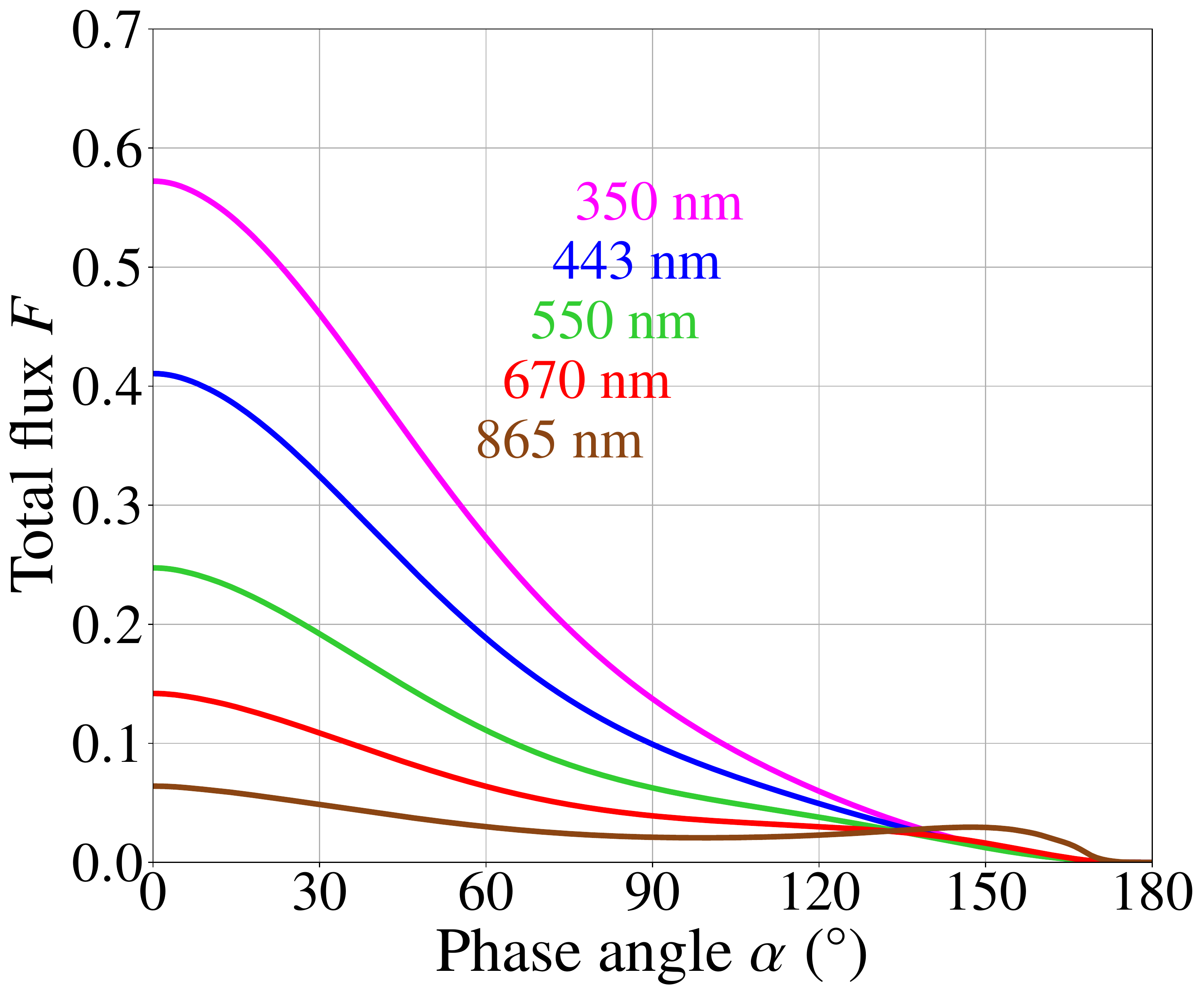}
	\includegraphics[width=0.33\textwidth]{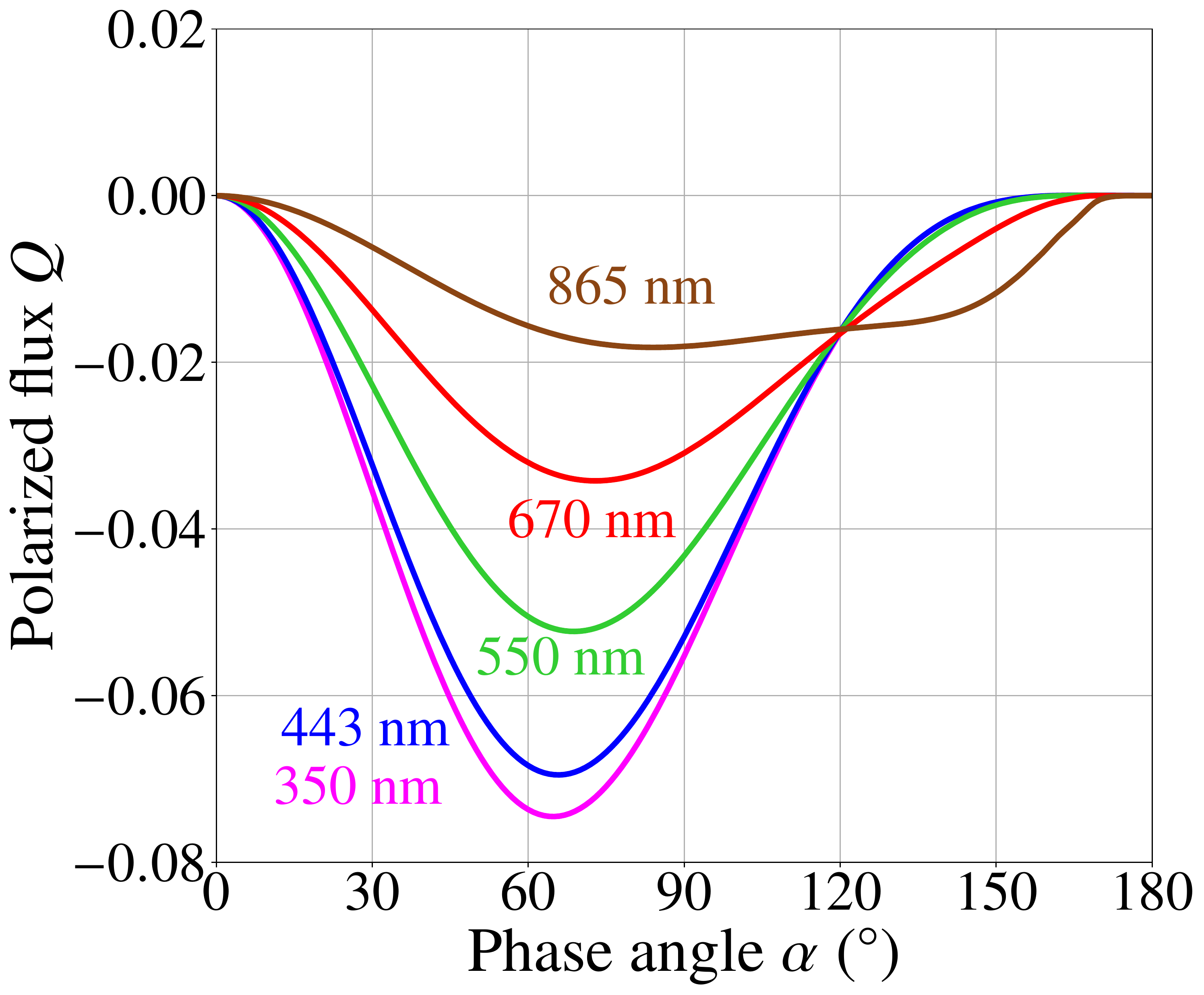}
	\includegraphics[width=0.33\textwidth]{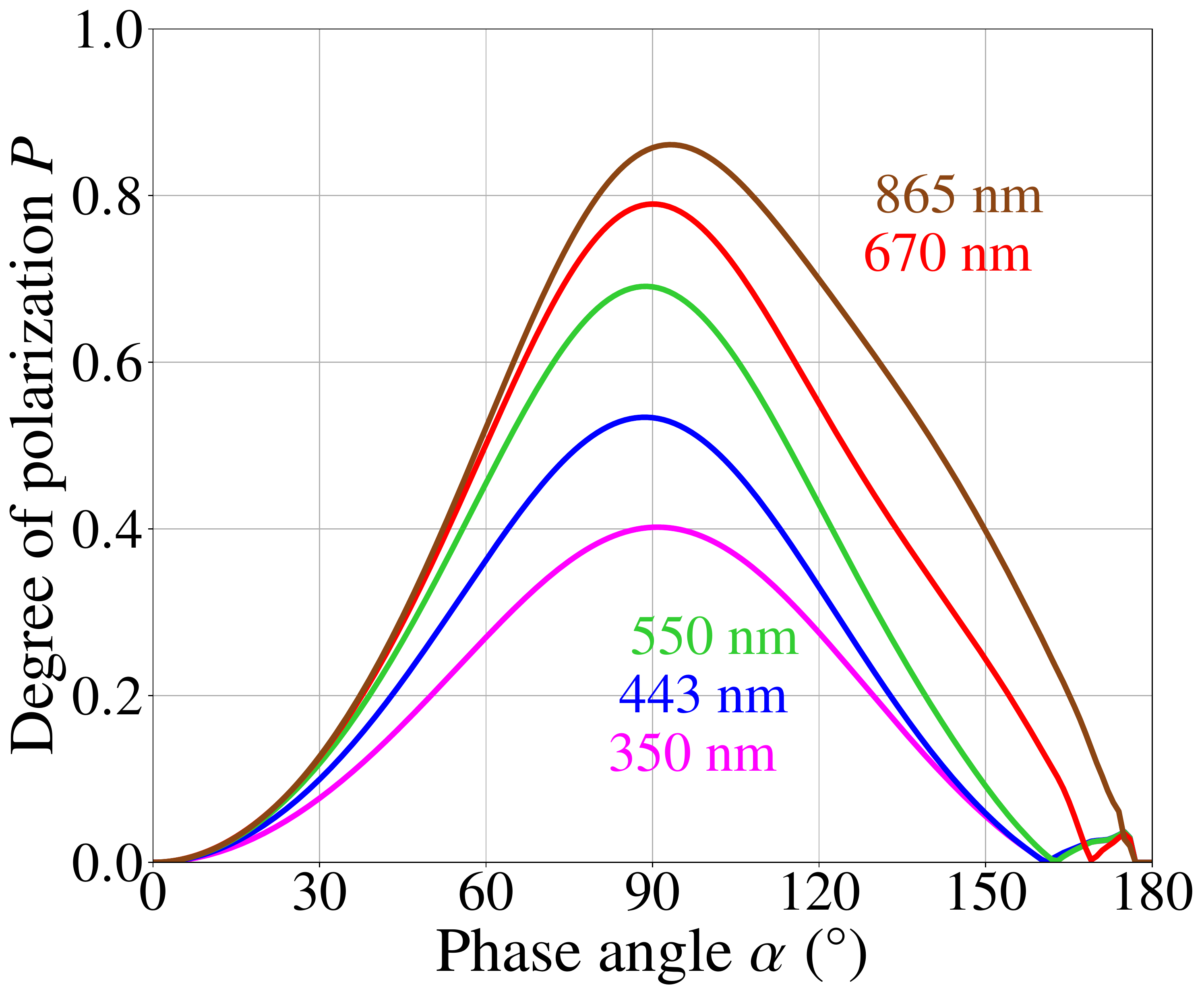}
	
	\includegraphics[width=0.33\textwidth]{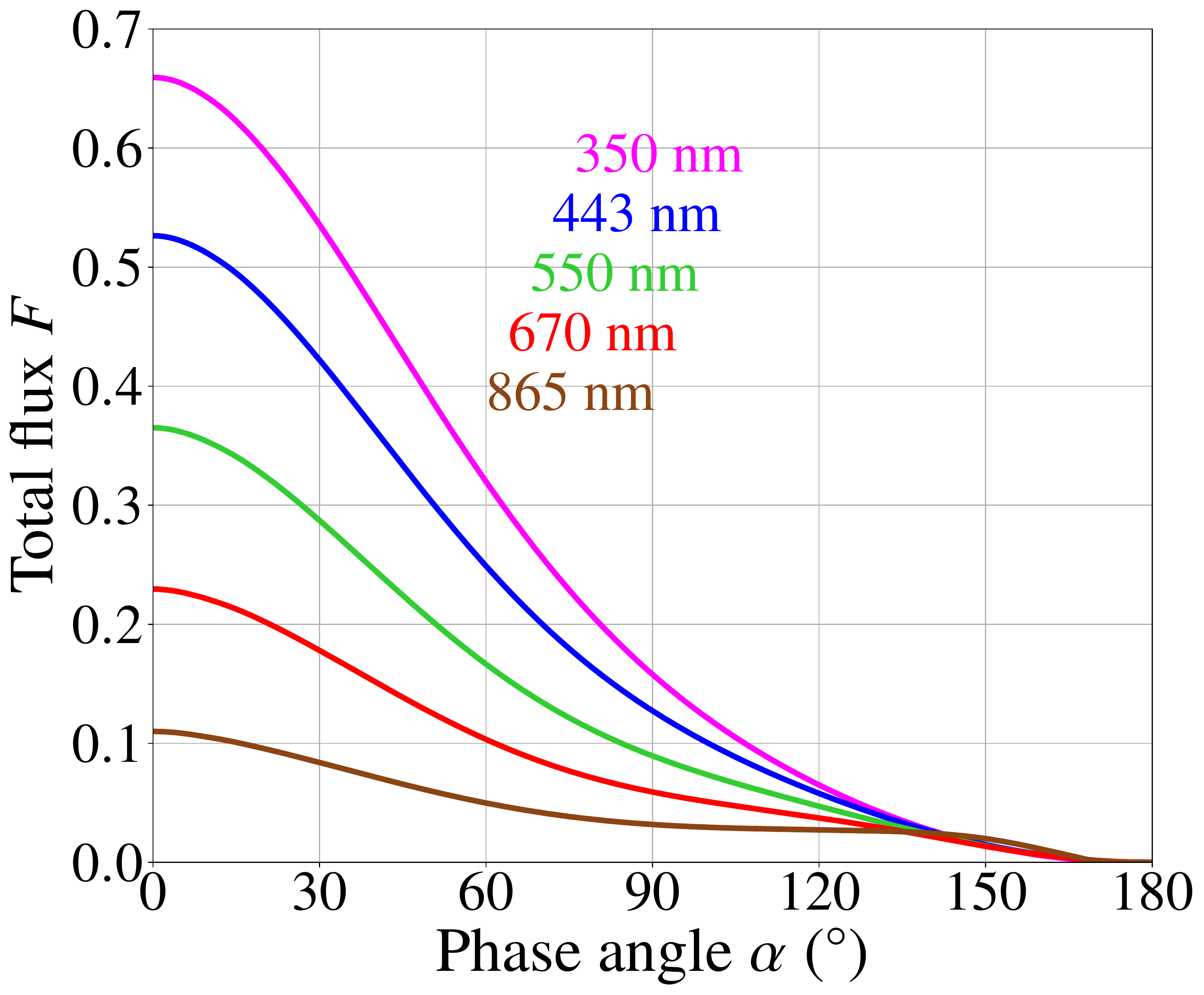}
	\includegraphics[width=0.33\textwidth]{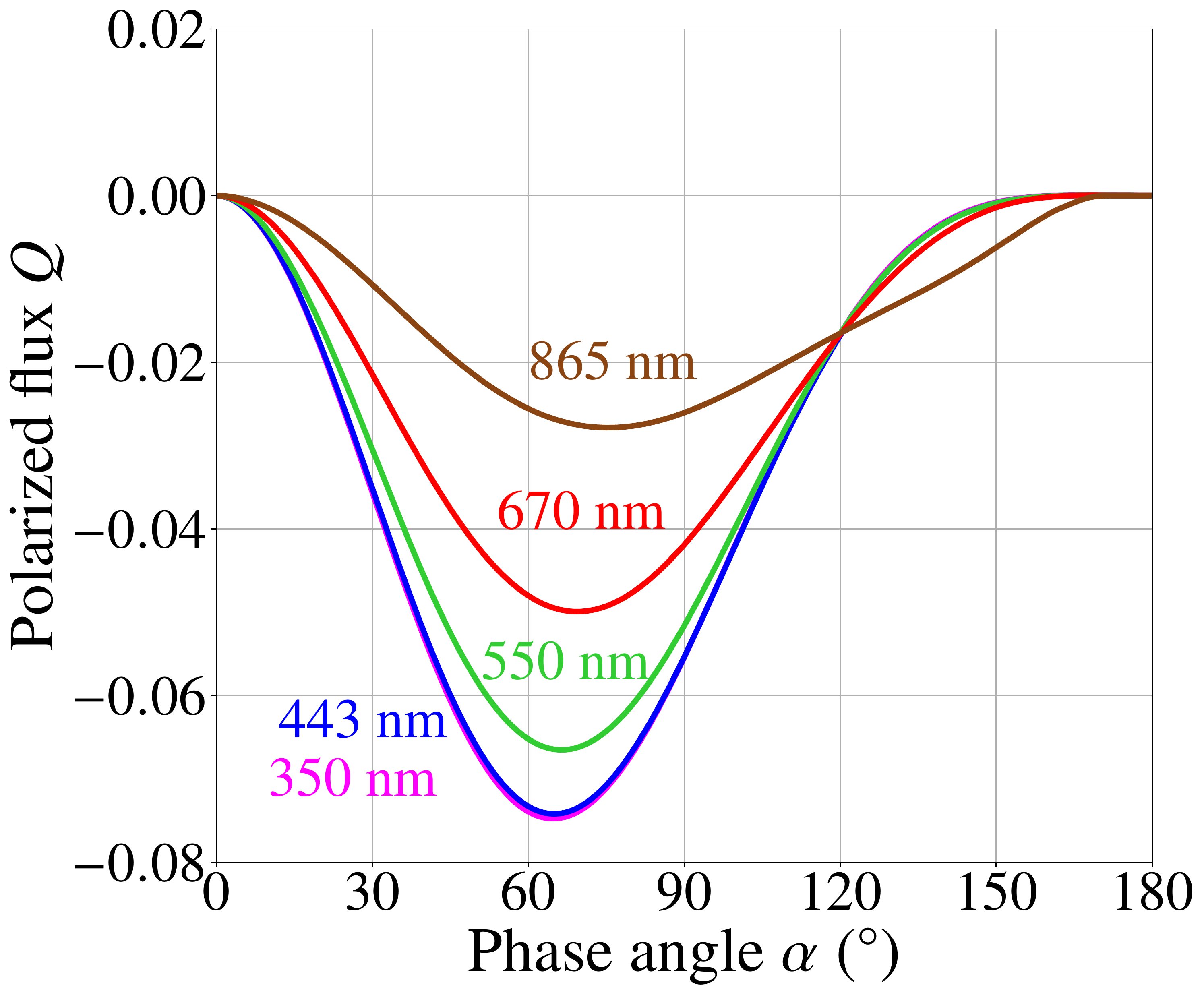}
	\includegraphics[width=0.33\textwidth]{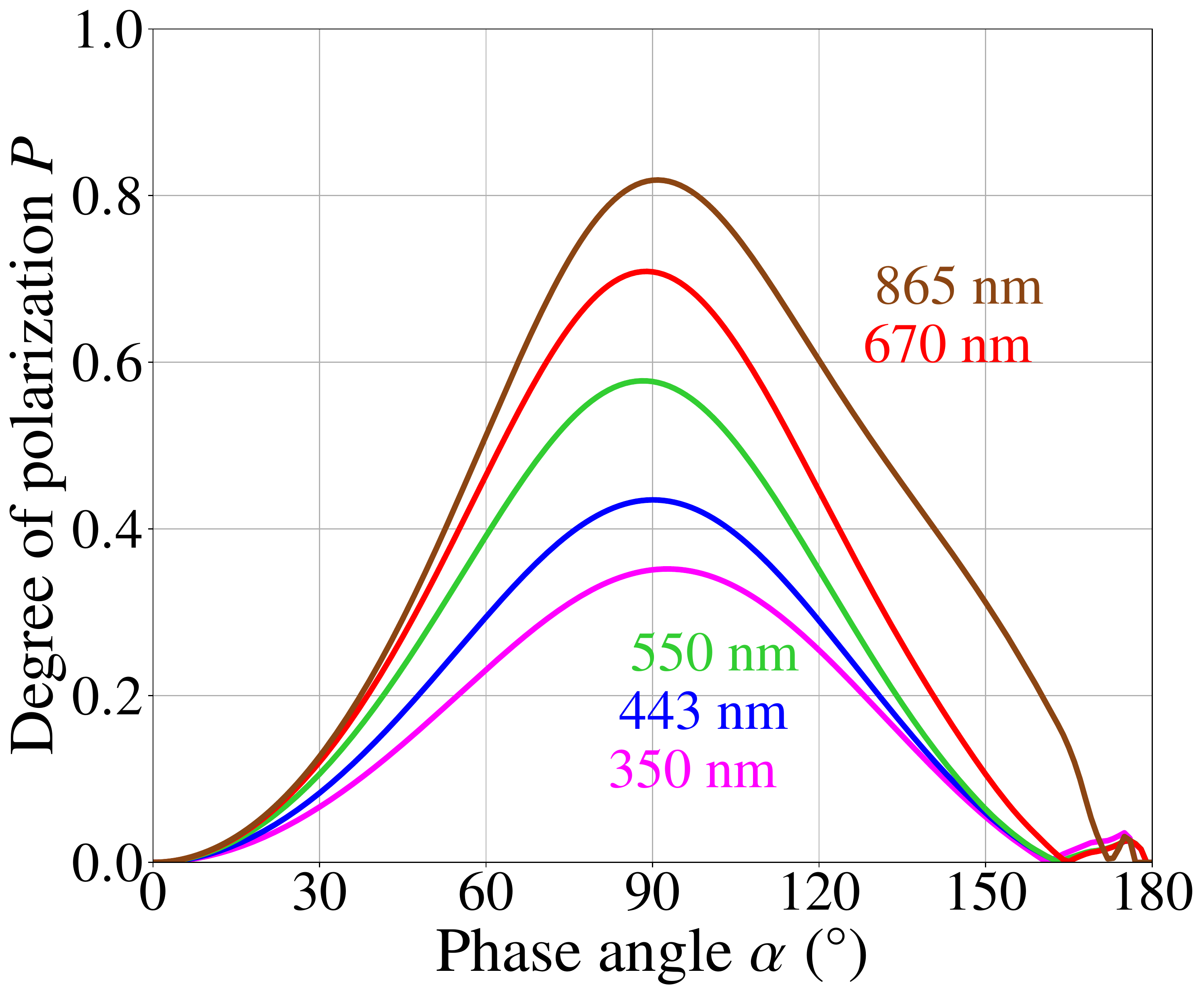}
\caption{Similar to Fig.~\ref{fig_HHplanets}, but for the ocean 
         planet, and for surface pressures equal to 0.5 bar (top row), 
         5 bar (middle row), and 10 bar (bottom row).}
\label{fig_HHsurfacepressure}
\end{figure*}

To have a better look at the color changing ocean planets, 
Fig.~\ref{fig_RGB_surfacepressure} shows the phase curves of $F$  
and $Q$ reflected by the ocean planet and the black surface planet for 
surface pressures ranging from 0.5 to 10~bar. The curves 
are shown in a Red-Green-Blue (RGB) color-scheme, in which the 
fluxes at various wavelengths are combined to mimic the actual 
perceived color. Figure~\ref{fig_HHsurfacepressure} shows a few 
cross-sections of Fig.~\ref{fig_RGB_surfacepressure} for $F$, $Q$, 
and $P$ (the latter is not shown in Fig.~\ref{fig_RGB_surfacepressure}) 
for surface pressures of 0.5, 5.0, and 10.0~bar. Note that the boiling 
temperatures of water at these pressures are, respectively, 
354~K (81~$^\circ$C), 425~K (152~$^\circ$C), and 452~K
(179~$^\circ$C).

Figure~\ref{fig_RGB_surfacepressure} clearly shows that a cloud-free, 
black surface planet will appear mostly blue in $F$ and $Q$ across this 
surface pressure range. Only at the largest phase angles and surface 
pressures, it will appear white.
The figure also shows the color change of an ocean planet
from blue, through white, to red, and back to white, with increasing
$\alpha$, in both $F$ and $Q$. 

The difference between the total fluxes $F$ reflected by the black surface planet 
and the ocean planet starts to disappear with increasing surface pressure, because
with increasing pressure, the contribution of the surface
reflection to the planetary signal decreases, as the scattering within 
the atmosphere increases and less light reaches the surface and 
subsequently the top of the atmosphere and the observer. 
The color change of the ocean planet thus also starts to disappear:
for surface pressure higher than about 8~bar, the ocean planet
will no longer be obviously red in any phase angle range
and the presence of liquid surface water would not be recognizable in $F$.
For surface pressures decreasing below 0.5 bar (not shown), 
the planet will become whitish at all phase angles because the influence 
of the atmosphere will disappear completely, while the Fresnel reflection is
wavelength independent.

Interestingly, in polarized flux $Q$, the ocean planet's color change 
from blue, through white, to red, and back to white with increasing $\alpha$,
remains strong,
and is even still present at a surface pressure $p_{\rm s}$ of 10~bar. 
At such high surface pressures, the best approach to detecting an ocean
appears to be using a combination of a short wavelength 
($\lambda \sim 550$~nm) and a long wavelength ($\lambda \sim 865$~nm)
filter, because the strength of the color change (i.e. not the relative 
but the absolute polarized fluxes) decreases with increasing $p_{\rm s}$. 
This can be seen in Fig.~\ref{fig_HHsurfacepressure}: with 
increasing pressure, $Q$ at intermediate 
phase angles increases for all $\lambda$. At phase angles beyond the 
crossing, $Q$ decreases because of the increase of the atmospheric path 
length, with $Q$ at longest (reddest) wavelengths remaining the largest,
because the red light is still more likely to reach the surface than the
blue light.
The phase angle where the crossing takes place is approximately conserved 
for surface pressures above 2 bar (see upper right plot in Fig. \ref{fig_RGB_surfacepressure}). In addition, the phase angle 
range across which the ocean planet would be red in $Q$ is much wider 
than the range across which the planet would be red in $F$. 
Indeed, with the color transition in $Q$ occurring around 
$\alpha=123^\circ$, the angular distance between the planet and the star would be larger than 
at $\alpha=134^\circ$, where the color transition in $F$ occurs, i.e. 0.83'' versus 0.72'' for a planet orbiting its star at 1~AU, with a 1~pc distance between 
the observer and the star.  
Note that $Q$ of the black surface planet also shows a color change 
from blue to red and back to blue/white at large phase angles,
but the actual values of $Q$ are very small here, for all
wavelengths, as can also be seen in Fig.~\ref{fig_HHplanets}.

\citet{Zugger2010} suggested that the shift of the maximum $P$ towards
larger $\alpha$ due to the Fresnel reflection 
(see Fig.~\ref{fig_HHplanets}) could be used for identifying an 
exo-ocean. However, as can be seen in Fig.~\ref{fig_HHsurfacepressure}, 
this shift of the maximum $P$ will depend on the surface pressure.
Indeed, with increasing pressure, the maximum $P$ of an ocean planet 
shifts towards $\alpha=90^\circ$. Thus, a shift of the maximum $P$ 
towards larger $\alpha$'s would reveal the presence of an exo-ocean and indicate, necessarily, a small surface pressure $p_{\rm s}$, as with a high surface pressure, there would be no observable phase 
angle shift of the maximum $P$. \vfill

\subsubsection{Influence of the wind speed $v$}
\label{sect_windspeed}

\begin{figure*}[ht]
	\includegraphics[width=0.33\textwidth]{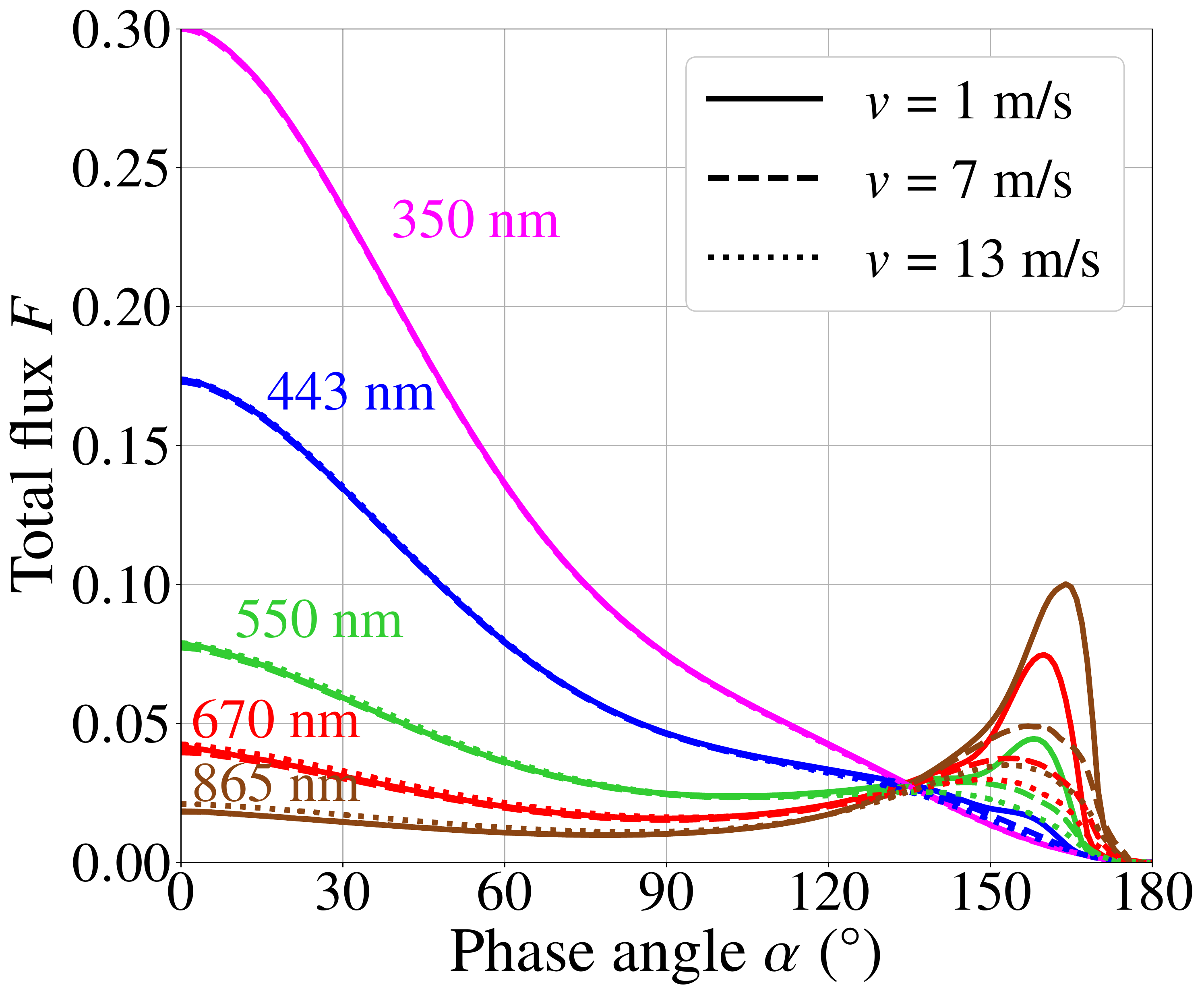}
	\includegraphics[width=0.33\textwidth]{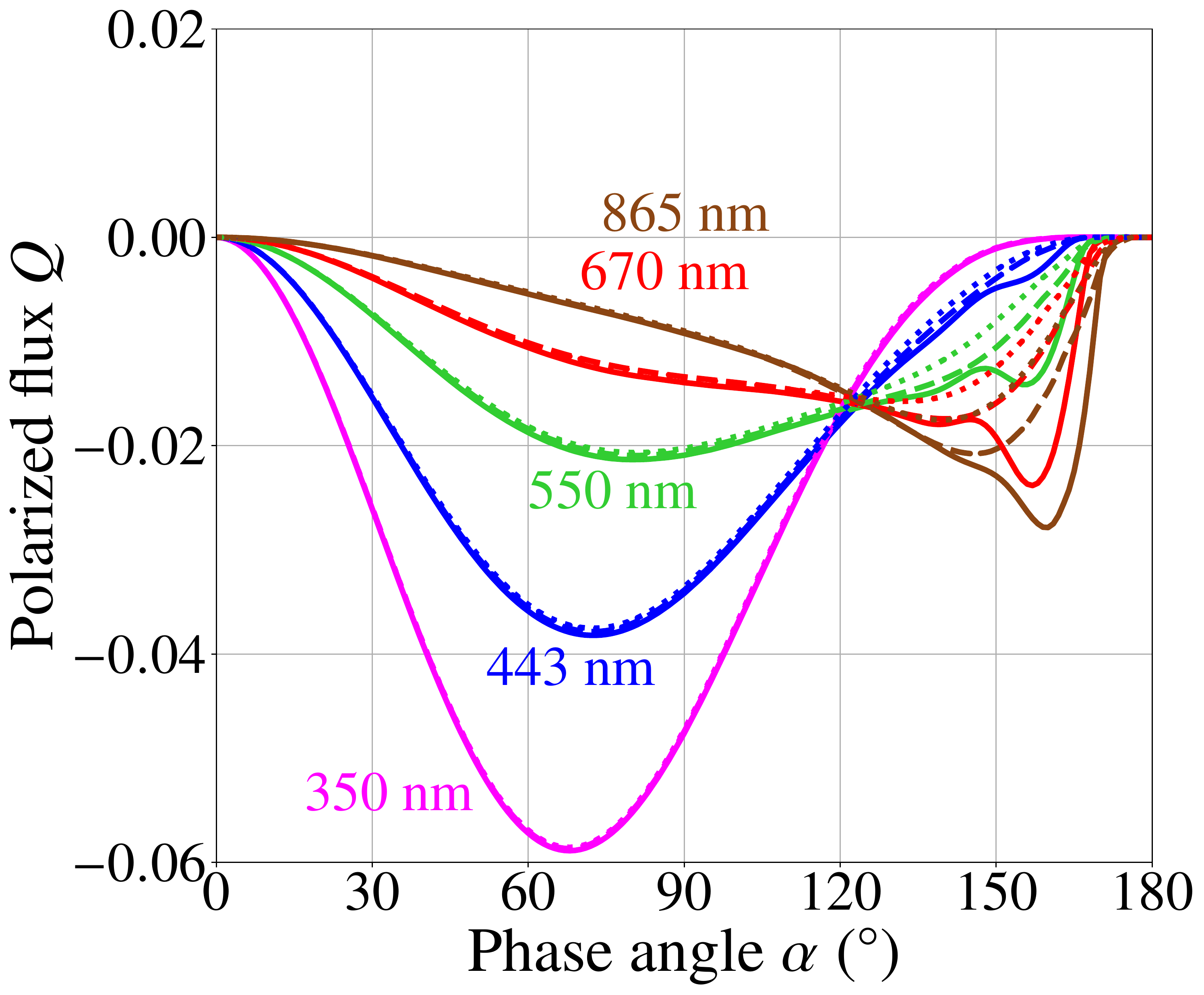}
	\includegraphics[width=0.33\textwidth]{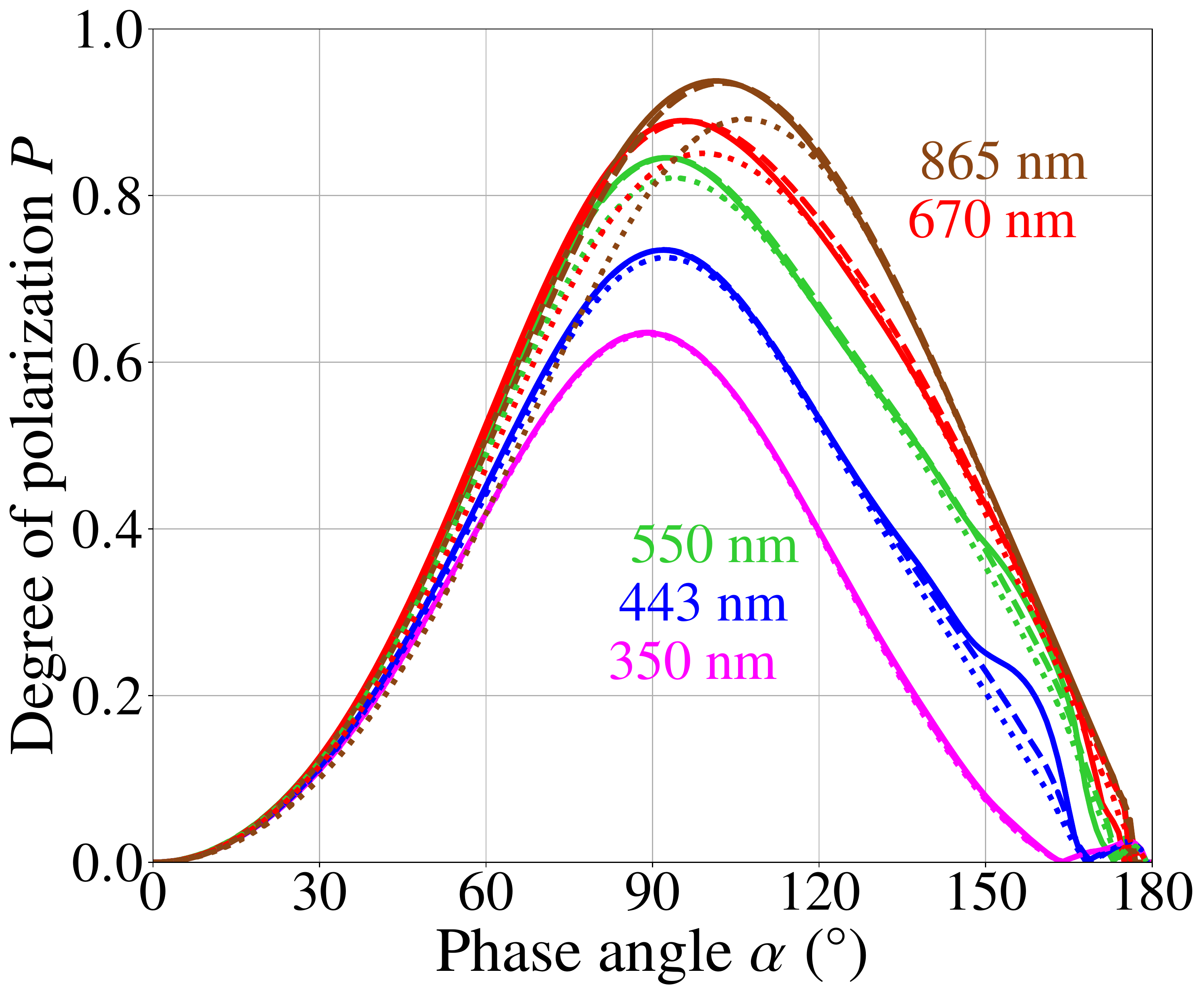}
\caption{Similar to Fig.~\ref{fig_HHplanets}, but for the
         ocean planet and different wind speeds.
         The wind speeds and associated white cap percentages are:
         1~m/s and 0.0003\% (solid lines),
         7~m/s and 0.28\% (dashed lines) (cf. Fig.~\ref{fig_HHplanets}), 
         13~m/s and 2.46\% (dotted lines).}	
\label{fig_HHplanets_wind}         
\end{figure*}

Figure~\ref{fig_HHplanets_wind} shows the phase curves of the ocean planet 
for various wavelengths and three wind speeds $v$: 1~m/s, 7~m/s (our baseline
wind speed), and 13~m/s. Larger wind speeds create higher waves and thus 
a wider glint pattern on the planet (see also the disk-resolved planets in 
Fig.~\ref{fig_diskresolved}) and more white caps \citep[see][]{Monahan1980}.  
Higher waves, however, also increase extent of shadows. 

First, as can be seen in the figure, for $\lambda=350$~nm, the curves 
for $F$ and $Q$ do not show the glint for any $v$ because of the 
large atmospheric gas optical thickness. For $\lambda=865$~nm, the glint
feature is the sharpest because at long wavelengths, this
optical thickness is very small and thus most of the incident sunlight will
reach the surface and, after being reflected, space without being scattered
in the atmosphere.
The wind speed does not affect the  
phase curves of $F$ and $Q$, except slightly at the largest wavelengths, 
up to at least the crossing phase angles,
where the curves for the different $\lambda$'s coincide, i.e.\
$\alpha=134^\circ$ for $F$ and $\alpha=123^\circ$ for $Q$. Indeed, the 
crossing phase angles themselves are virtually independent of $v$. 
At phase angles beyond the crossing, $F$ and $Q$ do depend on $v$:
the strength (amplitude) of the glint in $F$ and $Q$ increases with
decreasing $v$ at every $\lambda$.
The main reason is that the smaller $v$, the smaller the variation in 
wave angles, and hence the less diffuse the reflection, 
and the contribution of shadows will be smaller. 

Figure~\ref{fig_HHplanets_wind} also shows $P$ of the ocean planet. 
For $\lambda > 350$~nm, the maximum $P$ decreases slightly with 
increasing wind speed $v$. 
With increasing $\lambda$, the location of the maximum also shifts 
to slightly larger phase angles (i.e.\ from 100$^\circ$ at 865~nm and 1 m/s, 
to 108$^\circ$ at 865~nm and 13 m/s).
The decrease and shift of the maximum $P$ are {\em not directly} due 
to the increase of the surface roughness with increasing $v$. 
Indeed, in the absence of foam and without an atmosphere (or with an 
optically thin atmosphere), the phase curve of $P$ is actually independent 
of $v$ 
(see Appendix~\ref{app_equations}) because the probability 
distribution function describing the wave facet inclinations 
\citep{CoxandMunk1954,cox1956slopes} influences $F$ and $Q$ in the same 
way, thus leaving $P$ undisturbed.
In the presence of an atmosphere, however,
light that has been reflected by the Fresnel interface can
subsequently be scattered in the atmosphere, and such scattering processes 
can influence $F$ and $Q$ differently, and thereby influence $P$.

Another factor that influences $P$ and in particular the decrease and 
shift of the maximum $P$ with increasing $v$ is the increasing 
surface coverage by reflecting foam. 
Indeed, the white cap percentages are $0.0003$\%, $0.28$\% and $2.46$\% 
for 1~m/s, 7~m/s, and 13~m/s, respectively. An increase in white caps 
slightly increases $F$ (although in Fig.~\ref{fig_HHplanets_wind}, the
influence of the ocean roughness on $F$ is much larger at the large 
phase angles), and only slightly decreases $Q$ because the white cap
reflection is assumed to be Lambertian. In Appendix~\ref{app_whitecaps},
we provide some more results, such as for different white cap albedos.


\subsubsection{Influence of the ocean color}
\label{sect_oceancolor}

In Fig.~\ref{fig_HHplanets}, we showed that the flux $F$ that is reflected by
the ocean planet is larger than that reflected by the black surface 
planet at the same phase angle,
in particular at small phase angles. As described in
Sect.~\ref{sect_atmosphere_surface_models}, the water body below 
the Fresnel interface is not black (the surface below the water body is). 
Here, we will show the influence of the water body's color on the 
reflected signals. 

\begin{figure*}[t!]
	\includegraphics[width=0.33\textwidth]{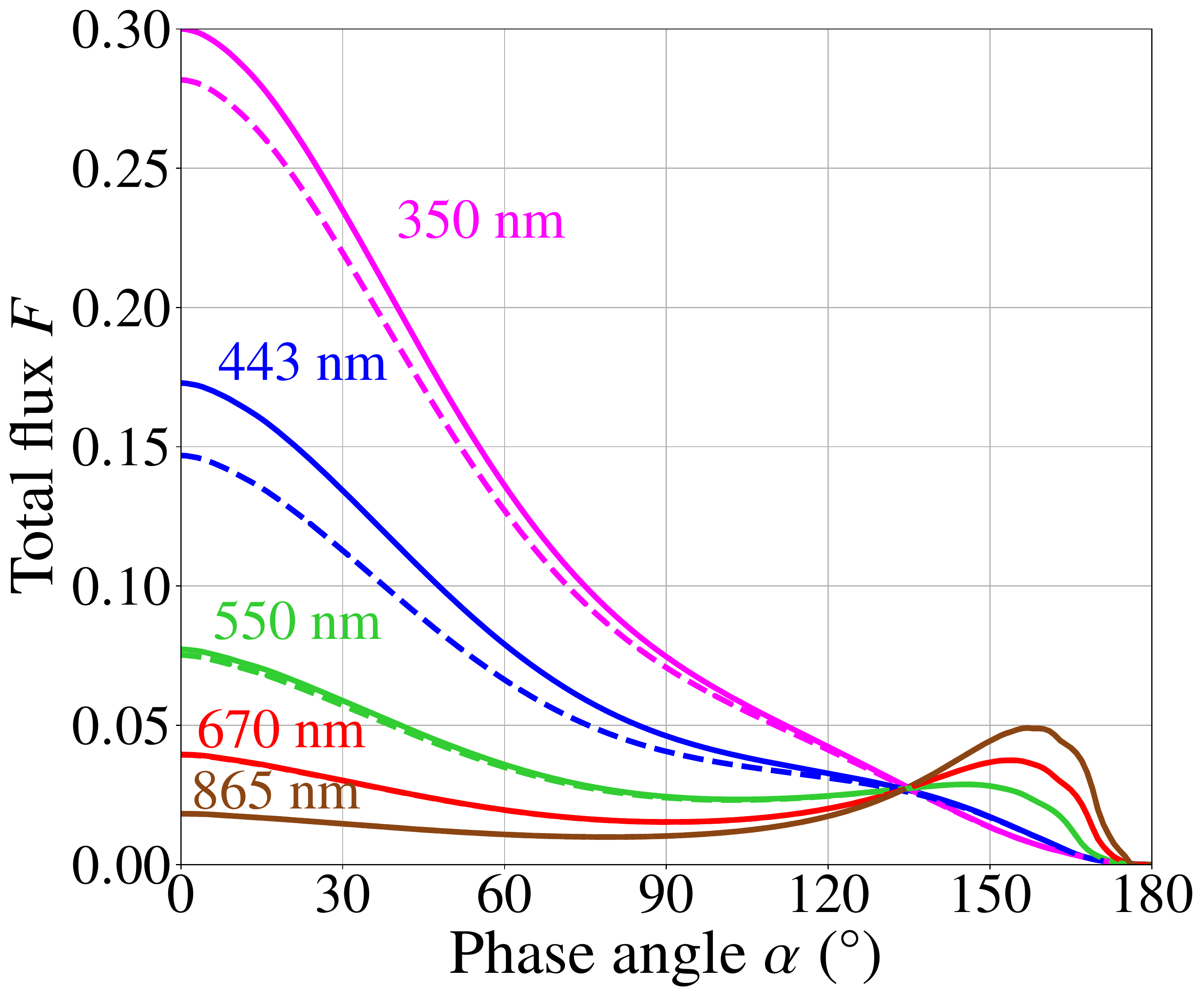}
	\includegraphics[width=0.33\textwidth]{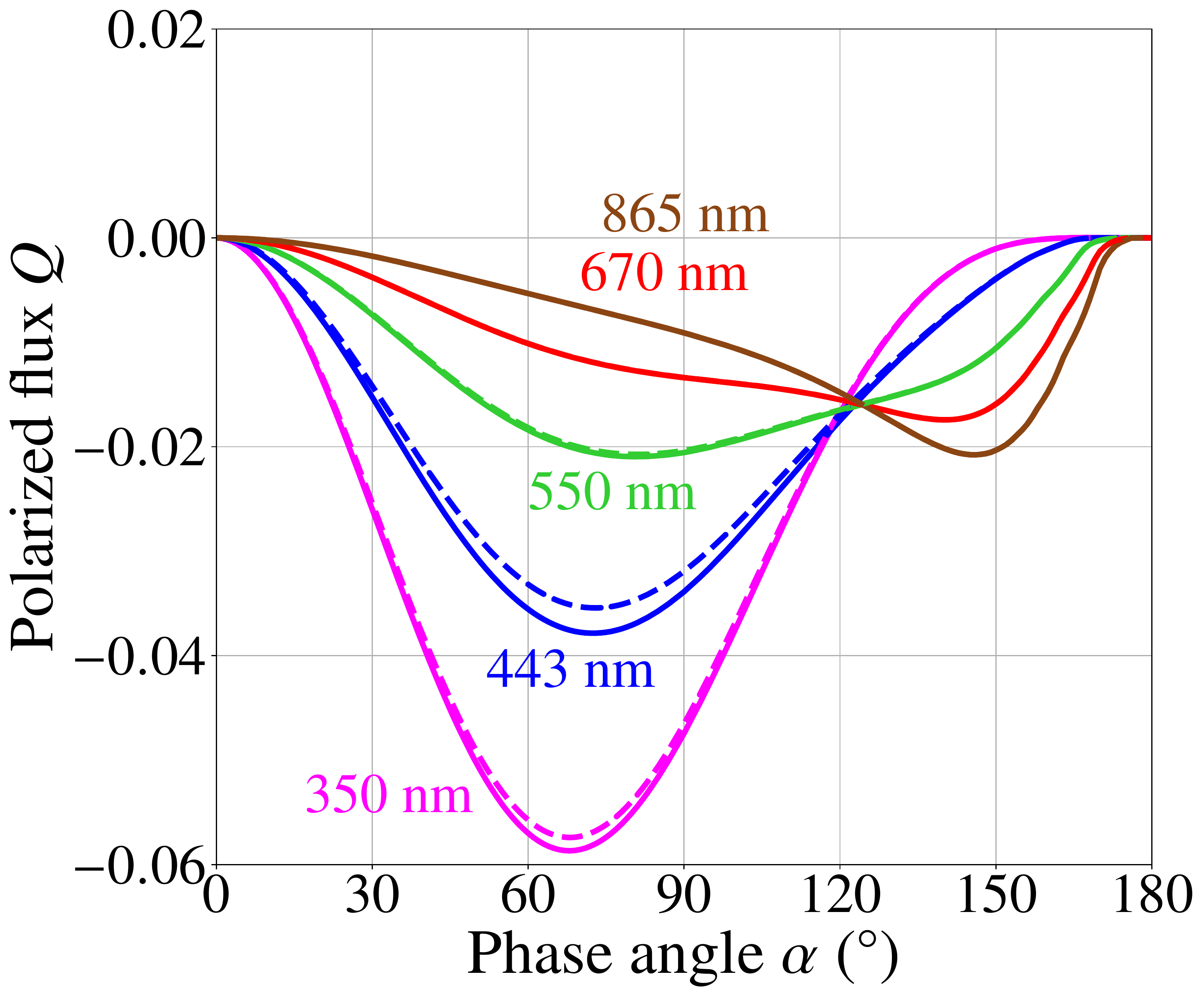}
	\includegraphics[width=0.33\textwidth]{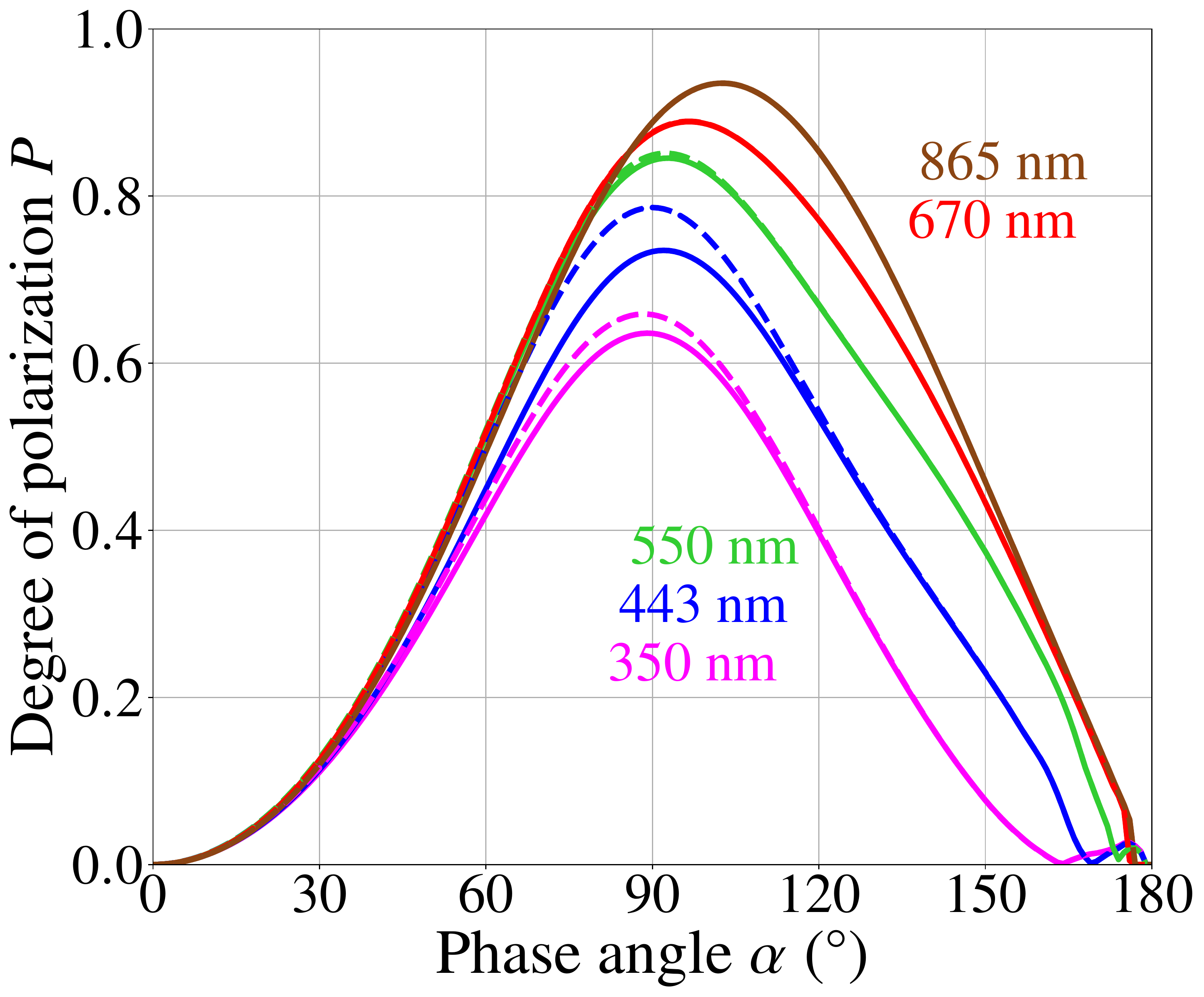}
\caption{Similar to Fig.~\ref{fig_HHplanets}, but for the ocean planet 
         (solid lines) and a planet with the same gaseous atmosphere and
         the same rough Fresnel interface, but {\em without} the water body 
         underneath (dashed lines). The wind speed $v$ is 7~m/s. }	
\label{fig_HHplanets_color}         
\end{figure*}

Figure~\ref{fig_HHplanets_color} shows the phase curves of the ocean
planet with the reflection of the water body below the Fresnel interface
described by Rayleigh
scattering, together with the curves for a planet with the same, 
cloud-free atmosphere and the same Fresnel interface, but with a 
black surface below the interface (thus without the water body). 
The difference between the curves shows how the scattering within the
water increases $F$ and $|Q|$ at the shorter wavelengths:  
the largest effect is seen for $\lambda=443$~nm, while there is virtually
no effect for $\lambda > 550$~nm. 
Concluding, the blueness of the ocean is a main contributor to the 
increase in $F$ at small phase angles and for $\lambda < 550$, 
as shown in Fig.~\ref{fig_HHplanets}.

In $F$, the influence of the water body decreases with $\alpha$: 
a global ocean makes a planet slightly more blue for $\alpha < 90^\circ$. 
In polarized flux $|Q|$, the influence is largest for 
$60^\circ < \alpha < 80^\circ$. The scattering in the water decreases
$P$ somewhat at the shorter wavelengths and intermediate phase angles.
\citet{Zugger2010} mention that the maximum $P$ is limited by scattering 
within the water. They, however, model the reflection by the water body
below the interface as a Lambertian surface (which scatters isotropically
and unpolarized), and their results cannot be directly compared against
ours that do include (polarized) scattering within the water body.

\subsection{Cloudy planets}
\label{sect_HIplanets}

The planets in the previous sections were cloud-free. 
A planet with an ocean on its surface, will, however, very likely also have 
clouds. Here, we will discuss the influence of the clouds on the detectability 
of the ocean, starting with the influence of a horizontally homogeneous 
cloud deck (Sect.~\ref{sect_homogeneousclouds}), followed by the influence
of broken clouds (Sect.~\ref{sect_brokenclouds}).

\begin{figure*}[t!]
	\includegraphics[width=0.33\textwidth]{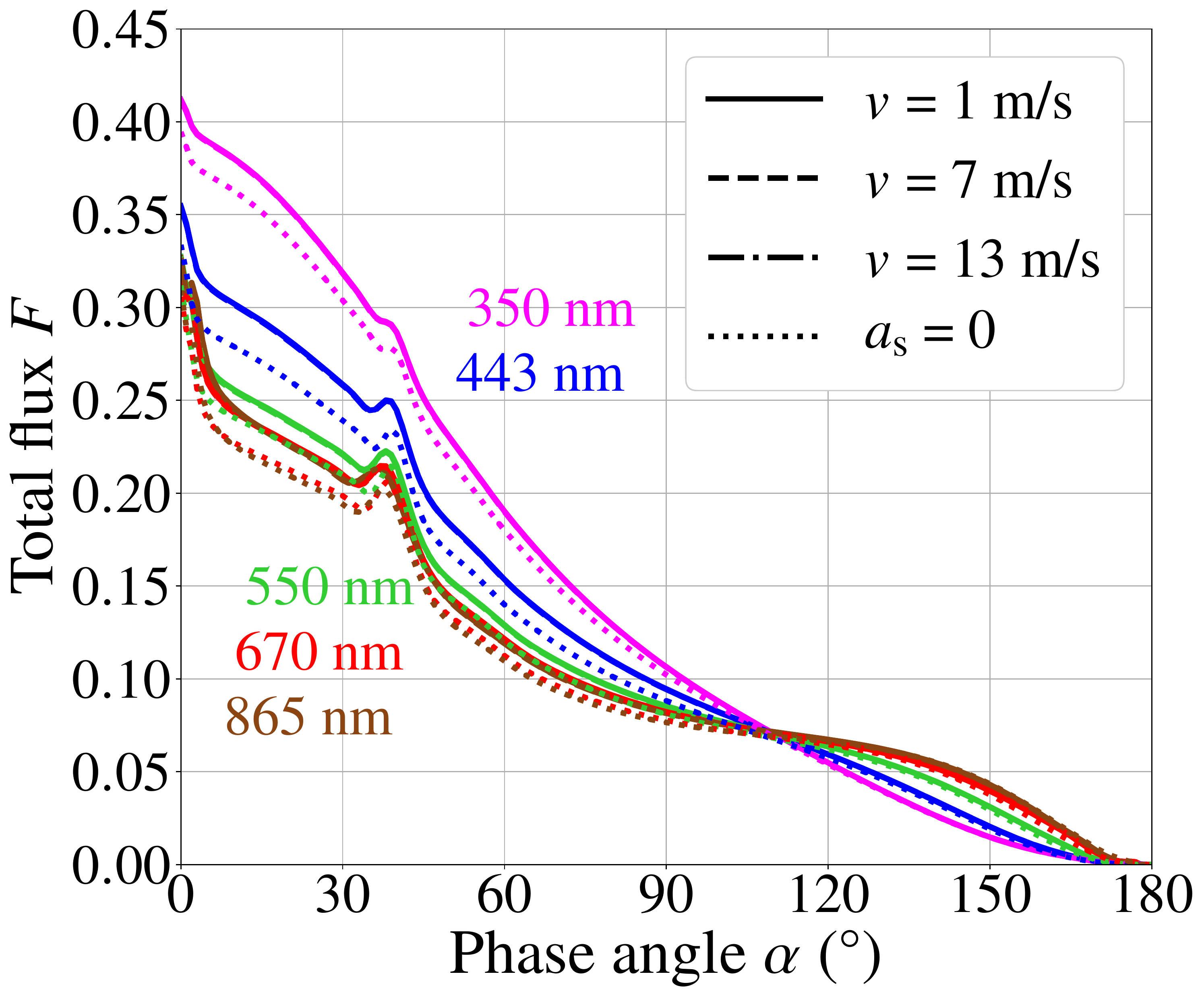}
		\label{fig:phasecurveoceancloudsflux}
	\includegraphics[width=0.33\textwidth]{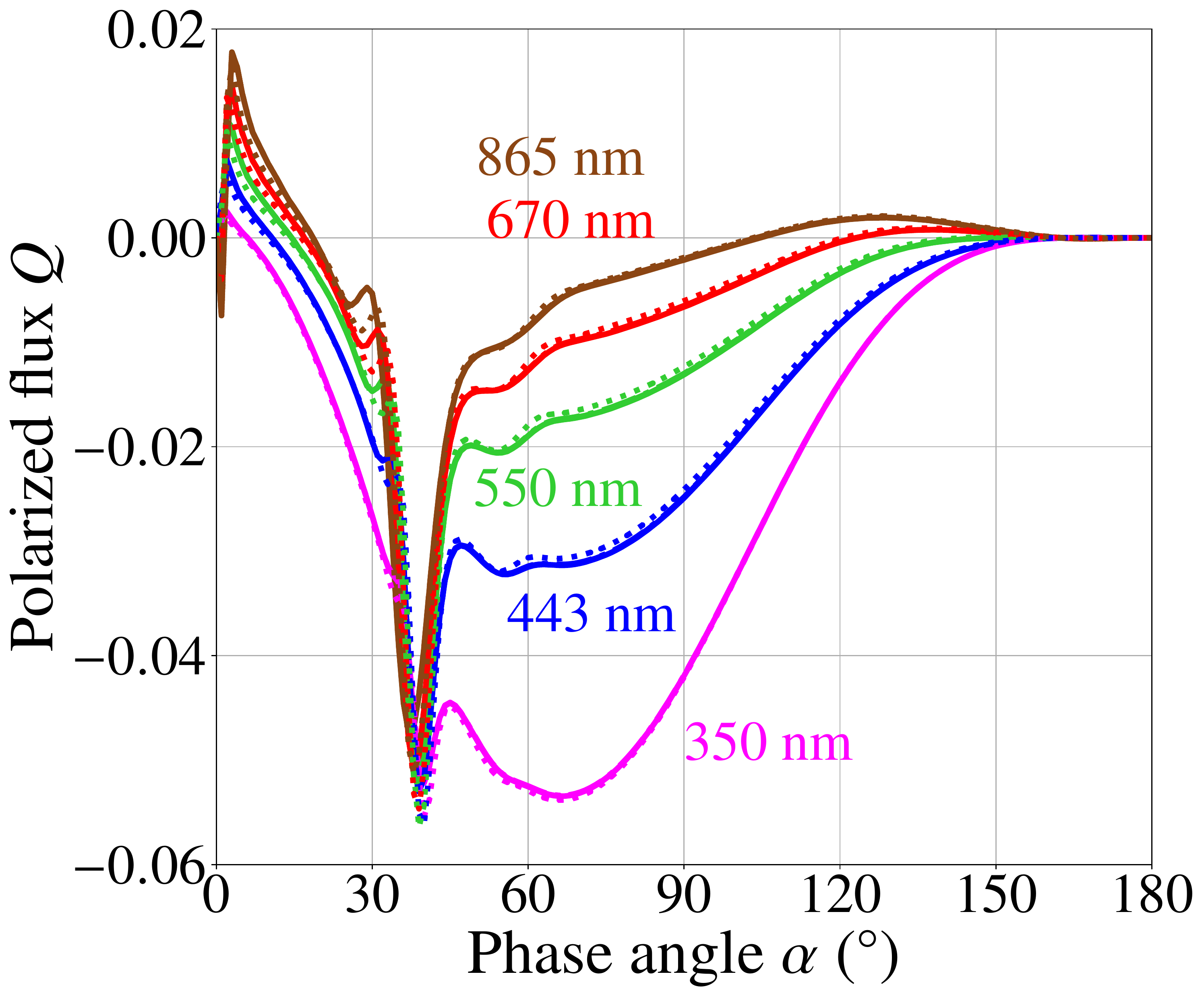}
		\label{fig:phasecurveoceancloudsQ}
	\includegraphics[width=0.33\textwidth]{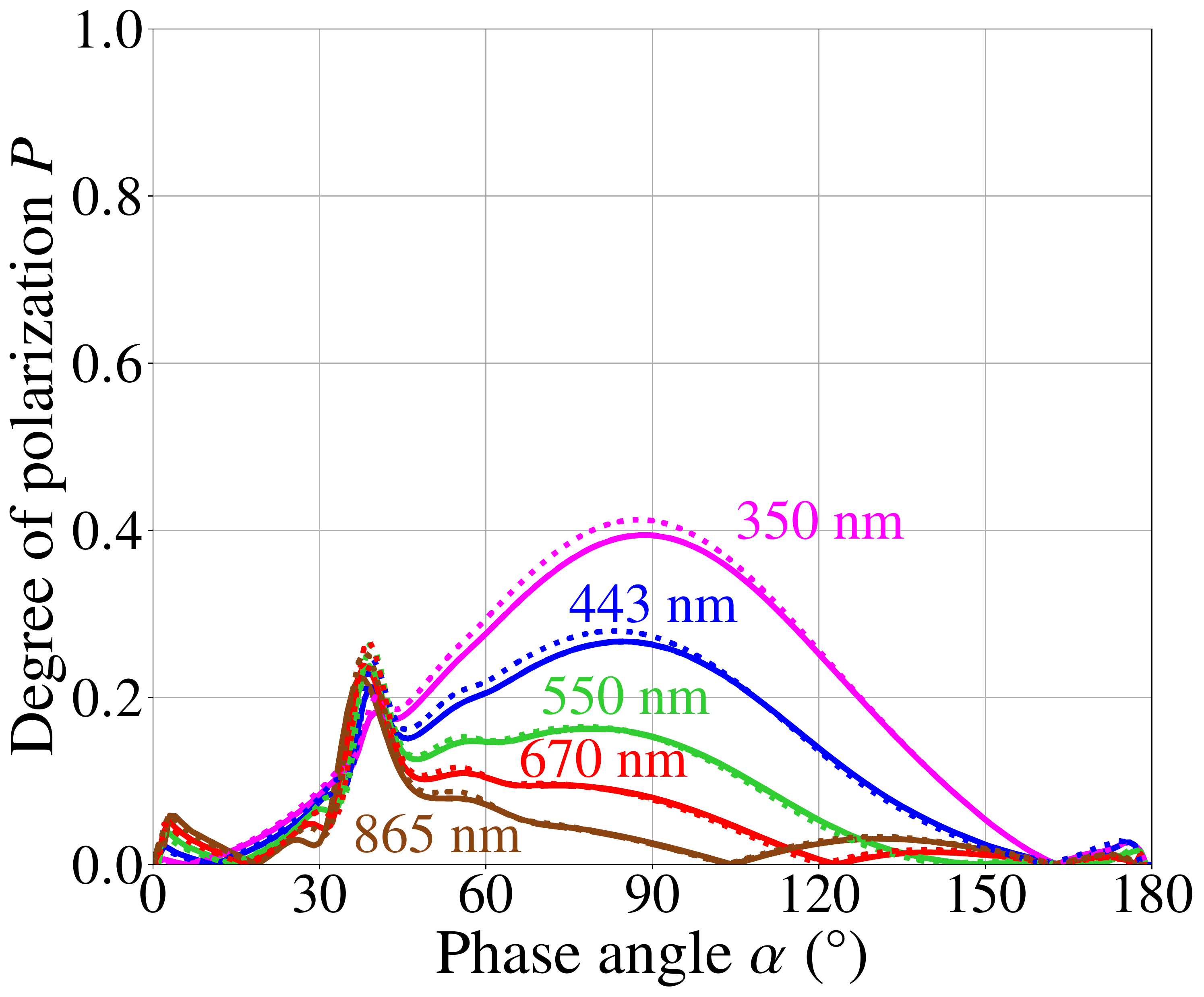}
		\label{fig:phasecurveoceancloudsdop}
    \caption{Similar to Fig.~\ref{fig_HHplanets}, but for the ocean
             planet with a horizontally homogeneous cloud deck. The wind speed 
             is 1~m/s (solid line), 7~m/s (dashed line), or 13~m/s (dashed-dotted 
             line). The dotted lines are the phase curves of a planet 
             with a black surface and the same cloud deck. }
\label{fig_HH_clouds}
\end{figure*}

\subsubsection{Influence of horizontally homogeneous clouds}
\label{sect_homogeneousclouds}

Figure~\ref{fig_HH_clouds} shows the phase curves of an ocean planet that is 
completely covered by a horizontally homogeneous cloud deck for different 
wavelengths and wind speeds $v$ (white caps are included). The details
on the cloud deck parameters can be found in Tab.~\ref{table_atmos}.
The curves for a completely cloudy planet with a black surface 
($a_{\rm s}=0$) are also shown. 

At all wavelengths, $F$ of the cloudy 
ocean planet is somewhat higher than that of the cloudy black surface planet,
thanks to the added reflection by the Fresnel interface and the ocean body.
As can be seen, $F$ of the cloudy ocean planet 
is virtually independent of $v$ at all phase angles. For $\alpha$ 
up to about 134$^\circ$, this was to be expected, as 
$F$ of the cloud-free ocean planet was also virtually independent of $v$
(see Fig.~\ref{fig_HHplanets_wind}). For the larger phase angles,
clouds apparently smooth out the flux differences that were due to
the wind on the cloud-free ocean planet.

While for the cloud-free ocean planets, the phase curves of $F$ crossed
at $\alpha=134^\circ$ (see Fig.~\ref{fig_HHplanets_wind}), those for 
the cloudy ocean and the cloudy black surface planet cross at $\alpha=108^\circ$. 
Analogous to the reflection
by the cloud-free ocean planets (Sect.~\ref{sect_HHplanets}), 
the crossing is due to the combination of Rayleigh scattering 
above the clouds and the reflection by the clouds, with more light with 
longer wavelengths reaching the clouds and subsequently, after
reflection by the clouds, the top of the atmosphere than light with 
shorter wavelengths. A completely cloudy ocean or black surface 
planet will thus change color 
from blue, through white, to red with increasing $\alpha$. 
With increasing cloud optical thickness and at $\alpha$'s below the
crossing, a planet's blueness decreases,
as the contribution of light that has been scattered by 
cloud particles increases. With increasing cloud top altitude 
(i.e.\ decreasing cloud top pressure) and at the largest $\alpha$'s, 
a planet's redness decreases, as more blue light will reach 
the cloud and the top of the atmosphere after reflection.

The bumps in the flux phase curves around $\alpha = 40^\circ$ are due to
the primary rainbow: light that has been reflected once within the spherical
water cloud droplets \citep{Hansen1974,Karalidi12,2007AsBio...7..320B}. 
There is only a very slight wavelength dependence in the position
of this rainbow, it will thus be mostly white. Indeed, as pointed out 
earlier by \citet{2011A&A...530A..69K}, the colors of this rainbow
due to scattering by the small cloud droplets 
(the blue rainbow occurs at a larger $\alpha$ than the red rainbow) are 
reverse from those of the rainbow in rain droplets (red at a larger 
$\alpha$ than blue). The phase curves of $P$ and $Q$ also show the primary 
rainbow feature and indeed, at $\alpha \sim 56^\circ$, also a small bump 
for the secondary rainbow. The $P$ and $Q$ phase curves 
for the completely cloudy ocean planet are very similar to those of the 
cloudy black surface planet. Also, $Q$ shows no color crossing point, and there 
is no shift of the maximum value of $P$ towards twice the Brewster angle, 
like for a cloud-free ocean planet (see Figs.~\ref{fig_HHplanets} - 
\ref{fig_HHplanets_wind}).
Our model clouds have an optical thickness of about 5.0 at 550~nm
(see Table~\ref{table_atmos}), which is a normal value for Earth-like clouds. 
With optically thin clouds, the ocean signal will be stronger.
However, we can conclude that it will be impossible to detect an exo-ocean 
when it is covered by a homogeneous, optically thick cloud deck. 

\subsubsection{Influence of an inhomogeneous cloud deck}
\label{sect_brokenclouds}

\begin{figure*}[ht!]
\begin{center}
\includegraphics[width=1.0\textwidth]{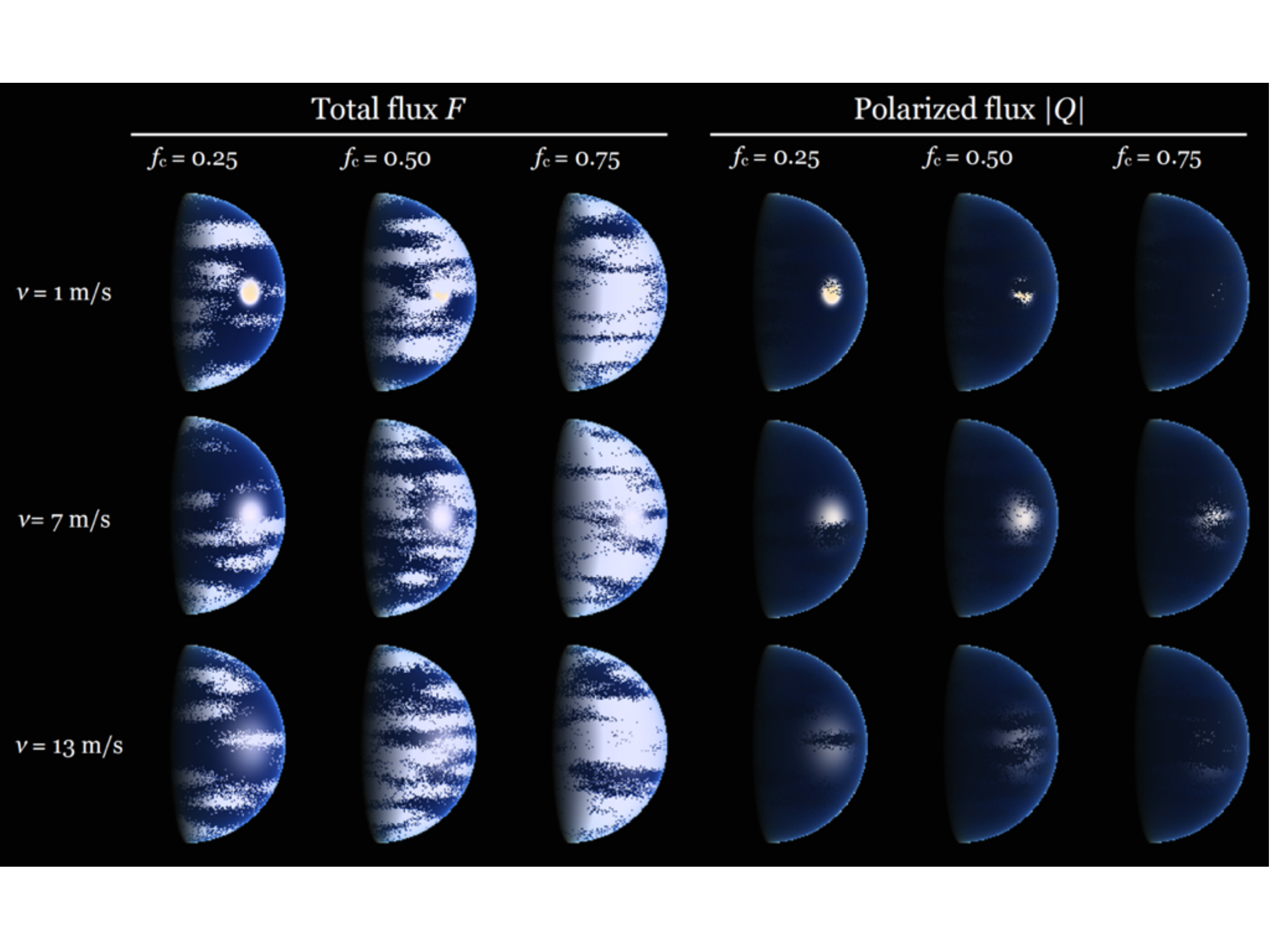}
\end{center}
\caption{Disk-resolved (160 x 160 pixels) model output RGB color images 
         in $F$ (column 1 to 3) and $|Q|$ (column 4 to 6) of ocean 
         planets with patchy clouds. The brightness of a pixel is scaled 
         with its average spectral $F$. Phase angle $\alpha=80^\circ$.
         The cloud coverage fraction $f_{\rm c}$ is 0.25 (column 1 and 4), 
         0.5 (column 2 and 5), or 0.75 (column 3 and 6), 
         and the wind speed $v$ is 1~m/s (top row), 7~m/s (middle row), or 
         13~m/s (bottom row).
         For every combination of $f_{\rm c}$ and $v$, 
         the cloud pattern across the disk is different.}
\label{fig_diskresolved}
\end{figure*}

With patchy clouds, both the location and the cloud coverage fraction
influences the reflected light signal (assuming for simplicity that
the clouds are all the same and at the same altitude).
Here, we show the influence of a horizontally inhomogeneous cloud deck on 
$F$, $Q$, and $P$ of light reflected by an ocean planet. Like on Earth, our
model cloud deck consists of patches of clouds, as described in 
Sect.~\ref{sect_atmosphere_model} with a cloud coverage fraction $f_{\rm c}$. 
Figure~\ref{fig_diskresolved} shows the disk-resolved RGB colors of the 
reflected $F$ and $|Q|$ for model planets with different cloud coverage 
fractions $f_{\rm c}$ and wind speeds $v$ (for every combination of
$f_{\rm c}$ and $v$, the model planet has a different cloud pattern)
at $\alpha=80^\circ$. The glint can be seen to appear between the 
clouds, and with increasing $v$, the width of the glint
pattern increases while its peak value decreases. 

\begin{figure*}[ht!]
\centering
\includegraphics[width=\textwidth]{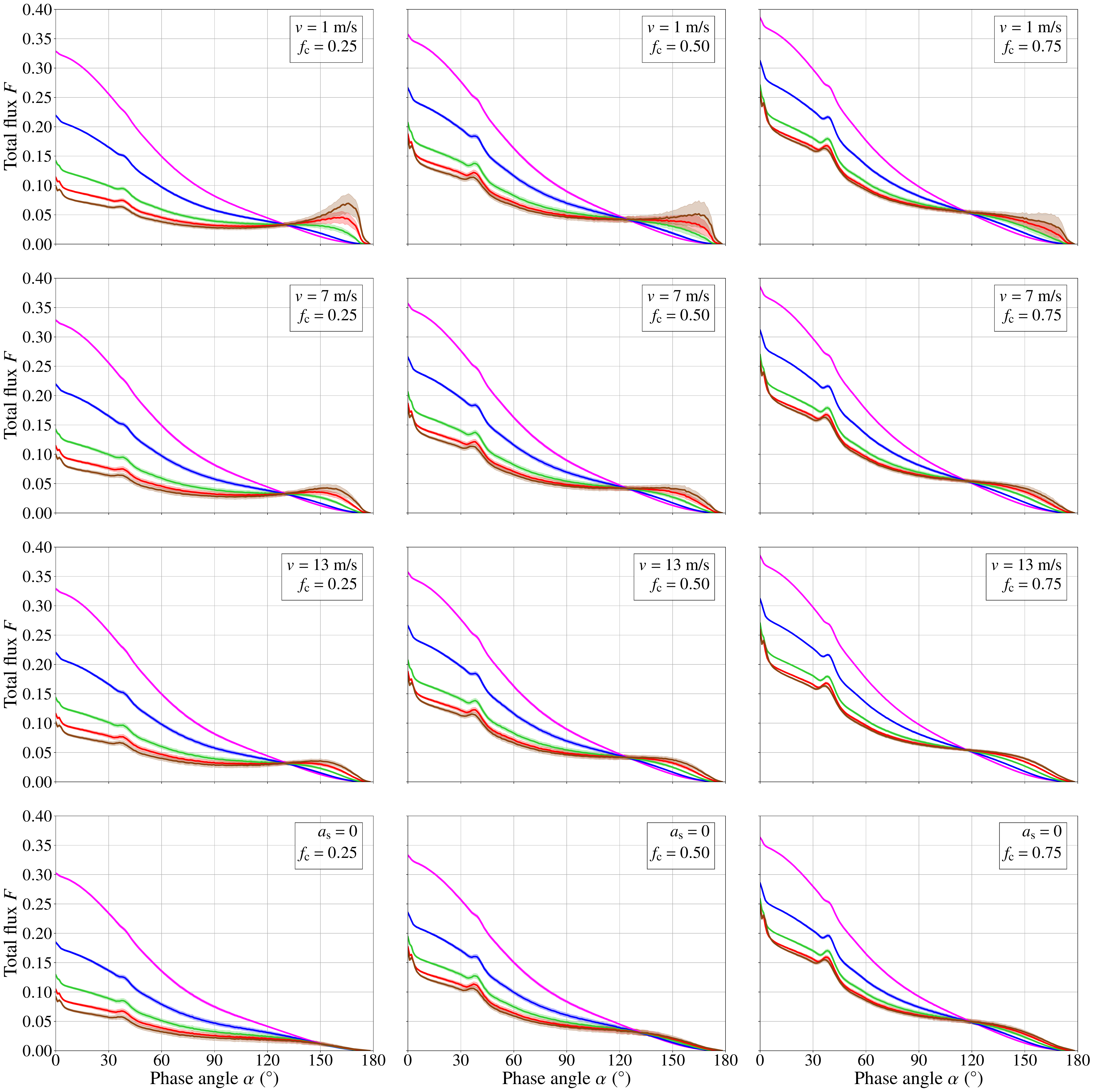}
\caption{The (disk-integrated) reflected $F$ of ocean planets with 
         patchy clouds 
         for 5 wavelengths: 350~nm (pink), 443~nm (blue), 550~nm (green), 
         670~nm (red), and 865~nm (brown), and for wind speeds of 
         1~m/s (first row), 7~m/s (second row), 
         and 13~m/s (third row), and cloud fractions $f_{\rm c}$ of 0.25 
         (left column), 0.50 (middle column), and 0.75 (right column).
         The fourth, bottom, row are the corresponding curves for planets with
         a black surface instead of an ocean. Each curve is the average of 300 curves computed for 300 different
         cloud patches configurations, which are different for each phase angle $\alpha$.
         The 1-$\sigma$ standard deviation is represented by the shaded areas 
         around each curve (except for the longest wavelengths and largest 
         phase angles, this deviation is very small). }
\label{fig_patchycloudsF}
\end{figure*}
\begin{figure*}[ht!]
\centering
\includegraphics[width=1.0\textwidth]{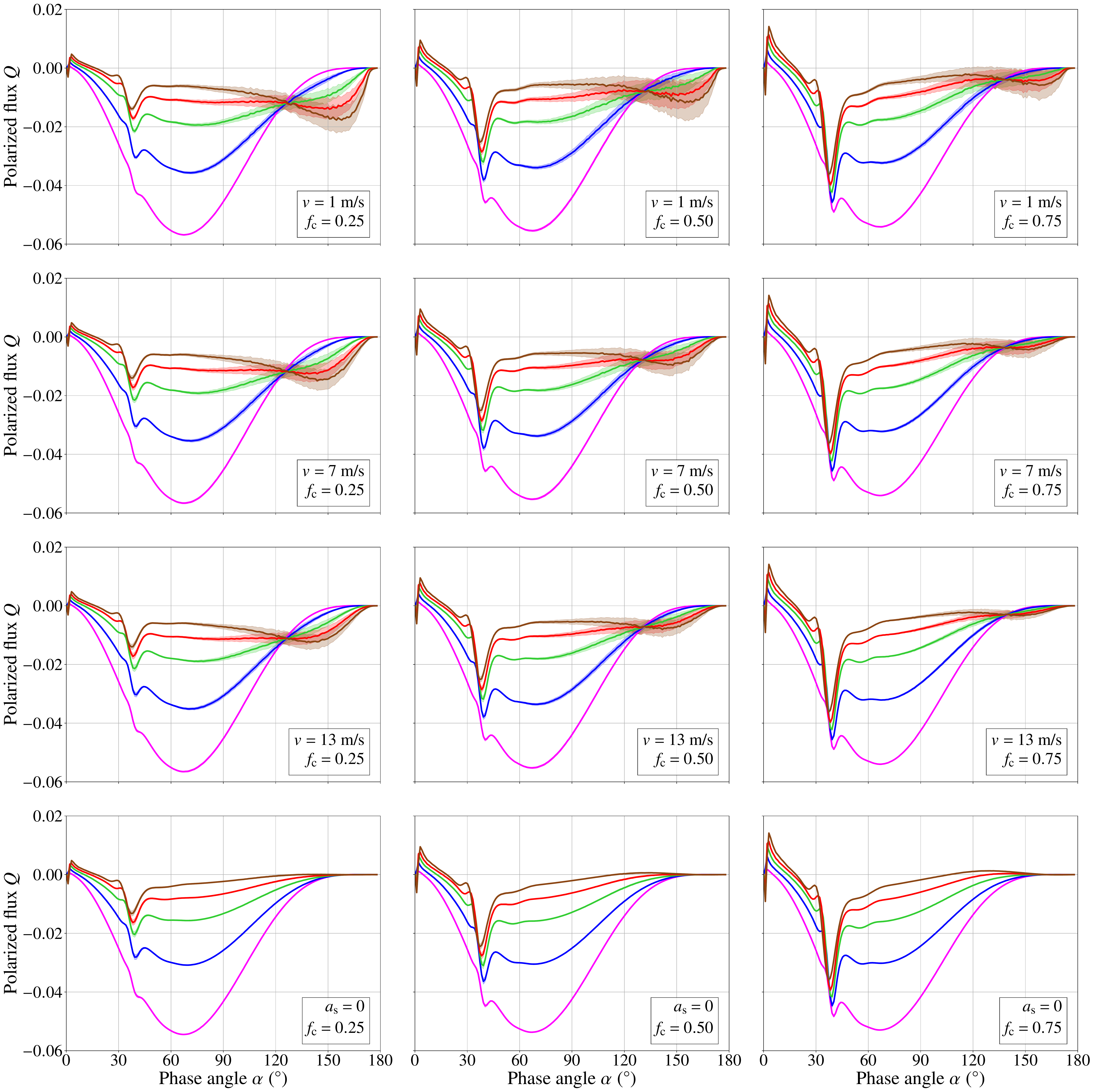}
\caption{Similar to Fig.~\ref{fig_patchycloudsF}, but for the linearly
         polarized flux $Q$.}
\label{fig_patchycloudsQ}
\end{figure*}
\begin{figure*}[ht!]
\centering
\includegraphics[width=\textwidth]{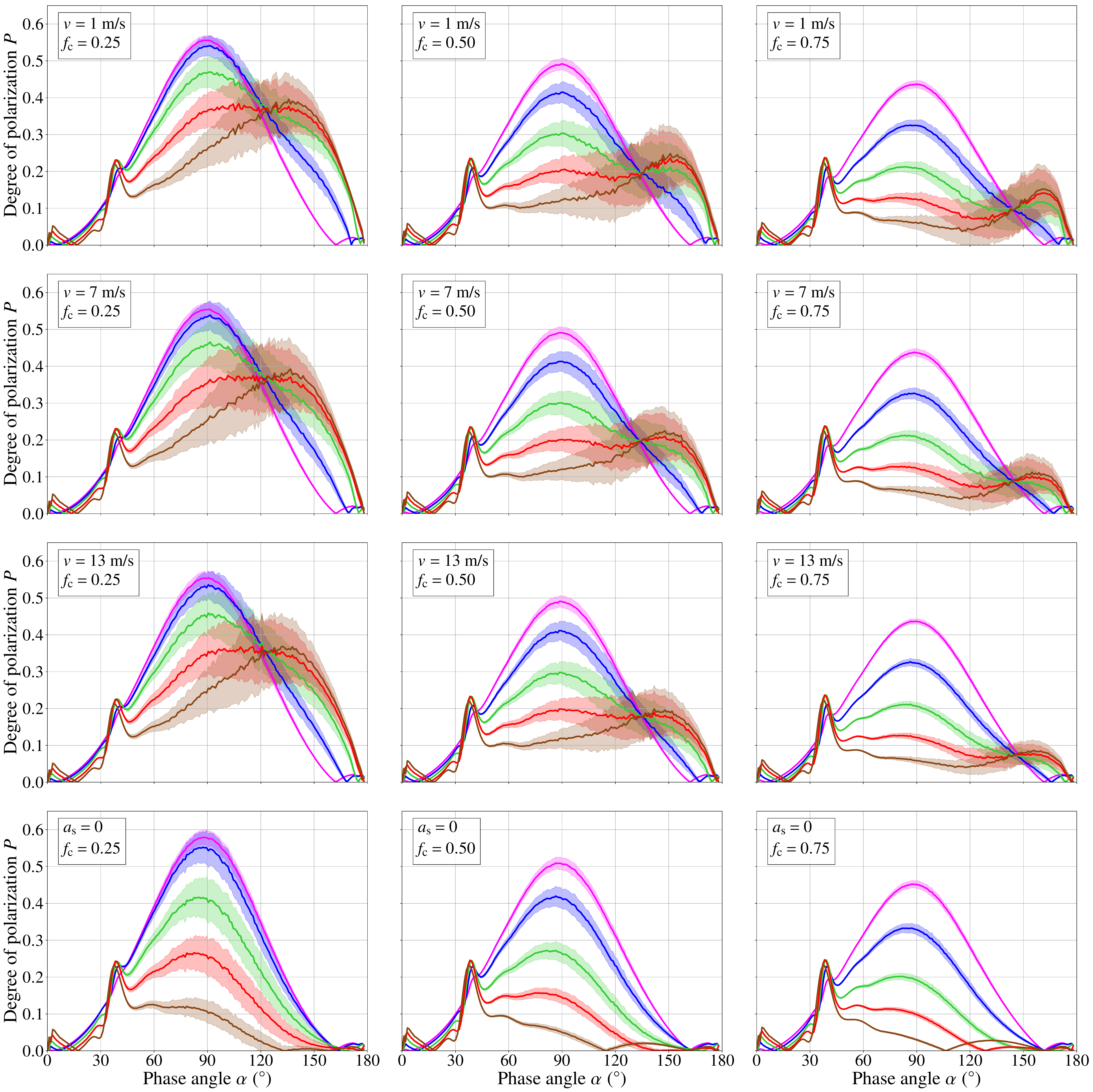}
\caption{Similar to Fig.~\ref{fig_patchycloudsF}, but for the 
         degree of linear polarization $P$.}
\label{fig_patchycloudsP}
\end{figure*}

Figures~\ref{fig_patchycloudsF}-\ref{fig_patchycloudsP} show
$F$, $Q$, and $P$ for the ocean planet for 
different values of $f_{\rm c}$ and $v$.
Also included are the curves for a black surface planet beneath the clouds. 
We used 160 pixels across the equator, just like in Fig.~\ref{fig_diskresolved}. 
For every $f_{\rm c}$, we computed $F$, $Q$, and $U$ (since
planets with patchy clouds are usually not symmetric with respect to
the reference plane) for 300 different patchy cloud patterns. The 
figures show the averages and the 1-$\sigma$ standard deviation 
(the variability).
Despite the non-symmetric cloud patterns, the disk-integrated 
linearly polarized flux $U$ appears to be smaller 
than 0.00003 for all model planets, across all phase angles. 
We thus do not show $U$ in the figures 
(even though it is very small, $U$ is included in the computation of $P$).

As can be seen in Fig.~\ref{fig_patchycloudsF}, total flux
$F$ generally increases 
with increasing $f_{\rm c}$, and the curves for the ocean planet are 
somewhat higher than for the black surface planet. The influence of the variability 
of the cloud coverage for a given value of $f_{\rm c}$ is small and only
apparent at the longest wavelengths and the largest $\alpha$'s. At the
smallest wavelengths, the significant Rayleigh scattering above and between 
the clouds subdues the contrast between the bright clouds and the dark
background, thus reducing the variability. 
At the largest phase angles, the surface area on the disk that 
provides the signal is smallest, and there is more variability in the
coverage of cloudy pixels across that area ($f_{\rm c}$ is defined on the
whole disk). A more in-depth discussion of the relation between 
$\alpha$ and variability can be found in \citet{2017A&A...607A..57R}.

Polarized flux $|Q|$ generally decreases
with increasing $f_{\rm c}$ (Fig.~\ref{fig_patchycloudsQ}), and 
$|Q|$ is slightly larger for the 
ocean planet than for the black surface planet. The curves for the black surface
planet show virtually no variability, while 
the variability in $Q$ of the ocean planet is significant for the
longer wavelengths and $\alpha$'s larger than 90$^\circ$, in particular 
for small wind speeds.

Like the $F$ and $Q$ phase curves of the cloud-free and completely cloudy 
ocean planets, the $F$ and $Q$ phase curves of the ocean planets with patchy
clouds for the different $\lambda$'s cross each other at a given 
phase angle $\alpha$
(see Figs.~\ref{fig_patchycloudsF} and~\ref{fig_patchycloudsQ}).
Around that $\alpha$ (which is different for $F$ and $Q$), the planets 
thus appear white in total and polarized fluxes.
The crossing is also present in $F$ for large cloud 
fractions ($f_{\rm c} \geq 0.5$) over a black surface planet, in agreement  
with the earlier results for completely cloudy black surface planets 
(Fig.~\ref{fig_HH_clouds}).
A black (or dark) surface planet with patchy clouds will, however, appear more 
whitish than red at the largest phase angles, because the spectral 
dispersion of $F$ is small.

\begin{figure*}[h!]
\begin{center}
\includegraphics[width=1.00\textwidth]{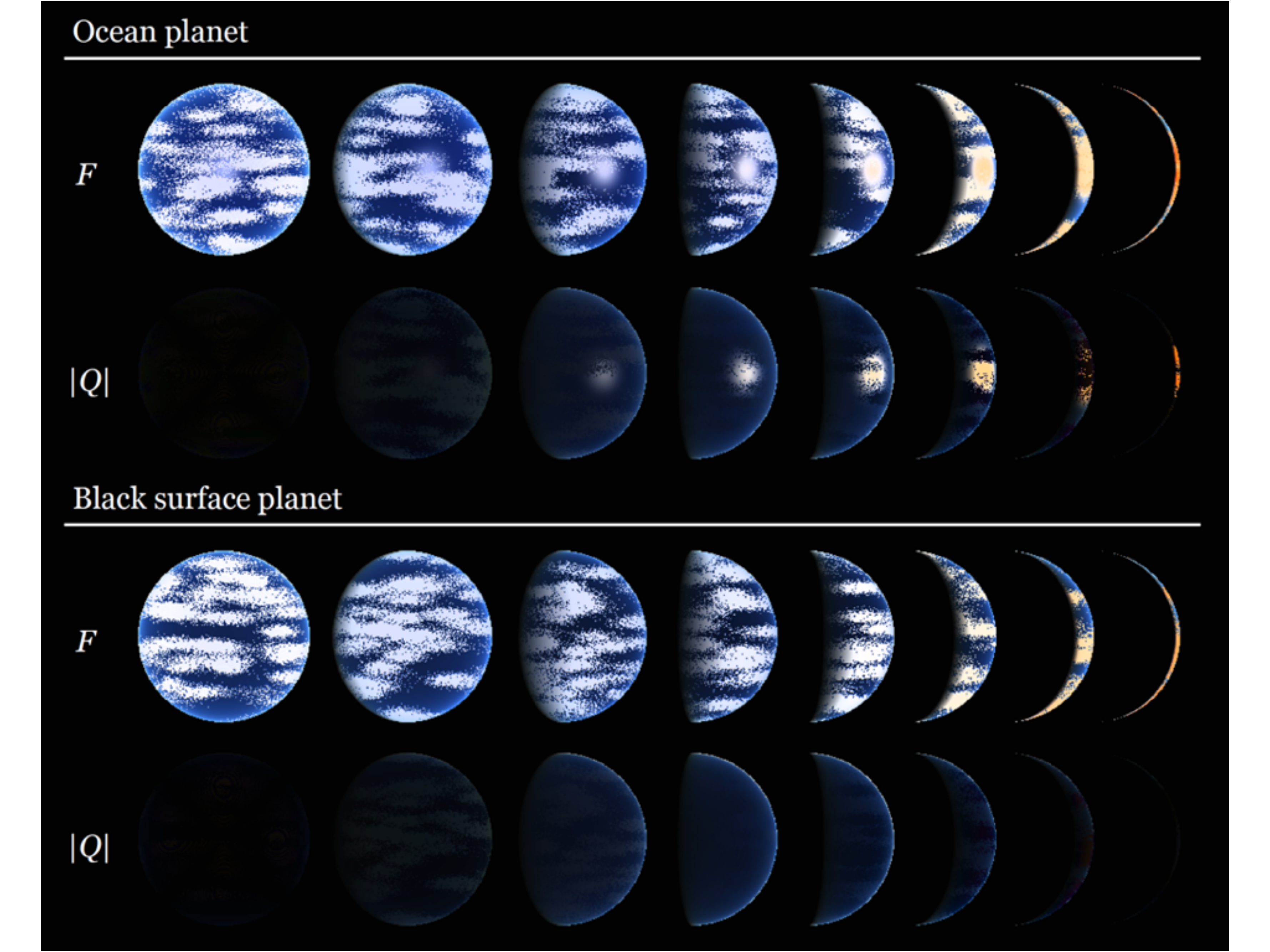}
\end{center}
\caption{Similar to Fig.~\ref{fig_diskresolved} for ocean planets (top)
         and black surface planets (bottom) covered by patchy clouds. The cloud 
         coverage fraction $f_{\rm c}$ is 0.50, and the wind speed $v$ 
         on the ocean planet is 7~m/s. Phase angles $\alpha$ range from
         $0^\circ$ (left), through $30^\circ$, $60^\circ$, $80^\circ$, 
         $100^\circ$, $120^\circ$, and $140^\circ$, to $160^\circ$
         (right).}
\label{fig_phasesocean}
\end{figure*}

Figure~\ref{fig_patchycloudsP} shows the $P$ phase 
curves of the ocean and black surface planets with patchy clouds.
The largest variation in $P$ is seen at the longest wavelengths, where
the contribution of the gaseous atmosphere is smallest and the
contribution of the surface is largest, and at intermediate 
to large $\alpha$'s, where the contribution of the glint is strongest, 
and thus where the location of a cloud is relatively more important than
at the smaller $\alpha$'s (cf.\ Fig.~\ref{fig_diskresolved}).
There appears to be little dependence of both the average curves
and the variability in $P$ on wind speed $v$, except for 
$\alpha > 100^\circ$.

For ocean planets with patchy clouds, the curves of $P$ show crossings,
i.e.\ the planet will show a color change, just like the curves for $F$ 
and $Q$. While this crossing
in $P$ does not occur for the completely cloudy planet ($f_{\rm c} = 1.0$), 
as was shown in Fig.~\ref{fig_HH_clouds}, 
it does show up for $f_{\rm c} =0.75$ (see Fig.~\ref{fig_patchycloudsP}).
The black surface planet with patchy clouds does not show a color change.
Measuring a color change in $P$ in the relevant phase angle range
would thus indicate the presence of an ocean, even in the presence of 
patchy clouds.

Apart from the crossing in the $P$ phase curve for a cloudy ocean planet, 
there are more differences from the phase curve of the cloudy black surface planet.
For example, Fig.~\ref{fig_patchycloudsP} shows that for 
$\lambda = 865$ nm and both $f_{\rm c}$ up to (at least) 0.50, the
slope of $P$ for the ocean planet is positive between 
$80^\circ < \alpha < 100^\circ$, while that for the black surface planet is negative.
Additionally, the variability in $P$ is much larger for the ocean planet
than for the black surface planet because of the appearance and disappearance 
of the polarized glint between the clouds. 

Figure~\ref{fig_phasesocean} shows the disk-resolved $F$ and $|Q|$
of the ocean planet and the black surface planet with patchy clouds 
($f_{\rm c}=0.50$) at a range of phase angles $\alpha$ to illustrate
the reddening of both planets in $F$ and the reddening of the ocean
planet in $|Q|$. It can also be seen that the glint makes the 
ocean planet somewhat lighter in blue than the black surface planet.
In the images of the black surface planet at the largest phase angles the
reddening of the planet in $F$ due to the clouds 
is visible. This cloud reddening also happens with the ocean
planet, but there the reddening of the glint is more significant.

\begin{figure*}[ht!]
\begin{center}
\includegraphics[width=0.48\textwidth]{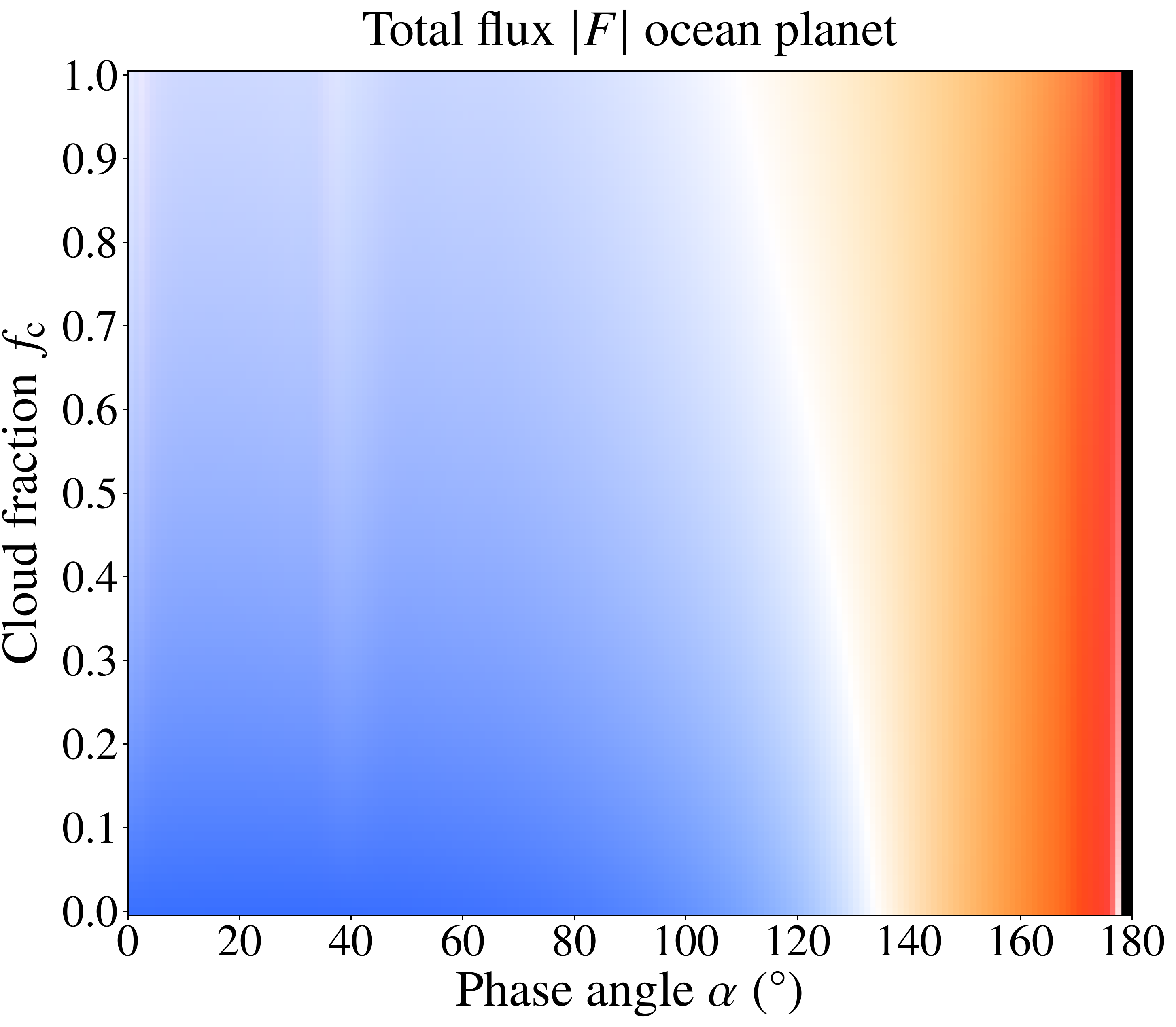}
\label{fig:colorcfsI}
\includegraphics[width=0.48\textwidth]{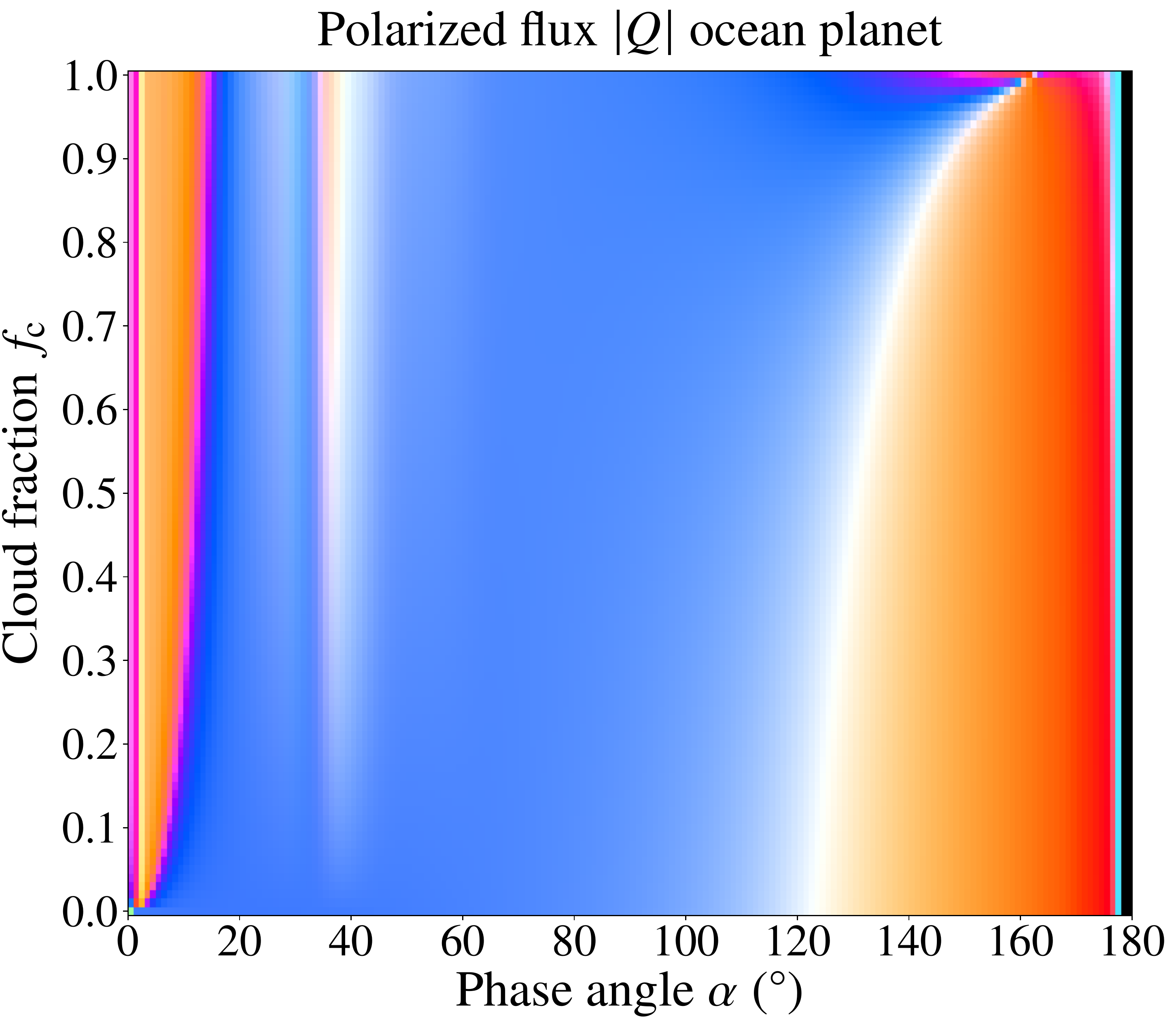}
\label{fig:colorcfsQ}
\includegraphics[width=0.48\textwidth]{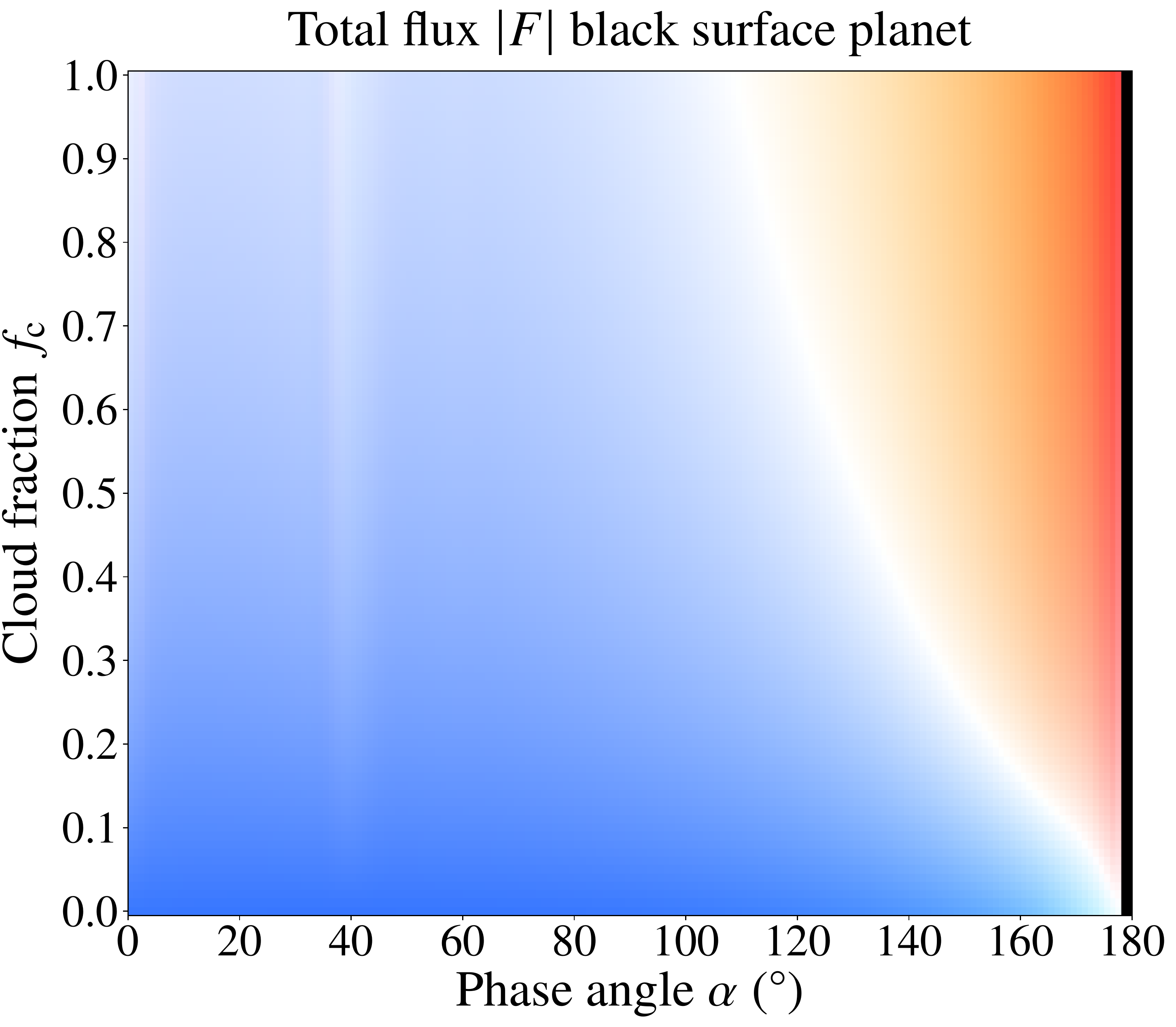}
\label{fig:colorcfsIlamb}
\includegraphics[width=0.48\textwidth]{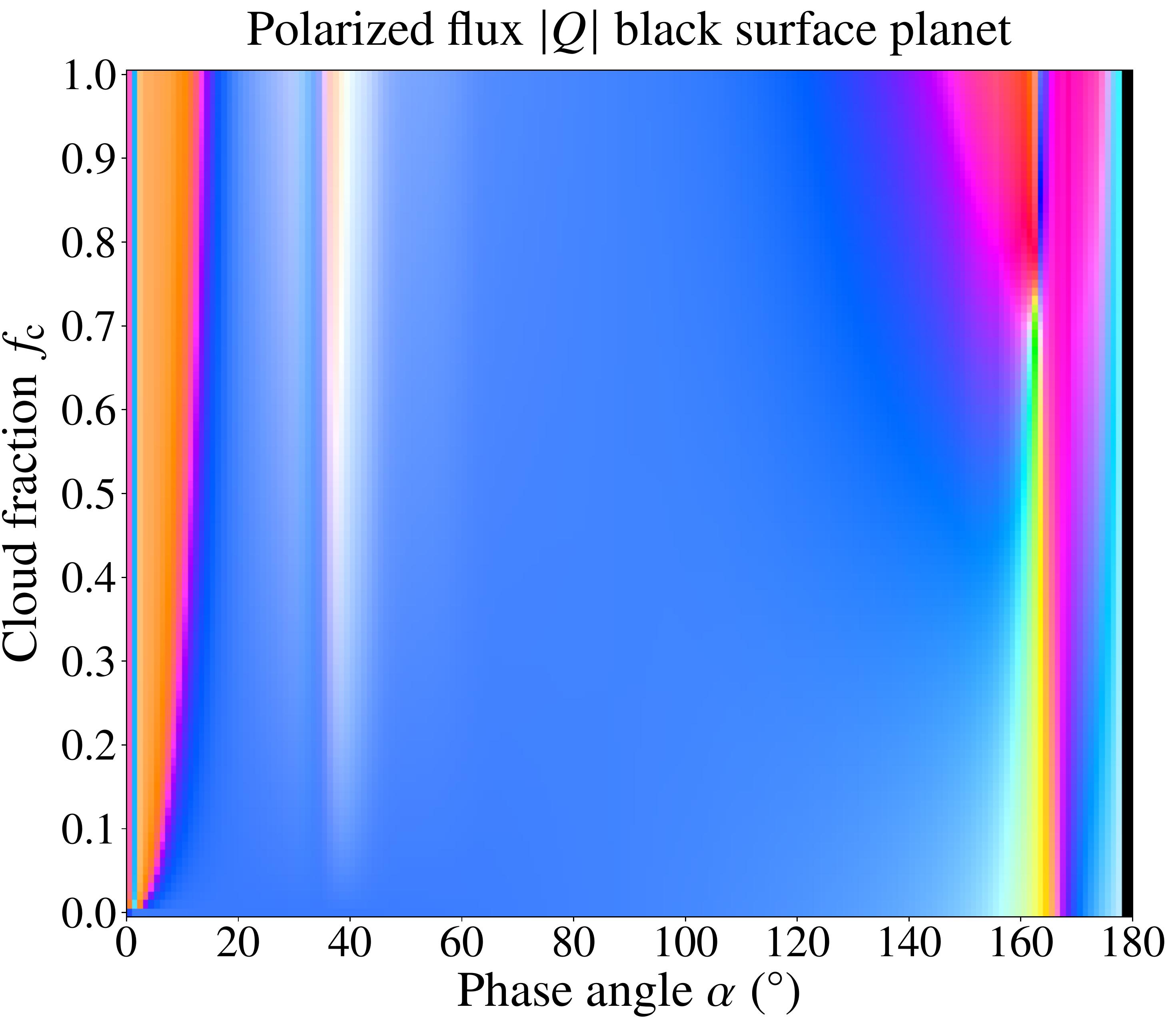}
\label{fig:colorcfsQlamb}
\end{center}
\vspace*{-6mm}
\caption{Similar to Fig.~\ref{fig_RGB_surfacepressure}, but for cloud 
         fractions ranging from 0.0 to 1.0 in steps of 0.01. The surface 
         pressure $p_{\rm s}$ is 1 bar. The results are shown for the 
         phase curves generated by the weighted sum of cloud-free 
         ($f_{\rm c}$ = 0.0) and and fully cloudy ($f_{\rm c}$ = 1.0) 
         planets (see App.~\ref{app_weightedsum}). The purple, yellow 
         and cyan colors at $\alpha > 140^\circ$ result from the high 
         RGB color sensitivity on the very small absolute polarized fluxes $|Q|$ for the different wavelengths at these phase angles 
(cf.\ Fig.~\ref{fig_patchycloudsQ}).}
\label{fig_RGB_cloudfractions}         
\end{figure*}

\section{Discussion of the color change}
\label{sect_discussion}

Our results show that the presence of an exo-ocean could be revealed 
through a color change in the polarized flux $Q$ of starlight that is 
reflected by a planet, except when the planet is completely covered
by a thick cloud deck. The total reflected flux $F$ and degree of
polarization $P$ can also show color changes.
The crossing $\alpha$'s, or $\alpha_{\rm cc}$'s, where the color change 
takes place, are generally
different for $F$, $Q$, and $P$, and depend on the cloud fraction 
$f_{\rm c}$. 

An indication of such a crossing in {\em measured phase curves of} 
$P$ of sunlight that is reflected by the Earth, has recently been 
presented by \citet{2019Sterzik}. 
They measured $P$ for $\alpha$ ranging from $\sim 50^\circ$ to 
$\sim 140^\circ$ as derived from Earth-shine 
data obtained with ESO's Very Large Telescopes in Chile. 
The phase angle range of these Earth-shine data is limited by 
the lunar phase during the observations\footnote{Small phase 
angles of the Earth require observations at a large lunar phase angle,
and thus with large background sky brightness, while large phase 
angles of Earth can only be done at small lunar phase angles, where
the dark part of the lunar surface is relatively small.}, so
unfortunately, the phase curves do not include the rainbow angle, which 
would have provided an interesting comparison with the strength of the 
rainbow in our simulations. 
In the Earth-shine measurements, $\alpha_{\rm cc}$ appears 
to vary from about 120$^\circ$ to 135$^\circ$. The observations 
span a period of about two years and during the observations
either South America and the Atlantic Ocean (east of Chile) 
were turned towards the moon or the Pacific Ocean (west of Chile)
\citep{2019Sterzik}. The $\alpha_{\rm cc}$ of about 
135$^\circ$ was measured over the Pacific Ocean.

\begin{figure*}[h!]
\begin{center}
\includegraphics[width=1.0\textwidth]{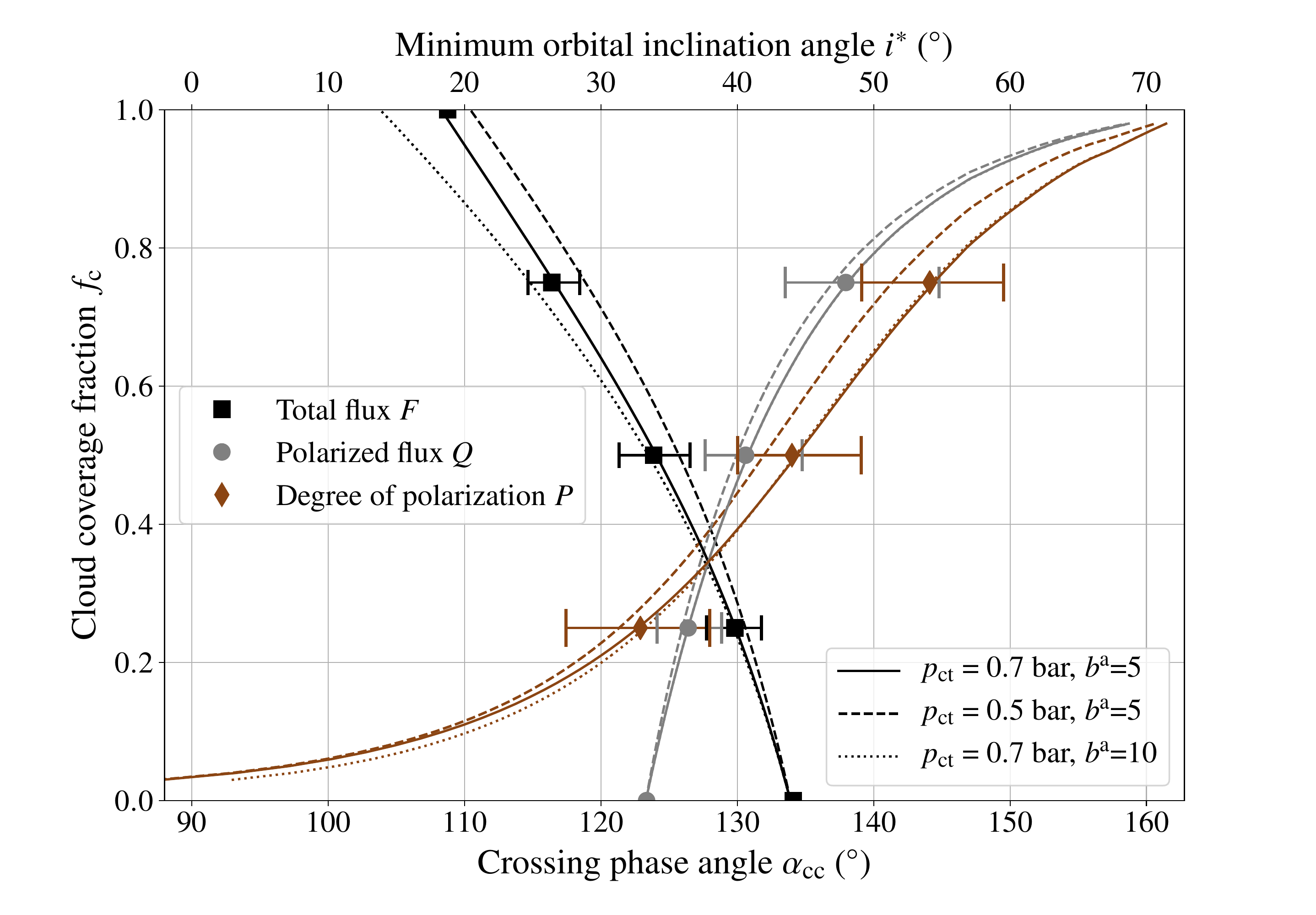}
\caption{The phase angle $\alpha_{\rm cc}$ where the phase curves of 
         the total flux $F$ (black), the (linearly) polarized flux 
         $Q$ (grey), and the degree of (linear) polarization $P$ (brown)
         of an ocean planet for the different wavelengths 
         cross each other, as functions of the cloud coverage fraction 
         $f_{\rm c}$. 
         The wind speed $v$ is 7~m/s, although $\alpha_{\rm cc}$ does
         not significantly depend on $v$.
         The symbols (squares, circles, diamonds) are the 
         $\alpha_{\rm cc}$'s shown before in 
         Figs.~\ref{fig_patchycloudsF}, \ref{fig_patchycloudsQ}, and 
         \ref{fig_patchycloudsP} as computed for horizontally 
         inhomogeneous planets, with the error bars representing the 
         1-$\sigma$ standard deviation of 300 iterations with different
         locations of the patchy clouds. 
         The solid, dashed, and dotted lines have been computed using 
         the weighted sum approach (see text),
         with the clouds at different altitudes ($p_{\rm ct}=0.5$ or 
         0.7~bar) 
         and for different cloud optical thicknesses 
         ($b^{\rm a}= 5$ or 10). 
         The horizontal axis at the top indicates 
         the minimum orbital inclination angle $i^*$ required to be able 
         to observe the exoplanet at the corresponding $\alpha_{\rm cc}$
         (an inclination angle of $90^\circ$ indicates a planetary 
         orbit that is 'seen' edge-on).
}
\label{fig_crossing}
\end{center}
\end{figure*}

Figure~\ref{fig_RGB_cloudfractions} shows RGB color phase curves 
of $F$ and $Q$ reflected by the ocean planet and the black surface 
planet for cloud fractions $f_{\rm c}$ ranging from 0.0 to 1.0 in steps of
0.01. 
The phase curves for each wavelength are computed using so-called quasi horizontally inhomogeneous 
ocean planets, in the weighted sum approach
as described by \citet{Stam08}. In this approach, $F$ and $Q$ of
a horizontally inhomogeneous planet are computed by taking a weighted
sum of the $F$'s and $Q$'s computed for horizontally homogeneous planets
with the atmosphere-surface combinations that occur on the
inhomogeneous planet. Each weighting factor equals the fraction 
of the planetary disk that is covered by a given atmosphere-surface 
combination. A careful comparison (see App.~\ref{app_weightedsum}) shows that 
while the weighted sum approach does not directly provide variances 
due to the spatial distribution of cloud patches across the disk,
it does indeed allow us to rapidly and accurately compute the 
{\em average} phase curves of $F$, $Q$, 
and hence the
respective values of $\alpha_{\rm cc}$'s, using a small step size for $f_{\rm c}$. 
Fig.~\ref{fig_RGB_cloudfractions} illustrates that the color change in 
$Q$ (i.e. where the RGB color changes from blue, through white, to red) is
unique for ocean planets for $f_{\rm c}$ up to 0.98. For $f_{\rm c} > 0.98$, 
the 
crossing of the phase curves at the RGB wavelengths is too small to detect.  

Figure~\ref{fig_crossing} shows $\alpha_{\rm cc}$ for 
$F$, $Q$, and $P$ as functions of the cloud fraction $f_{\rm c}$, using the weighted sum approach. The $P$ of the inhomogeneous planet is computed
from the summed total $F$ and $Q$ ($U$ of such a quasi 
horizontally inhomogeneous planet is always zero). The figure also includes $\alpha_{\rm cc}$'s for the ocean planets
as shown in Figs.~\ref{fig_patchycloudsF} - \ref{fig_patchycloudsP},
and thus computed for actual horizontally inhomogeneous planets.
A comparison between these data points (the symbols) 
and the data computed assuming quasi horizontally inhomogeneous planets
(the lines) in Fig.~\ref{fig_crossing}, 
shows that the weighted sum approach predicts $\alpha_{\rm cc}$ with 
a maximum deviation of 0.35$^\circ$ (in $P$ for $f_{\rm c} = 0.50$). 
Also shown in Fig.~\ref{fig_crossing}, is $\alpha_{\rm cc}$ 
computed for higher clouds (with a cloud top pressure, $p_{\rm ct}$, 
of 0.5 bar) and optically thicker clouds (with $b^{\rm a}$ = 10). 
Note that we did not include curves for planets with black surfaces in 
Fig.~\ref{fig_crossing},
because there are no color changes in $Q$ and $P$ for such planets
(cf.\ the bottom rows of Figs.~\ref{fig_patchycloudsQ} 
and~\ref{fig_patchycloudsP}), while the color change in $F$ is
not very strong due to the small dispersion. The horizontal axis at the top indicates the
minimum inclination angle orbital $i^*$ that is required to
be able to observe an exoplanet at that particular phase angle;
i.e.\ a color change expected to take place at 
$\alpha_{\rm cc}=140^\circ$ can be observed if the inclination
angle of the planetary orbit is at least 50$^\circ$.
 
Increasing the cloud coverage $f_{\rm c}$ on an ocean planet 
decreases $\alpha_{\rm cc}$ for the total flux $F$, as can be seen
in the figure, because more
long wavelength light will be reflected at a given $\alpha$: 
$\alpha_{\rm cc}$ ranges from $\sim 134^\circ$ for 
cloud-free planets ($f_{\rm c}=0.0$) to $\sim 108^\circ$ for 
a fully cloudy planet ($f_{\rm c}=1.0$).
Increasing the cloud top altitude for a given $f_{\rm c}$, increases 
$\alpha_{\rm cc}$, because the higher the clouds, the smaller the
contribution 
of short wavelength (Rayleigh scattered) light at a given $\alpha$. 
An optically thicker cloud reflects more light, and will thus decrease
$\alpha_{\rm cc}$ in $F$. 

Increasing $f_{\rm c}$ increases $\alpha_{\rm cc}$ for the polarized
flux $Q$, because more light will be reflected by the clouds, and
less by the ocean at a given $\alpha$: $\alpha_{\rm cc}$ ranges from $\sim 123^\circ$ for cloud-free
planets ($f_{\rm c}=0.0$) to $\sim 157^\circ$, where the color change
starts to disappear for an almost fully cloudy planet ($f_{\rm c}=0.98$). 
Increasing the cloud top altitude (i.e.\ decreasing $p_{\rm ct}$)
at a given $f_{\rm c}$, increases $\alpha_{\rm cc}$ because with
higher clouds, a longer atmospheric path length, thus a more slanting
incoming direction, is required for the short wavelength light
to be sufficiently scattered out of the incident beam.
Increasing the cloud optical thickness, $b^{\rm a}$, has virtually
no effect on $\alpha_{\rm cc}$ of $Q$, because a thicker cloud
will add mostly unpolarized light to the reflected signal.

For the degree of polarization $P$, Fig.~\ref{fig_crossing} shows that 
increasing $f_{\rm c}$, increases $\alpha_{\rm cc}$. Indeed, the
range of $\alpha_{\rm cc}$ appears to be largest for $P$: 
for $f_{\rm c}$ as small as 0.03, $\alpha_{\rm cc} \sim 88^\circ$,
while for $f_{\rm c}= 0.98$, the color change starts to disappear at 
$\alpha_{\rm cc} \sim 161^\circ$ (cf.\ Fig.~\ref{fig_HH_clouds}).
There is no crossing in $P$ for $f_{\rm c} < 0.03$,
which is consistent with our conclusion that a crossing in $P$ 
indicates an ocean {\em that is partly covered by clouds}: without clouds,
there is no color change.
Increasing the cloud top altitude, decreases $\alpha_{\rm cc}$ for $P$,
while increasing the cloud optical thickness $b^{\rm a}$, increases
$\alpha_{\rm cc}$ slightly and only for $f_{\rm c} < 0.3$.

The crossings in the phase curves of $P$ derived from Earth-shine
measurements by \citet{2019Sterzik} occurred between
about 120$^\circ$ to 135$^\circ$. Taking into account the order of magnitude 
of the variation in $\alpha_{\rm cc}$ with the $\pm$ 1-$\sigma$ 
standard deviation of the phase curve, the cloud coverage fraction would
have ranged between 0.2 and 0.6. The later value pertains 
specifically to observations with the Pacific Ocean (west of Chile)
turned towards the moon. Observations taken when South America
and the Atlantic Ocean were turned towards the moon (east of Chile)
would be influenced not only by the reflection by the ocean but
also by South America. The influence of land surface 
reflection (albedo, reflection properties and coverage fraction) 
on $\alpha_{\rm cc}$ will be part of a future study.

Finally, our model clouds consist of liquid water particles only,
but on Earth, high altitude clouds can of course also contain ice particles. 
If these particles are flat and horizontally oriented, they will produce 
their own glint feature 
\citep[examples from Earth-observations, are described in][]{2017GeoRL..44.5197M,2004JAtS...61.2888B,
1999JQSRT..63..521C,2004JAtS...61.2073N}. Whether or not the polarization
signature of the glint due to such (usually optically thin) ice clouds 
will mimic and/or influence that of a surface ocean should be part of a 
future study.
In particular, the influence of bright surface areas, 
such as high planetary latitudes that could be covered by ice and/or snow,
on the disk--integrated fluxes and polarization would be interesting
to investigate further. \citet{Cowanfalseglint2012} postulated that the
larger relative contribution of light reflected by polar 
regions at a planet's crescent phase (assuming the light equator 
coincides more or less with the planet's true equator), could yield
a false positive for the glint of light reflected by an ocean. 
However, unless such snow and ice covered surfaces are very clean, 
we would mostly expect an increase in total flux $F$, not in 
$Q$ and/or $U$, and thus a decrease of $P$. 
Such an increase in $F$ and decrease in $P$ would mainly happen
at the longest wavelengths, and would thus not result in a
color change in $Q$ and/or $P$ and a false-positive exo-ocean detection.

\section{Summary and outlook}
\label{sect_summary}

We have presented computed total and (linearly) polarized fluxes, 
$F$ and $Q$,
and the degree of (linear) polarization, $P$, of starlight 
that is reflected by ocean planets, i.e.\ planets with a liquid ocean
ruffled by the wind
on their surfaces, with the aim of identifying observables for 
the presence of an ocean, by comparing their signals against those
of so-called black surface planets, with the same atmosphere but with a 
black surface.\footnote{For planets that are symmetric with respect to the 
reference plane of the polarized fluxes (the plane through the centers of
the star, the planet and the observer) only the polarized flux $Q$ will
be non-zero. While planets with patchy clouds will usually not be symmetric 
with respect to the reference plane, the disk-integrated polarized flux
$U$ appears to be very small and while it is included in the computation
of $P$, we do not explicitly discuss it.}
Below, we summarize and reflect on our results.

{\em The total flux $F$} 
of light reflected by an ocean planet will be
slightly larger than that of a black surface planet, up till phase angles 
$\alpha$ of about 100$^\circ$. 
Across these phase angles, both ocean and black surface planets will appear 
blue, for every cloud coverage fraction $f_{\rm c}$.
With $\alpha$ increasing from about $100^\circ$, the color of a 
black surface planet will change from blue to white,
while that of an ocean planet will change from blue to white, 
then to red, and then back to white (but extremely dark) at 
the largest phase angles.

The crossing phase angle $\alpha_{\rm cc}$, i.e.\ the $\alpha$ 
where the planet's color changes from blue, through white, to red, increases with 
increasing surface
pressure: at 1~bar, $\alpha_{\rm cc} \sim 134^\circ$, 
while at 6~bar, $\alpha_{\rm cc} \sim 140^\circ$.
The phase angle where the cloud-free ocean planet changes
back from red to white, 
decreases with increasing surface pressure $p_{\rm s}$.
For $p_{\rm s}$ increasing up from about 6~bars, the 
planet's reddish phase fades and with that the difference 
between the ocean planet and the black surface planet starts to disappear.

Angle $\alpha_{\rm cc}$ decreases with increasing $f_{\rm c}$:
while for a cloud--free planet ($f_{\rm c}= 0.0$) and 
$p_{\rm s} \sim 1$~bar, $\alpha_{\rm cc} \sim 134^\circ$, 
for a fully cloudy planet ($f_{\rm c}=1.0$), 
$\alpha_{\rm cc} \sim 108^\circ$.
Angle $\alpha_{\rm cc}$ is insensitive to the wind 
speed $v$ over the ocean, although the strength of the color
change does depend on $v$: the smaller $v$, and thus the lower
the waves and the less whitecaps, the redder the ocean planet would
appear.

{\em The polarized flux $Q$}
of light reflected by an ocean planet shows a similar color change 
from blue, through white, to red with increasing $\alpha$ as
the total flux $F$. Crucially, for $Q$, the color change does {\em not} 
occur for black surface planets, {\em only} for ocean planets, 
independent of the cloud coverage fraction $f_{\rm c}$.
Detecting a color change in $Q$ would thus indicate the presence of
an ocean on the surface of a planet.
Angle $\alpha_{\rm cc}$ for $Q$ increases with increasing $f_{\rm c}$. 
For $f_{\rm c}=0.0$, $\alpha_{\rm cc} \sim 123^\circ$, while for almost completely cloudy planets ($f_{\rm c}=0.98$), 
the color change starts to disappear at $\alpha_{\rm cc} \sim 157^\circ$.
Also, the color change in $Q$ remains strong at higher surface pressures, and is even still present 
(somewhat) at a surface pressure $p_{\rm s}$ of 10~bar. 
Furthermore, for $p_{\rm s} \gtrapprox 2$~bar, the crossing angle 
$\alpha_{\rm cc}$ is more or less independent of both $p_{\rm s}$.
Angle $\alpha_{\rm cc}$ also appears to be independent of $v$, 
although, like with $F$, the smaller $v$, the redder the ocean 
planet would appear in $Q$.     

For ocean planets with patchy clouds ($0.0 < f_{\rm c} < 1.0$), the
variability of $Q$ due to variation in the location of the clouds
is, at each $\alpha$, much larger than a black surface planet with a similar
value of $f_{\rm c}$, in particular at longer wavelengths. 
This variability is caused by the highly polarized glint appearing 
and disappearing from behind the patchy clouds.  

{\em The degree of polarization $P$} of light reflected by a cloud-free
ocean planet is slightly lower at $\lambda = 350$~nm, 443~nm, and 550~nm
compared to that of a cloud-free black surface planet, mostly due to the larger 
flux $F$ that is reflected by the ocean. 
Because the probability distribution of the slopes of the wave facets 
and the shadow function affect fluxes $F$ and $Q$ in the same way,
the dependence of $P$ on the wind speed $v$ is only due to the 
white cap distribution and scattering within the atmosphere, and
thus relatively small.
Indeed, for $v=13$~m/s, the maximum $P$ is about 5\% smaller than
for $v=7$~m/s at $\lambda=865$~nm and using a whitecap albedo of 0.22.
The phase angle where $P$ is maximum, increases from around $90^\circ$ 
to almost twice the Brewster angle of 106$^\circ$ with increasing 
$\lambda$, as the optical thickness of the atmosphere decreases
and the contribution of the ocean to the reflected signal becomes more
dominant. The location of the maximum $P$, however, depends on the
surface pressure $p_{\rm s}$. 
Indeed, an ocean-induced shift of the phase angle where $P$ is maximum
towards larger values only occurs if $p_{\rm s}$ is small
(if $p_{\rm s}$ is large, the presence of an ocean does not lead to
a phase angle shift).

Like in $Q$, the variability in $P$ of an ocean planet, 
especially at the longer wavelengths and for all wind speeds,
due to the glint appearing and disappearing from behind the clouds,
would be much larger than the variability in $P$ of a black surface planet.
A measurement of such variability could thus be used to identify
an exo-ocean. 

An ocean planet will change color in $P$, but only if there are clouds.
The color change phase angle $\alpha_{\rm cc}$ strongly increases with 
$f_{\rm c}$, from about 88$^\circ$ for $f_{\rm c}=0.03$ 
to about
161$^\circ$ for $f_{\rm c} = 0.98$. 
Note that this range of $\alpha_{\rm cc}$ is much broader than that for
$F$ or $Q$, and it includes phase angles with a favorite planet-star
separation. For $f_{\rm c} < 0.03$, an ocean planet does not change 
color in $P$. It would be interesting to investigate whether such low
cloud fractions were to be expected in the presence of an ocean. 
The $P$ phase curves as derived from Earth-shine measurements by
\citep{2019Sterzik} show crossings that indicate cloud
coverage fractions $f_{\rm c}$ between 0.2 and 0.6.

While these cloud coverage fractions appear to be realistic for the Earth,
the Earth-shine measurements covered a time period of several years, 
and two regions on Earth:
the Pacific ocean (west of Chile) and the Atlantic ocean and South
America (east of Chile). In addition, although \citet{2019Sterzik}
took careful measures to correct their observations for the depolarizing
reflection by the lunar surface, the Earth's polarization phase functions
might still include some unknown, and possibly wavelength dependent 
residual of the reflection by the moon, that could influence 
$\alpha_{\rm cc}$ as derived from the curves.
Indeed, flux and polarization measurements of sunlight that is 
reflected by the whole Earth, across the visible spectrum, 
covering our planet's daily rotation and all phase angles, 
would allow the confirmation of using the color change of, 
in particular, $Q$ for identifying the presence of an ocean.
Such measurements (which would also show the rainbow feature
that is not covered in the Earth-shine data) could be performed
from the lunar surface or from a lunar orbiter 
\citep[for measurement concepts, see][]{2012P&SS...74..202K,2016OExpr..2421435H}.

Our simulations only include scattering by the atmospheric gas
and by clouds consisting of liquid water droplets. Scattering and 
absorption by aerosol particles has not
been included. Over the ocean, most aerosol particles are expected 
to be sea salt particles. Note that most of these aerosol particles 
will reside in the boundary layer, below the clouds. 
The influence of aerosols on the color change of a
planet would depend on the aerosol column number density, and the
aerosol microphysical properties, such as size, shape, and composition.
The observation of the color change in $P$ in Earth-shine measurements
by \citet{2019Sterzik} strongly suggests that the influence
of aerosol particles will be minor, but a follow-up study with 
realistic ocean-type aerosols added to our atmosphere model, will 
give more insight into their role. In such a follow-up study, also the influence of oriented ice crystals
can be investigated.

In our numerical simulations, we assume that the incident starlight is
wavelength independent. This assumption is irrelevant for our  
results, as measured fluxes $F$ and $Q$ could straightforwardly be
normalized to the color of the incident stellar 
light\footnote{By doing that, inelastic scattering processes, such as 
rotational Raman scattering, would be ignored, but this would only
affect high spectral resolution features, such
as absorption lines and Fraunhofer lines.}.
The degree of polarization $P$ is of course independent of any
wavelength dependence of the incident stellar 
spectrum\footnote{Again, apart from high
spectral resolution features due to e.g.\ rotational Raman scattering.}. 

Our simulations provide yet another example of the value of polarimetry
as a tool for exoplanet characterization: especially the detection of a color change of the polarized flux $Q$ would uniquely identify
a liquid surface ocean, 
and a color
change in $P$ would indicate an ocean that is at least partly covered
by clouds.
The phase angle $\alpha_{\rm cc}$ at which the color change in $Q$ 
takes place depends on the cloud coverage fraction $f_{\rm c}$: 
$\alpha_{\rm cc} \sim 123^\circ$ for a
cloud-free planet ($f_{\rm c}=0.0$) and 
$\alpha_{\rm cc} \sim 157^\circ$ for an almost 
fully clouded planet ($f_{\rm c}=0.98$) where the color change starts to disappear.
The color change in $P$ takes place at phase angles from 
about $88^\circ$ for $f_{\rm c}= 0.03$ to $\sim 161^\circ$ 
for $f_{\rm c} = 0.98$.
Whether or not an exoplanet ever attains the phase angle at which a
color change takes place, depends on the inclination angle
$i$ of the planetary orbit, as can be seen in Fig.~\ref{fig_crossing}. 
A phase angle of 123$^\circ$ is attained
along orbits with an inclination angle of at least 33$^\circ$, while 
for a phase angle of  161$^\circ$, the orbital inclination angle has
to be at least 71$^\circ$. For intermediate cloud fractions, such
as that on Earth, inclination angles larger than about 40$^\circ$
are required. 

Current and near future space missions such as 
NASA's TESS \citep{2015JATIS...1a4003R} and ESA's PLATO mission
\citep{2014ExA....38..249R} will provide transiting exoplanets,
thus with $i$ close to 90$^\circ$,
around bright, nearby stars. These planets will be excellent targets
for searching for a color change in the polarized light signal. 
This color change can be measured at smaller phase angles,
where the angular separation between the star and the planet is larger,
than the glint, i.e.\ the increased reflected flux $F$ of starlight on
the ocean surface, that was proposed as exo-ocean identifier by 
e.g. \citet{Williams2008} and \citet{Robinsonetal2010} 
(our Fig.~\ref{fig_patchycloudsF} shows the glint in $F$ around
$\alpha=160^\circ$ for a wind speed of 1 m/s). Thus, using 
polarization to identify exo-oceans is not limited to crescent planetary
phase angles.

We have shown that the color change in total flux $F$ can also occur
with a zero surface albedo planet, with clouds below a Rayleigh scattering 
gas layer, while color changes in $Q$ and $P$ only occur for the ocean
planets. We have also shown that the lack of a color change in $F$ could
mistakenly be interpreted as a negative for an exo-ocean in case of high
surface pressures. The color change in $Q$ remains present at high 
surface pressures (see Fig. \ref{fig_RGB_surfacepressure}). 
Furthermore, as shown by \citet{Cowanfalseglint2012}, the relative
contribution of high planetary latitudes, which are more likely to be 
covered by ice and/or snow, to the disk-integrated fluxes increases 
with increasing phase angle towards the crescent phase, 
creating a false-positive detection of the glint. 
However, unless the ice is perfectly clean, we would not expect this increased contribution to yield a 
color change in $P$ and/or $Q$. 
Polarimetry would thus be a more robust tool against the 
false-positive identification of exo-oceans. 
Dedicated numerical simulations would be required to confirm this.

The crucial role that liquid water plays for life on Earth
should spur feasibility studies into measuring the color change 
of an exoplanet in polarized flux and/or degree of polarization
by future telescopes that are designed to measure starlight that
is reflected by exoplanets. Apart from EPOL \citep[]{2010SPIE.7735E..6GK}, 
the imaging polarimeter for EPICS, the Exoplanet Imaging Camera and 
Spectrograph for the European Extremely Large Telescope
\citep[]{2010SPIE.7735E..2EK}, current space telescope concepts,
such as LUVOIR \citep[]{2018arXiv180909668T} and HabEx 
\citep[]{2018arXiv180909674G} do not include polarimetry across
the visible,\footnote{LUVOIR does include POLLUX, a polarimeter 
at ultraviolet wavelengths \citep[]{2018SPIE10699E..3BB}}
and will thus not be able to measure the color changes.

While our numerical simulations pertain to the planet-signal
only, and do not include any background starlight, the color
change in polarized flux $Q$ and/or degree of polarization $P$ 
could also be searched for in the combined light of the star
and the planet. Obviously, $P$ of this combined light would
be very small because of the added, mostly unpolarized stellar flux, but extracting this very small planet signal from the 
background stellar signal and noise could possibly be facilitated using the 
variation of $P$ along the planetary orbit, and the knowledge 
that the color change would be limited to the planetary signal.
 
\begin{acknowledgements}
The authors thank Johan de Haan, Jacek Chowdhary, and Mike Zugger for 
sharing their knowledge and insights on the topic of this paper, and they 
thank the anonymous referee and the editor for their constructive 
comments.
\end{acknowledgements}


\begin{thebibliography}{79}

\bibitem[{{Bailey}(2007)}]{2007AsBio...7..320B}
{Bailey}, J. 2007, Astrobiology, 7, 320

\bibitem[{{Berdyugina} {et~al.}(2011){Berdyugina}, {Berdyugin}, {Fluri}, \&
  {Piirola}}]{2011ApJ...728L...6B}
{Berdyugina}, S.~V., {Berdyugin}, A.~V., {Fluri}, D.~M., \& {Piirola}, V. 2011,
  \apjl, 728, L6

\bibitem[{{Bouret} {et~al.}(2018){Bouret}, {Neiner}, {G{\'o}mez de Castro},
  {Evans}, {Gaensicke}, {Shore}, {Fossati}, {Gry}, {Charlot}, {Marin},
  {Noterdaeme}, \& {Chaufray}}]{2018SPIE10699E..3BB}
{Bouret}, J.-C., {Neiner}, C., {G{\'o}mez de Castro}, A.~I., {et~al.} 2018, in
  Society of Photo-Optical Instrumentation Engineers (SPIE) Conference Series,
  Vol. 10699, Space Telescopes and Instrumentation 2018: Ultraviolet to Gamma
  Ray, 106993B

\bibitem[{{Bre\'on}(1993)}]{Breon1993}
{Bre\'on}, F.~M. 1993, Remote Sensing of Environment, 43, 179

\bibitem[{{Br{\'e}on} \& {Dubrulle}(2004)}]{2004JAtS...61.2888B}
{Br{\'e}on}, F.-M. \& {Dubrulle}, B. 2004, Journal of Atmospheric Sciences, 61,
  2888

\bibitem[{Buenzli \& Schmid(2009)}]{buenzli2009grid}
Buenzli, E. \& Schmid, H.~M. 2009, \aap, 504, 259

\bibitem[{{Chepfer}(1999)}]{1999JQSRT..63..521C}
{Chepfer}, H. 1999, \jqsrt, 63, 521

\bibitem[{{Chowdhary}(1999)}]{Chowdhardy1999PhDthesis}
{Chowdhary}, J. 1999, PhD thesis, Columbia University

\bibitem[{{Chowdhary} {et~al.}(2006){Chowdhary}, {Cairns}, \&
  {Travis}}]{Chowdharyetal2006}
{Chowdhary}, J., {Cairns}, B., \& {Travis}, L.~D. 2006, Applied Optics, 45,
  5542

\bibitem[{{Cotton} {et~al.}(2017){Cotton}, {Marshall}, {Bailey},
  {Kedziora-Chudczer}, {Bott}, {Marsden}, \& {Carter}}]{2017MNRAS.467..873C}
{Cotton}, D.~V., {Marshall}, J.~P., {Bailey}, J., {et~al.} 2017, \mnras, 467,
  873

\bibitem[{{Cowan} {et~al.}(2012){Cowan}, {Abbot}, \&
  {Voigt}}]{Cowanfalseglint2012}
{Cowan}, N.~B., {Abbot}, D.~S., \& {Voigt}, A. 2012, \apj, 752, L3

\bibitem[{{Cowan} {et~al.}(2009){Cowan}, {Agol}, {Meadows}, {Robinson},
  {Livengood}, {Deming}, {Lisse}, {A'Hearn}, {Wellnitz}, {Seager},
  {Charbonneau}, \& {EPOXI Team}}]{Cowanetal2009}
{Cowan}, N.~B., {Agol}, E., {Meadows}, V.~S., {et~al.} 2009, \apj, 700, 915

\bibitem[{{Cox} \& {Munk}(1954)}]{CoxandMunk1954}
{Cox}, C. \& {Munk}, W. 1954, Journal of the Optical Society of America
  (1917-1983), 44, 838

\bibitem[{Cox \& Munk(1956)}]{cox1956slopes}
Cox, C. \& Munk, W. 1956, Slopes of the sea surface deduced from photographs of
  sun glitter, Bulletin of the Scripps Institution of Oceanography of the
  University of California (University of California Press \& Cambridge
  University Press)

\bibitem[{{de Haan} {et~al.}(1987){de Haan}, {Bosma}, \& {Hovenier}}]{deHaan87}
{de Haan}, J.~F., {Bosma}, P.~B., \& {Hovenier}, J.~W. 1987, AAP, 183, 371

\bibitem[{De~Rooij \& van~der Stap(1984)}]{de1984expansion}
De~Rooij, W.~A. \& van~der Stap, C. C. A.~H. 1984, \aap, 131, 237

\bibitem[{{Deming} {et~al.}(2013){Deming}, {Wilkins}, {McCullough}, {Burrows},
  {Fortney}, {Agol}, {Dobbs-Dixon}, {Madhusudhan}, {Crouzet}, {Desert},
  {Gilliland}, {Haynes}, {Knutson}, {Line}, {Magic}, {Mandell}, {Ranjan},
  {Charbonneau}, {Clampin}, {Seager}, \& {Showman}}]{Deming2013}
{Deming}, D., {Wilkins}, A., {McCullough}, P., {et~al.} 2013, \apj, 774, 95

\bibitem[{{Fraine} {et~al.}(2014){Fraine}, {Deming}, {Benneke}, {Knutson},
  {Jord{\'a}n}, {Espinoza}, {Madhusudhan}, {Wilkins}, \&
  {Todorov}}]{Fraine2014}
{Fraine}, J., {Deming}, D., {Benneke}, B., {et~al.} 2014, \nat, 513, 526

\bibitem[{{Garcia}(2012{\natexlab{a}})}]{Garcia2012}
{Garcia}, R.~D.~M. 2012{\natexlab{a}}, Journal of Quantitative Spectroscopy and
  Radiative Transfer, 113, 306

\bibitem[{{Garcia}(2012{\natexlab{b}})}]{Garcia2012ResponsetoComment}
{Garcia}, R.~D.~M. 2012{\natexlab{b}}, Journal of Quantitative Spectroscopy and
  Radiative Transfer, 113, 2251

\bibitem[{{Garc{\'{\i}}a Mu{\~n}oz}(2015)}]{2015IJAsB..14..379G}
{Garc{\'{\i}}a Mu{\~n}oz}, A. 2015, International Journal of Astrobiology, 14,
  379

\bibitem[{{Gaudi} {et~al.}(2018){Gaudi}, {Seager}, {Mennesson}, {Kiessling},
  {Warfield}, {Kuan}, {Cahoy}, {Clarke}, {Domagal-Goldman}, {Feinberg},
  {Guyon}, {Kasdin}, {Mawet}, {Robinson}, {Rogers}, {Scowen}, {Somerville},
  {Stapelfeldt}, {Stark}, {Stern}, {Turnbull}, {Martin}, {Alvarez-Salazar},
  {Amini}, {Arnold}, {Balasubramanian}, {Baysinger}, {Blais}, {Brooks},
  {Calvet}, {Cormarkovic}, {Cox}, {Danner}, {Davis}, {Dorsett}, {Effinger},
  {Eng}, {Garcia}, {Gaskin}, {Harris}, {Howe}, {Knight}, {Krist}, {Levine},
  {Li}, {Lisman}, {Mandic}, {Marchen}, {Marrese-Reading}, {McGowen},
  {Miyaguchi}, {Morgan}, {Nemati}, {Nikzad}, {Nissen}, {Novicki}, {Perrine},
  {Redding}, {Richards}, {Rud}, {Scharf}, {Serabyn}, {Shaklan}, {Smith},
  {Stahl}, {Stahl}, {Tang}, {Van Buren}, {Villalvazo}, {Warwick}, {Webb},
  {Wofford}, {Woo}, {Wood}, {Ziemer}, {Douglas}, {Faramaz}, {Hildebrand t},
  {Meshkat}, {Plavchan}, {Ruane}, \& {Turner}}]{2018arXiv180909674G}
{Gaudi}, B.~S., {Seager}, S., {Mennesson}, B., {et~al.} 2018, arXiv e-prints,
  arXiv:1809.09674

\bibitem[{{Hale} \& {Querry}(1973)}]{hale1973optical}
{Hale}, G.~M. \& {Querry}, M.~R. 1973, Applied Optics, 12, 555

\bibitem[{{Hansen} \& {Hovenier}(1974)}]{1974JAtS...31.1137H}
{Hansen}, J.~E. \& {Hovenier}, J.~W. 1974, Journal of Atmospheric Sciences, 31,
  1137

\bibitem[{Hansen \& Travis(1974)}]{Hansen1974}
Hansen, J.~E. \& Travis, L.~D. 1974, Space Science Reviews, 16, 527

\bibitem[{{Hoeijmakers} {et~al.}(2016){Hoeijmakers}, {Arts}, {Snik}, {Keller},
  {Stam}, \& {Kuiper}}]{2016OExpr..2421435H}
{Hoeijmakers}, H.~J., {Arts}, M.~L.~J., {Snik}, F., {et~al.} 2016, Optics
  Express, 24, 21435

\bibitem[{{Hovenier} {et~al.}(2004){Hovenier}, {van der Mee}, \&
  {Domke}}]{2004Hovenier}
{Hovenier}, J.~W., {van der Mee}, C., \& {Domke}, H. 2004, {Transfer of
  Polarized Light in Planetary Atmospheres; Basic Concepts and Practical
  Methods} (Kluwer, Dordrecht; Springer, Berlin)

\bibitem[{{Hovenier} \& {van der Mee}(1983)}]{1983A&A...128....1H}
{Hovenier}, J.~W. \& {van der Mee}, C.~V.~M. 1983, \aap, 128, 1

\bibitem[{{Hsiung}(1986)}]{Hsiung1986}
{Hsiung}, J. 1986, Journal of Geophysical Research, 91, 10,585

\bibitem[{{Kaltenegger} \& {Traub}(2009)}]{2009ApJ...698..519K}
{Kaltenegger}, L. \& {Traub}, W.~A. 2009, \apj, 698, 519

\bibitem[{{Karalidi} \& {Stam}(2012)}]{KaralidiStam2012}
{Karalidi}, T. \& {Stam}, D.~M. 2012, \aap, 546, A56

\bibitem[{{Karalidi} {et~al.}(2011){Karalidi}, {Stam}, \&
  {Hovenier}}]{2011A&A...530A..69K}
{Karalidi}, T., {Stam}, D.~M., \& {Hovenier}, J.~W. 2011, \aap, 530, A69

\bibitem[{{Karalidi} {et~al.}(2012{\natexlab{a}}){Karalidi}, {Stam}, \&
  {Hovenier}}]{Karalidi12}
{Karalidi}, T., {Stam}, D.~M., \& {Hovenier}, J.~W. 2012{\natexlab{a}}, \aap,
  548, A90

\bibitem[{{Karalidi} {et~al.}(2012{\natexlab{b}}){Karalidi}, {Stam}, {Snik},
  {Bagnulo}, {Sparks}, \& {Keller}}]{2012P&SS...74..202K}
{Karalidi}, T., {Stam}, D.~M., {Snik}, F., {et~al.} 2012{\natexlab{b}},
  \planss, 74, 202

\bibitem[{{Kasper} {et~al.}(2010){Kasper}, {Beuzit}, {Verinaud}, {Gratton},
  {Kerber}, {Yaitskova}, {Boccaletti}, {Thatte}, {Schmid}, {Keller}, {Baudoz},
  {Abe}, {Aller-Carpentier}, {Antichi}, {Bonavita}, {Dohlen}, {Fedrigo},
  {Hanenburg}, {Hubin}, {Jager}, {Korkiakoski}, {Martinez}, {Mesa}, {Preis},
  {Rabou}, {Roelfsema}, {Salter}, {Tecza}, \& {Venema}}]{2010SPIE.7735E..2EK}
{Kasper}, M., {Beuzit}, J.-L., {Verinaud}, C., {et~al.} 2010, in \procspie,
  Vol. 7735, Ground-based and Airborne Instrumentation for Astronomy III,
  77352E--77352E--9

\bibitem[{{Keller} {et~al.}(2010){Keller}, {Schmid}, {Venema}, {Hanenburg},
  {Jager}, {Kasper}, {Martinez}, {Rigal}, {Rodenhuis}, {Roelfsema}, {Snik},
  {Verinaud}, \& {Yaitskova}}]{2010SPIE.7735E..6GK}
{Keller}, C.~U., {Schmid}, H.~M., {Venema}, L.~B., {et~al.} 2010, in \procspie,
  Vol. 7735, Ground-based and Airborne Instrumentation for Astronomy III,
  77356G

\bibitem[{{Kemp} {et~al.}(1987){Kemp}, {Henson}, {Steiner}, \&
  {Powell}}]{Kemp1987}
{Kemp}, J.~C., {Henson}, G.~D., {Steiner}, C.~T., \& {Powell}, E.~R. 1987,
  \nat, 326, 270

\bibitem[{{Koepke}(1984)}]{koepke1984effective}
{Koepke}, P. 1984, Applied Optics, 23, 1816

\bibitem[{Kostogryz {et~al.}(2015)Kostogryz, Yakobchuk, \&
  Berdyugina}]{kostogryz2015polarization}
Kostogryz, N., Yakobchuk, T., \& Berdyugina, S. 2015, \apj, 806, 97

\bibitem[{{Kreidberg} {et~al.}(2014){Kreidberg}, {Bean}, {D{\'e}sert}, {Line},
  {Fortney}, {Madhusudhan}, {Stevenson}, {Showman}, {Charbonneau},
  {McCullough}, {Seager}, {Burrows}, {Henry}, {Williamson}, {Kataria}, \&
  {Homeier}}]{Kreidberg2014}
{Kreidberg}, L., {Bean}, J.~L., {D{\'e}sert}, J.-M., {et~al.} 2014, \apj, 793,
  L27

\bibitem[{{Lustig-Yaeger} {et~al.}(2018){Lustig-Yaeger}, {Meadows}, {Tovar
  Mendoza}, {Schwieterman}, {Fujii}, {Luger}, \&
  {Robinson}}]{Lustig-Yaeger2018}
{Lustig-Yaeger}, J., {Meadows}, V.~S., {Tovar Mendoza}, G., {et~al.} 2018,
  \apj, 156, 301

\bibitem[{{Marshak} {et~al.}(2017){Marshak}, {V{\'a}rnai}, \&
  {Kostinski}}]{2017GeoRL..44.5197M}
{Marshak}, A., {V{\'a}rnai}, T., \& {Kostinski}, A. 2017, \grl, 44, 5197

\bibitem[{{Mishchenko} {et~al.}(2000){Mishchenko}, {Hovenier}, \&
  {Travis}}]{BookMishchenko2000}
{Mishchenko}, M.~I., {Hovenier}, J.~W., \& {Travis}, L.~D. 2000, {Light
  scattering by nonspherical particles : theory, measurements, and
  applications} (Sand Diego: Academic Press)

\bibitem[{{Mishchenko} \& {Travis}(1997)}]{MishchenkoandTravis1997}
{Mishchenko}, M.~I. \& {Travis}, L.~D. 1997, Journal of Geophysical Research,
  102, No.D14:16989

\bibitem[{{Monahan} \& {{\'O} Muircheartaigh}(1980)}]{Monahan1980}
{Monahan}, E.~C. \& {{\'O} Muircheartaigh}, I. 1980, Journal of Physical
  Oceanography, 10, 2094

\bibitem[{{Morel}(1974)}]{morel1974optical}
{Morel}, A. 1974, Optical Aspects of Oceanography, 1, 1

\bibitem[{{Morel} \& {Maritorena}(2001)}]{MorelandMaritorena2001}
{Morel}, A. \& {Maritorena}, S. 2001, Journal of Geophysics Research, 106, 7163

\bibitem[{{Nakajima}(1983)}]{NakajimaandTanaka1983}
{Nakajima}, T. 1983, Journal of Quantitative Spectroscopy and Radiative
  Transfer, 29, 521

\bibitem[{{Noel} \& {Chepfer}(2004)}]{2004JAtS...61.2073N}
{Noel}, V. \& {Chepfer}, H. 2004, Journal of Atmospheric Sciences, 61, 2073

\bibitem[{{Oakley} \& {Cash}(2009)}]{OakleyandCash2009}
{Oakley}, P.~H.~H. \& {Cash}, W. 2009, \apj, 700, 1428

\bibitem[{{Pope} \& {Fry}(1997)}]{popeandfry1997}
{Pope}, R.~M. \& {Fry}, E.~S. 1997, Applied Optics, 36, 8710

\bibitem[{{Rauer} {et~al.}(2014){Rauer}, {Catala}, {Aerts}, {Appourchaux},
  {Benz}, {Brandeker}, {Christensen-Dalsgaard}, {Deleuil}, {Gizon}, {Goupil},
  {G{\"u}del}, {Janot-Pacheco}, {Mas-Hesse}, {Pagano}, {Piotto}, {Pollacco},
  {Santos}, {Smith}, {Su{\'a}rez}, {Szab{\'o}}, {Udry}, {Adibekyan}, {Alibert},
  {Almenara}, {Amaro-Seoane}, {Eiff}, {Asplund}, {Antonello}, {Barnes},
  {Baudin}, {Belkacem}, {Bergemann}, {Bihain}, {Birch}, {Bonfils}, {Boisse},
  {Bonomo}, {Borsa}, {Brand{\~a}o}, {Brocato}, {Brun}, {Burleigh}, {Burston},
  {Cabrera}, {Cassisi}, {Chaplin}, {Charpinet}, {Chiappini}, {Church},
  {Csizmadia}, {Cunha}, {Damasso}, {Davies}, {Deeg}, {D{\'{\i}}az}, {Dreizler},
  {Dreyer}, {Eggenberger}, {Ehrenreich}, {Eigm{\"u}ller}, {Erikson}, {Farmer},
  {Feltzing}, {de Oliveira Fialho}, {Figueira}, {Forveille}, {Fridlund},
  {Garc{\'{\i}}a}, {Giommi}, {Giuffrida}, {Godolt}, {Gomes da Silva},
  {Granzer}, {Grenfell}, {Grotsch-Noels}, {G{\"u}nther}, {Haswell}, {Hatzes},
  {H{\'e}brard}, {Hekker}, {Helled}, {Heng}, {Jenkins}, {Johansen},
  {Khodachenko}, {Kislyakova}, {Kley}, {Kolb}, {Krivova}, {Kupka}, {Lammer},
  {Lanza}, {Lebreton}, {Magrin}, {Marcos-Arenal}, {Marrese}, {Marques},
  {Martins}, {Mathis}, {Mathur}, {Messina}, {Miglio}, {Montalban}, {Montalto},
  {Monteiro}, {Moradi}, {Moravveji}, {Mordasini}, {Morel}, {Mortier},
  {Nascimbeni}, {Nelson}, {Nielsen}, {Noack}, {Norton}, {Ofir}, {Oshagh},
  {Ouazzani}, {P{\'a}pics}, {Parro}, {Petit}, {Plez}, {Poretti}, {Quirrenbach},
  {Ragazzoni}, {Raimondo}, {Rainer}, {Reese}, {Redmer}, {Reffert},
  {Rojas-Ayala}, {Roxburgh}, {Salmon}, {Santerne}, {Schneider}, {Schou},
  {Schuh}, {Schunker}, {Silva-Valio}, {Silvotti}, {Skillen}, {Snellen}, {Sohl},
  {Sousa}, {Sozzetti}, {Stello}, {Strassmeier}, {{\v S}vanda}, {Szab{\'o}},
  {Tkachenko}, {Valencia}, {Van Grootel}, {Vauclair}, {Ventura}, {Wagner},
  {Walton}, {Weingrill}, {Werner}, {Wheatley}, \&
  {Zwintz}}]{2014ExA....38..249R}
{Rauer}, H., {Catala}, C., {Aerts}, C., {et~al.} 2014, Experimental Astronomy,
  38, 249

\bibitem[{{Ricker} {et~al.}(2015){Ricker}, {Winn}, {Vanderspek}, {Latham},
  {Bakos}, {Bean}, {Berta-Thompson}, {Brown}, {Buchhave}, {Butler}, {Butler},
  {Chaplin}, {Charbonneau}, {Christensen-Dalsgaard}, {Clampin}, {Deming},
  {Doty}, {De Lee}, {Dressing}, {Dunham}, {Endl}, {Fressin}, {Ge}, {Henning},
  {Holman}, {Howard}, {Ida}, {Jenkins}, {Jernigan}, {Johnson}, {Kaltenegger},
  {Kawai}, {Kjeldsen}, {Laughlin}, {Levine}, {Lin}, {Lissauer}, {MacQueen},
  {Marcy}, {McCullough}, {Morton}, {Narita}, {Paegert}, {Palle}, {Pepe},
  {Pepper}, {Quirrenbach}, {Rinehart}, {Sasselov}, {Sato}, {Seager},
  {Sozzetti}, {Stassun}, {Sullivan}, {Szentgyorgyi}, {Torres}, {Udry}, \&
  {Villasenor}}]{2015JATIS...1a4003R}
{Ricker}, G.~R., {Winn}, J.~N., {Vanderspek}, R., {et~al.} 2015, Journal of
  Astronomical Telescopes, Instruments, and Systems, 1, 014003

\bibitem[{{Robinson} {et~al.}(2010){Robinson}, {Meadows}, \&
  {Crisp}}]{Robinsonetal2010}
{Robinson}, T.~D., {Meadows}, V.~S., \& {Crisp}, D. 2010, \apj, 721, L67

\bibitem[{{Rossi} {et~al.}(2018){Rossi}, {Berzosa-Molina}, \&
  {Stam}}]{rossi2018pymiedap}
{Rossi}, L., {Berzosa-Molina}, J., \& {Stam}, D.~M. 2018, \aap, 616, A147

\bibitem[{Rossi \& Stam(2017)}]{Rossi2017}
Rossi, L. \& Stam, D.~M. 2017, \aap

\bibitem[{{Rossi} \& {Stam}(2017)}]{2017A&A...607A..57R}
{Rossi}, L. \& {Stam}, D.~M. 2017, \aap, 607, A57

\bibitem[{{Rossi} \& {Stam}(2018)}]{2018A&A...616A.117R}
{Rossi}, L. \& {Stam}, D.~M. 2018, \aap, 616, A117

\bibitem[{{Sancer}(1969)}]{Sancer1969}
{Sancer}, M. 1969, IEEE Transactions on Antennas and Propagation, 17, 577

\bibitem[{{Smith}(1967)}]{Smith1967}
{Smith}, B. 1967, IEEE Transactions on Antennas and Propagation, 15, 668

\bibitem[{{Smith} \& {Baker}(1981)}]{SmithandBaker1981}
{Smith}, R.~C. \& {Baker}, K.~S. 1981, Applied Optics, 20, 177

\bibitem[{{Sogandares} \& {Fry}(1997)}]{SogandaresandFry1997}
{Sogandares}, F.~M. \& {Fry}, E.~S. 1997, Applied Optics, 36, 8699

\bibitem[{Stam(2008)}]{Stam08}
Stam, D.~M. 2008, A\&A, 482, 989

\bibitem[{{Stam} {et~al.}(2006){Stam}, {de Rooij}, {Cornet}, \&
  {Hovenier}}]{Stam2006}
{Stam}, D.~M., {de Rooij}, W.~A., {Cornet}, G., \& {Hovenier}, J.~W. 2006,
  \aap, 452, 669

\bibitem[{Stam \& Hovenier(2005)}]{stametal05}
Stam, D.~M. \& Hovenier, J.~W. 2005, \aap, 444, 275

\bibitem[{Sterzik} {et~al.}(2019){Sterzik}, {Bagnulo}, {Stam}, {Emde}, \&
  {Manev}]{2019Sterzik}
{Sterzik}, M.~F., {Bagnulo}, S., {Stam}, D.~M., {Emde}, C., \& {Manev}, M.
  2019, \aap, 662, A41

\bibitem[{{Sun} \& {Lukashin}(2013)}]{SunLukashin2013}
{Sun}, W. \& {Lukashin}, C. 2013, Atmospheric Chemistry \& Physics, 13, 10303

\bibitem[{{Talens} {et~al.}(2017){Talens}, {Spronck}, {Lesage}, {Otten},
  {Stuik}, {Pollacco}, \& {Snellen}}]{2017A&A...601A..11T}
{Talens}, G.~J.~J., {Spronck}, J.~F.~P., {Lesage}, A.~L., {et~al.} 2017, \aap,
  601, A11

\bibitem[{{The LUVOIR Team}(2018)}]{2018arXiv180909668T}
{The LUVOIR Team}. 2018, arXiv e-prints, arXiv:1809.09668

\bibitem[{{Tinetti} {et~al.}(2007){Tinetti}, {Vidal-Madjar}, {Liang},
  {Beaulieu}, {Yung}, {Carey}, {Barber}, {Tennyson}, {Ribas}, {Allard},
  {Ballester}, {Sing}, \& {Selsis}}]{Tinetti2007}
{Tinetti}, G., {Vidal-Madjar}, A., {Liang}, M.-C., {et~al.} 2007, \nat, 448,
  169

\bibitem[{Tsang {et~al.}(1985)Tsang, Kong, \& Shin}]{tsang1985theory}
Tsang, L., Kong, J.~A., \& Shin, R.~T. 1985, Theory of microwave remote sensing
  (New York: Wiley Interscience)

\bibitem[{{Wiktorowicz} \& {Stam}(2015)}]{2015psps.book..439W}
{Wiktorowicz}, S.~J. \& {Stam}, D.~M. 2015, {Exoplanets}, ed. L.~{Kolokolova},
  J.~{Hough}, \& A.-C. {Levasseur-Regourd}, 439

\bibitem[{{Williams} \& {Gaidos}(2008)}]{Williams2008}
{Williams}, D.~M. \& {Gaidos}, E. 2008, \icarus, 195, 927

\bibitem[{{Xu} {et~al.}(2016){Xu}, {Dubovik}, {Zhai}, {Diner}, {Kalashnikova},
  {Seidel}, {Litvinov}, {Bovchaliuk}, {Garay}, {van Harten}, \&
  {Davis}}]{Xuetal2016}
{Xu}, F., {Dubovik}, O., {Zhai}, P.-W., {et~al.} 2016, Atmospheric Measurement
  Techniques, 9, 2877

\bibitem[{{Zhai} {et~al.}(2010){Zhai}, {Hu}, {Chowdhary}, {Trepte}, {Lucker},
  \& {Josset}}]{Zhaietal2010}
{Zhai}, P.-W., {Hu}, Y., {Chowdhary}, J., {et~al.} 2010, Journal of
  Quantitative Spectroscopy and Radiative Transfer, 111, 1025

\bibitem[{{Zhai} {et~al.}(2012){Zhai}, {Kattawar}, \& {Hu}}]{Zhaietal2012}
{Zhai}, P.-W., {Kattawar}, G.~W., \& {Hu}, Y. 2012, Journal of Quantitative
  Spectroscopy and Radiative Transfer, 113, 1981

\bibitem[{{Zugger} {et~al.}(2010){Zugger}, {Kasting}, {Williams}, {Kane}, \&
  {Philbrick}}]{Zugger2010}
{Zugger}, M.~E., {Kasting}, J.~F., {Williams}, D.~M., {Kane}, T.~J., \&
  {Philbrick}, C.~R. 2010, \apj, 723, 1168

\bibitem[{{Zugger} {et~al.}(2011{\natexlab{a}}){Zugger}, {Kasting}, {Williams},
  {Kane}, \& {Philbrick}}]{ZuggerErratum2011}
{Zugger}, M.~E., {Kasting}, J.~F., {Williams}, D.~M., {Kane}, T.~J., \&
  {Philbrick}, C.~R. 2011{\natexlab{a}}, \apj, 739, 55

\bibitem[{{Zugger} {et~al.}(2011{\natexlab{b}}){Zugger}, {Kasting}, {Williams},
  {Kane}, \& {Philbrick}}]{Zugger2011}
{Zugger}, M.~E., {Kasting}, J.~F., {Williams}, D.~M., {Kane}, T.~J., \&
  {Philbrick}, C.~R. 2011{\natexlab{b}}, \apj, 739, 12

\end{thebibliography}

\clearpage
\begin{appendix}

\section{The reflection matrix of the ocean}
\label{app_equations}

Because we fully include the polarization of light, we describe 
scattering, reflection and transmission processes by $4 \times 4$ 
matrices\footnote{When ignoring circular polarization, the matrices
will be $3 \times 3$.} instead of by scalars.  
The reflection of light by a gas-liquid interface 
(in our case, an air-liquid water interface), with waves shaped by 
randomly oriented, flat facets, is described by the following
matrix ${\bf R}_{\rm I}$ \citep[see][]{MishchenkoandTravis1997,Zhaietal2010}
\begin{eqnarray}
  {\bf R}_{\rm I}(\mu,\mu_0,\phi-\phi_0) & = & 
     S(\mu,\mu_0,\sigma) \hspace{0.1cm} \frac{\pi}{\mu_0}
     \frac{{\rm PDF}(\mu_n,\sigma)}{4 \mu \mu_n} \cdot \nonumber  \\
     & & \hspace{0.5cm} \cdot \hspace{0.3cm} {\bf L}(-\xi_2) \hspace{0.1cm} 
     {\bf R}_{\rm F}(\mu_i,n_1,n_2) 
     \hspace{0.1cm}
     {\bf L}(-\xi_1)
\label{eq_A1}     
\end{eqnarray}
with ${\bf R}_{\rm F}$ the matrix describing the Fresnel reflection
by a flat wave facet, with $n_1$ and $n_2$ the refractive indices
of air and water, respectively \citep[see][Eq.~47]{Zhaietal2010}. In Eq.~\ref{eq_A1}, 
$\mu=\cos\theta$, $\mu_0 =\cos\theta_0$, $\mu_n=\cos\theta_n$,
and $\mu_i = \cos\theta_i$,
with $\theta$, $\theta_0$, $\theta_n$, and $\theta_i$ the angles 
between the local
zenith direction and the direction of propagation of the reflected
light, the incident light (not only the direct beam but also the diffuse 
incident light), the normal on the wave facet, and the Fresnel
reflection angle, respectively.

The matrices ${\bf L}$ are rotation matrices that are used to rotate
the reference plane for the incident light (through the local zenith 
direction and the direction of incidence) to the reference plane for
the reflection by the wave facet (through the local normal on the facet, 
and the directions of incidence and reflection), and next to the 
reference plane of the reflected light (through the local zenith
direction and the direction of reflection)
\citep[see][]{1983A&A...128....1H}, given by
\begin{equation}
   {\bf L}(\xi)= \left[ \begin{array}{cccc}
             1 & 0 & 0 & 0 \\
             0 & \cos 2\xi & \sin 2\xi & 0 \\
             0 & -\sin 2\xi & \cos 2\xi & 0 \\
             0 & 0 & 0 & 1 \\
              \end{array}
              \right],
\label{eq_L}
\end{equation}
with $\xi$ the angle between the two reference planes, measured rotating in
the anti-clockwise direction from the old to the new reference plane when
looking in the direction of propagation  (thus clockwise for an observer that looks towards the planet) ($\xi \geq 0^\circ$). The first rotation in
Eq.~\ref{eq_A1} is over angle $-\xi_1$ and the second rotation 
over angle $-\xi_2$.
 
Furthermore in Eq.~\ref{eq_A1}, ${\rm PDF}$ is the probability 
density function 
of the wave facet inclination angles, as follows 
\citep[see][]{Breon1993,Zhaietal2010}
\begin{equation}
   {\rm PDF}(\mu_n,\sigma) = \frac{1}{\pi\sigma^2 \mu_n^3} 
   {\rm exp}\left(-\frac{1-\mu_n^2}{\sigma^2\mu_n^2}\right).
\end{equation}
The variance of the slopes of the wave facets, $\sigma$, 
depends on the wind speed $v$
(in m/s) and is given by \citep[see][]{CoxandMunk1954}
\begin{equation}
\label{eq:coxmunkrelationv}
   \sigma^2 = 0.003 + 0.00512 \hspace{0.1cm} v.
\end{equation}
A larger wind speed $v$ thus increases the variance in the slopes.

$S$ in Eq.~\ref{eq_A1} describes the shadowing, i.e.\  
the blockage of incident light and reflected light due to inclined
wave facets. For this function we use \citep[see][]{Smith1967,Sancer1969} 
\begin{equation}
   S(\mu,\mu_0,\sigma)= \frac{1}{1+\Lambda(\mu,\sigma) + \Lambda(\mu_0,\sigma)},
\label{eq:shadowingfuction}
\end{equation}
where
\begin{eqnarray*}
   \Lambda(\gamma,\sigma) & = & \frac{1}{2} \left\{ \frac{\sigma}{\gamma} 
        \left[ \frac{1-\gamma^2}{\pi}\right]^{1/2} \right. \\ 
        & & \hspace*{0.6cm} 
        \left. {\rm exp}\left[-\frac{\gamma^2}{\sigma^2(1-\gamma^2)}\right] -  
        {\rm erfc}\left[\frac{\gamma}{\sigma\sqrt{1-\gamma^2}} \right] \right\}.
\label{eq:shadowingterms}
\end{eqnarray*}
Here, ${\rm erfc}$ is the complementary error function. 

Part of the light that is incident on the Fresnel reflecting interface
will be transmitted into the liquid ocean. Once in the ocean, this light
can be scattered, or absorbed, or it can be absorbed by the black 
surface that we assume below the water body. The transmission of 
incident light from above (the star/sun or the sky) through the rough interface
into the water, is described by the following transmission matrix:
\begin{eqnarray}
    {\bf T}_{\rm I}(\mu,\mu_0,\phi-\phi_0) & = & 
        S(\mu,\mu_0,\sigma) \frac{\pi}{\mu_0} 
        \frac{{\rm PDF}(\mu_n,\sigma)}{\mu \mu_n} \cdot \nonumber \\
        & & \hspace{0.5cm} \cdot \hspace{0.3cm} {\bf L}(-\xi_2) \hspace{0.1cm}
        {\bf T}_{\rm F}(\mu_i,n_1,n_2) \hspace{0.1cm} 
        {\bf L}(-\xi_1) \cdot \nonumber \\
        & & \hspace{1.0cm} \cdot \hspace{0.3cm} 
        \frac{n_2^2 \hspace{0.1cm} \mu_t \mu_i}{(n_2 \mu_t - n_1 \mu_i)^2}
\label{eq_transmissionmatrix}
\end{eqnarray}
where $\mu_t = \cos \theta_t$, with $\theta_t$ the transmission angle 
at the wave facet, and ${\bf T}_{\rm F}$ is the Fresnel transmission matrix 
for a flat interface. Note that ${\bf T}_{\rm F}$ contains a factor 
$(n_2/n_1)(\mu_t/\mu_i)$ to account for the changing solid angle of 
the light beam as it is transmitted through the air-water interface
and for the changing flux because of the different permittivity of 
water compared to that of air \citep[see Eq.~48 of][]{Zhaietal2010}. 
The interface reflection and transmission matrices for illumination 
from below, ${\bf R}_{\rm I}^*$ and ${\bf T}_{\rm I}^*$, respectively,
may readily be derived following a similar approach 
\citep[see][]{Zhaietal2010}.

The matrix describing the reflection by the water below
the interface, ${\bf R}_{\rm W}$, is computed using the adding-doubling
algorithm (see Sect.~\ref{sect_surface_model}). The water body is 
bounded below by a black surface (the algorithm can also handle 
non-black surfaces, for example, when modeling
reflection by shallow waters above lightly colored sand).

We compute the reflection matrix, ${\bf R}_{\rm CO}$, of the 'clean' ocean 
body as a whole, i.e.\ of the rough interface and the water body, but 
without white caps, according to \citep{Xuetal2016} 
\begin{eqnarray}
& & {\bf Q}_1 = {\bf R}_{\rm I}^* \hspace{0.1cm} {\bf R}_{\rm W} 
   \label{eq:Q1extended} \\
& & {\bf Q}_{p+1} = {\bf Q}_1 \hspace{0.1cm} {\bf Q}_{p} \\
& & {\bf Q} = \sum_{p=1}^\infty {\bf Q}_p \\
& & {\bf D} = {\bf T}_{\rm I} + {\bf Q} \hspace{0.1cm} {\bf T}_{\rm I}   
   \label{eq:Dextended} \\
& & {\bf U} = {\bf R}_{\rm W} \hspace{0.1cm} {\bf D} \label{eq:Uextended}\\
& & {\bf R}_{\rm WL} = {\bf T}_{\rm I}^* \hspace{0.1cm} {\bf U} 
   \label{eq:Rwlextended} \\
& & {\bf R}_{\rm CO} = {\bf R}_{\rm WL} + {\bf R}_{\rm I}
\end{eqnarray}
where ${\bf D}$, ${\bf T}_{\rm I}$, ${\bf U}$ and ${\bf T}_{\rm I}^*$ are 
\textit{rectangular} super-matrices (see Sect.~\ref{sect_surface_model}) 
and ${\bf R}_{\rm WL}$ describes the reflection by the sub-interface
water body (including the reflections between the water body and the 
interface). 

Finally, we compute ${\bf R}_{\rm O}$, the reflection matrix of the entire 
ocean (interface plus water body plus black surface below the water), and
including white caps as follows \citep[][]{koepke1984effective,SunLukashin2013}
\begin{equation}
   {\bf R}_{\rm O} = q \hspace{0.1cm} {\bf R}_{\rm WC} + (1-q) \hspace{0.1cm} 
                     {\bf R}_{\rm CO}
\label{eq:Rwtot}
\end{equation}
where ${\bf R}_{\rm WC}$ is the matrix describing the Lambertian reflection 
by the white caps with a foam albedo $a_{\rm foam}$ and $q$ is the 
fraction of white caps as determined by the empirical relation of 
\citet{Monahan1980}
\begin{equation}
   q = 2.95 \cdot 10^{-6} v^{3.52}.
\label{eq:whitecapfraction}
\end{equation}
The influence of the white caps and the white cap albedo will be discussed
in Appendix~\ref{app_whitecaps}.

\section{The reflection by white caps}
\label{app_whitecaps}

\begin{figure*}[h!]
\includegraphics[width=0.33\textwidth]{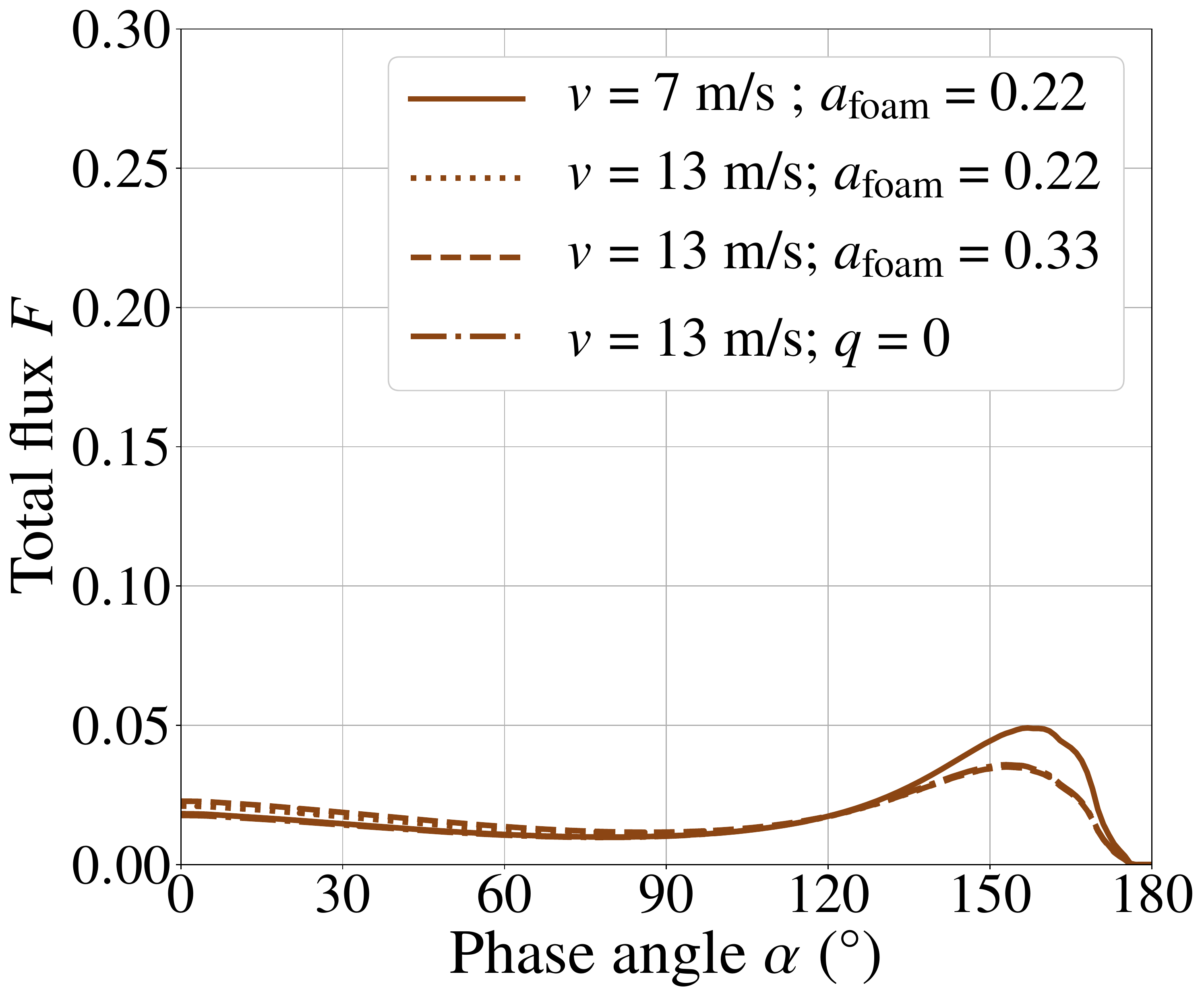}
\includegraphics[width=0.33\textwidth]{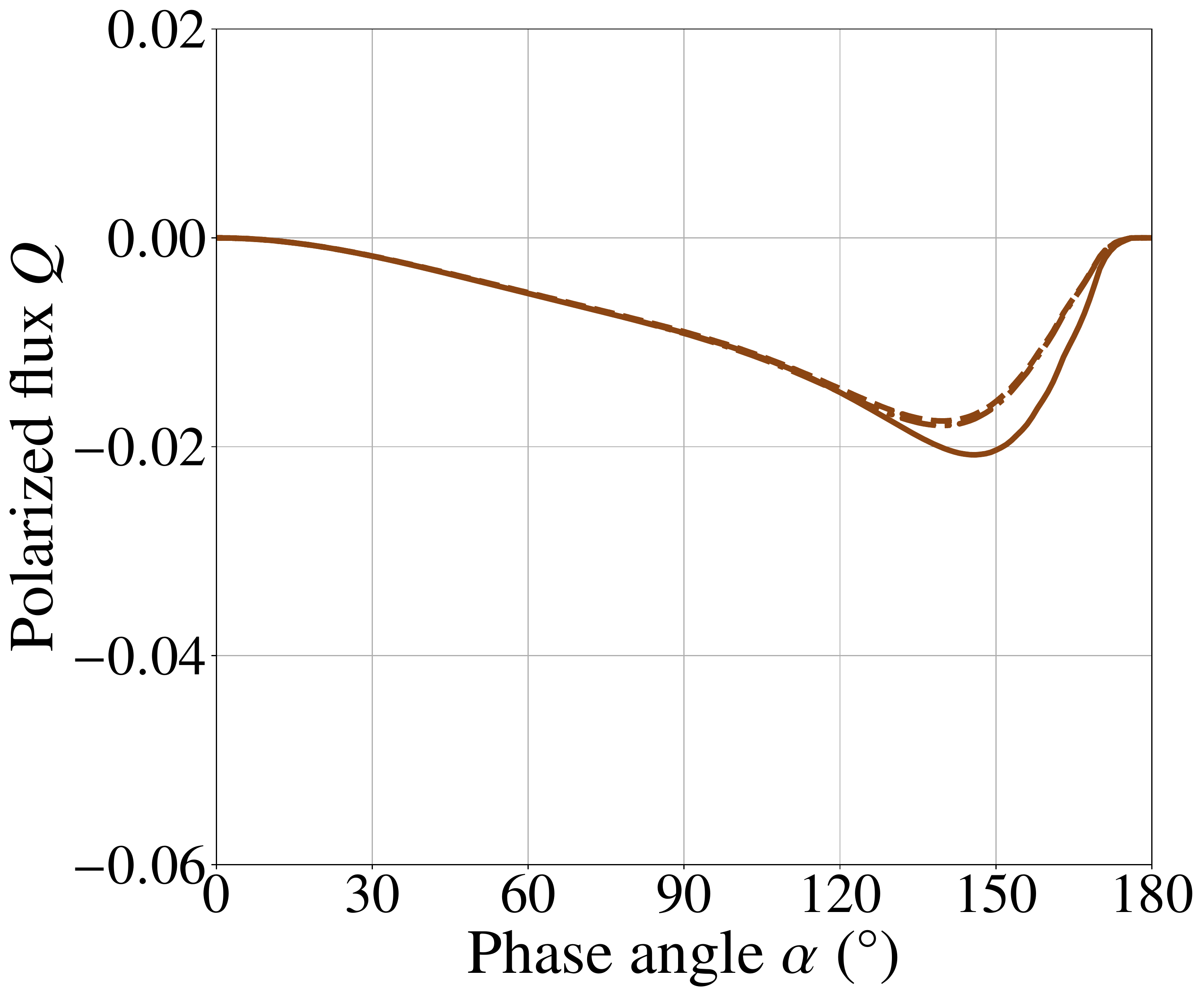}
\includegraphics[width=0.33\textwidth]{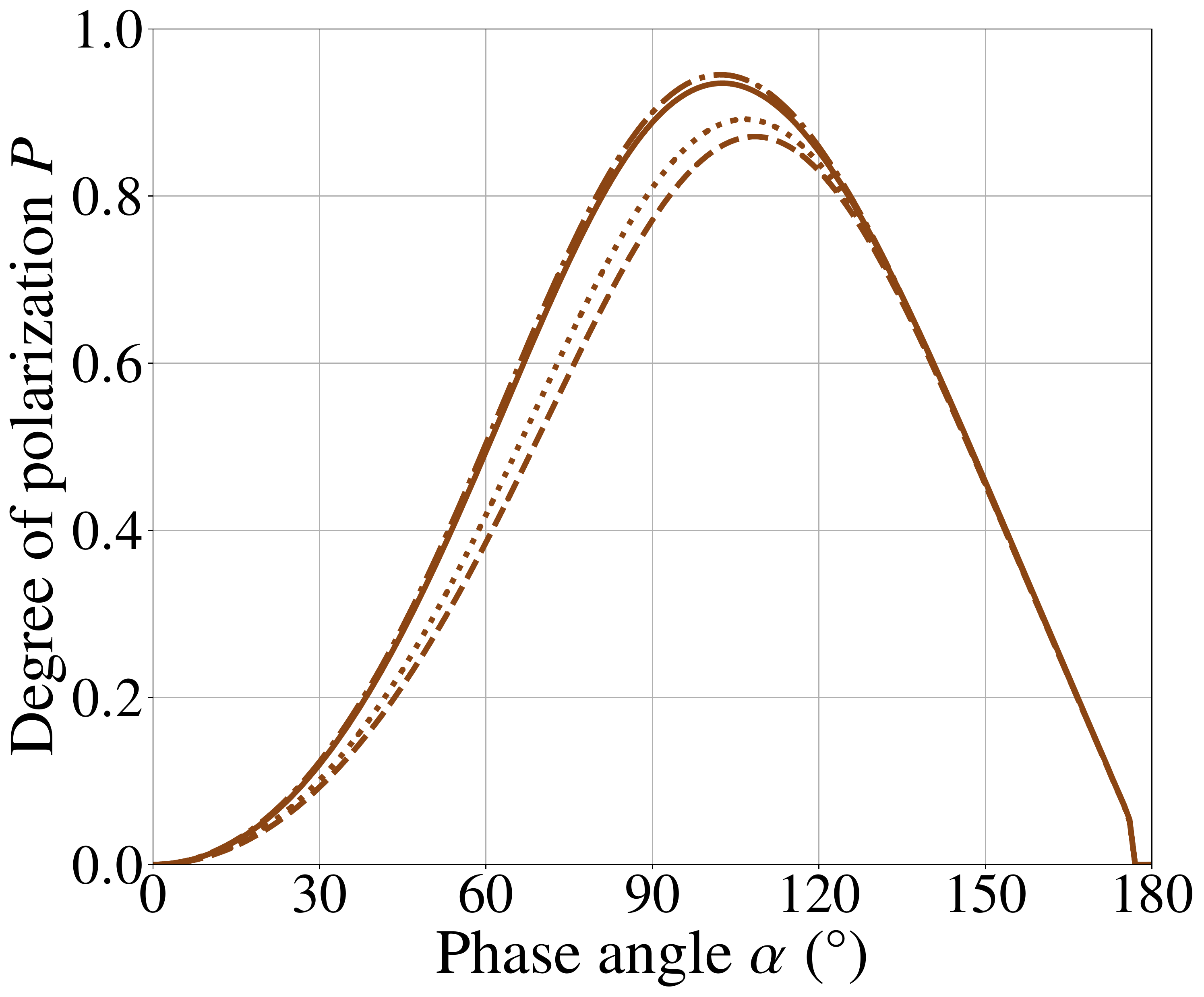}
\caption{Similar to Fig.~\ref{fig_HHplanets_wind}, but for 865~nm
         and different wind speeds and whitecap properties.
         The speeds and the respective whitecap properties are:
         7~m/s and $a_{\rm foam}=0.22$ (solid lines) (cf. Fig.~\ref{fig_HHplanets_wind}),
         13~m/s and $a_{\rm foam}=0.22$ (dotted lines), 13~m/s and $a_{\rm foam}=0.33$ (dashed lines), and 13~m/s but no whitecaps, i.e. $q=0$ (dashed-dotted lines).}	
\label{fig_HHwhitecaps}         
\end{figure*}

Wind blowing across an ocean generates waves on the surface and with
that, white caps or foam. The reflection matrix of the ocean as a whole 
(interface, water body, surface below the water) 
is computed using a weighted sum of the reflection matrix of the ocean 
without white caps, with its surface described by wave facets reflecting 
as Fresnel surfaces, and the reflection matrix of the white caps (see Eq.~\ref{eq:Rwtot}). 

The reflection by the white caps or foam is assumed to be Lambertian,
thus isotropic and non-polarizing, and the surface that is covered 
by white caps is flat. 
As baseline, we assume an effective foam albedo, $a_{\rm foam}$, 
of 0.22, as empirically derived by 
\citet{koepke1984effective}. Figure~\ref{fig_HHwhitecaps} 
shows the total and polarized fluxes and the degree of
polarization of the ocean planet at 865~nm, for a wind speed $v$ 
of 7~m/s
and our standard whitecap albedo $a_{\rm foam}$ of 0.22, 
for $v= 13$~m/s and $a_{\rm foam}=0.22$, for $v=13$~m/s
and $a_{\rm foam}= 0.33$ (the latter value is 
based on the uncertainty in the foam albedo value proposed 
by \citet{koepke1984effective}), 
and for $v=13$~m/s and no whitecaps. The curves clearly
show that the effect of the whitecap reflection is very small when 
$v=7$~m/s, as only 0.0003\% of the surface is covered by
whitecaps; the Lambertian reflection by the foam only slightly
($\lesssim$ 1\%) decreases the maximum $P$ as compared to the case 
without whitecaps.
For $v=13$~m/s, the whitecaps cover 2.46\% of the surface,
and decrease the maximum $P$ by about 5 and 10\% for 
$a_{\rm foam}= 0.22$ and 0.33, respectively. 
The whitecaps shift the maximum $P$ from $\alpha \sim 100^\circ$ 
to about 103$^\circ$ and 108$^\circ$, respectively. 
We conclude that the smaller $P$ at $v= 13$~m/s for 
$\lambda$ = 865 nm in Fig.~\ref{fig_HHplanets_wind}, 
compared to $P$ at $v=7$~m/s, is indeed caused by the whitecaps.

If white caps are ignored ($q=0$ in Eq.~\ref{eq:Rwtot}), 
the degree of polarization $P$ of light that is reflected by the ocean 
is independent of the wind speed $v$. The reason is that both the 
shadow function $S$ and the probability density function for the wave
facet inclination angles ${\rm PDF}$, that 
depend on $v$ (see Eqs.~\ref{eq_A1}-\ref{eq_transmissionmatrix}), 
influence all elements of the ocean's reflection matrix ${\bf R}_{\rm CO}$
(and, if $q=0$, matrix ${\bf R}_{\rm O}$) equally. 
The degree of polarization of an ocean planet as a whole can, however,
depend (slightly) on $v$, because when $v$ is large enough
such that white caps cannot be ignored, light that 
is reflected by the foam and that is subsequently scattered 
within the atmosphere, will influence the total and polarized
fluxes, and hence $P$ of the light that emerges from the top of 
the atmosphere.

\section{Horizontally inhomogeneous planets and the weighted sum approach}
\label{app_weightedsum}

A fast way to estimate the average flux and polarization phase curves 
of a partly cloudy planet, for any cloud coverage fraction $f_{\rm c}$, 
is to compute the weighted sum of horizontally homogeneous planets 
\citep{Stam08}. That is, the total and polarized fluxes of a partly cloudy 
planet are derived from the respective total and polarized fluxes 
of a completely cloudy planet, $\mathbf{F_{\rm cloudy}}$, and a 
cloud-free planet, $\mathbf{F_{\rm uncloudy}}$, according to
\begin{equation}
\mathbf{F}(\lambda,\alpha) = f_{\rm c} 
               \mathbf{F}_{\rm cloudy}(\lambda,\alpha) + 
               (1 - f_{\rm c}) \mathbf{F}_{\rm uncloudy}(\lambda,\alpha).
\label{eq_weightedsum}
\end{equation}
Because ${\mathbf F}_{\rm cloudy}$ and ${\mathbf F}_{\rm uncloudy}$
have to be computed only once, Eq.~\ref{eq_weightedsum} can be evaluated 
very fast for arbitrary cloud coverage fractions $f_{\rm c}$.

When using a weighted sum, the locations of cloud patches on a planet
are not taken into account, while the locations can influence the
total and polarized fluxes, and hence $P$.
In this appendix, we compare results of Eq.~\ref{eq_weightedsum} 
to the average phase curves of planets with various distributions
of horizontally inhomogeneous 
patchy clouds (for which the locations of the patches are thus 
fully accounted for) (see Sect.~\ref{sect_brokenclouds}).

\begin{figure*}[t!]
\includegraphics[width=0.33\textwidth]{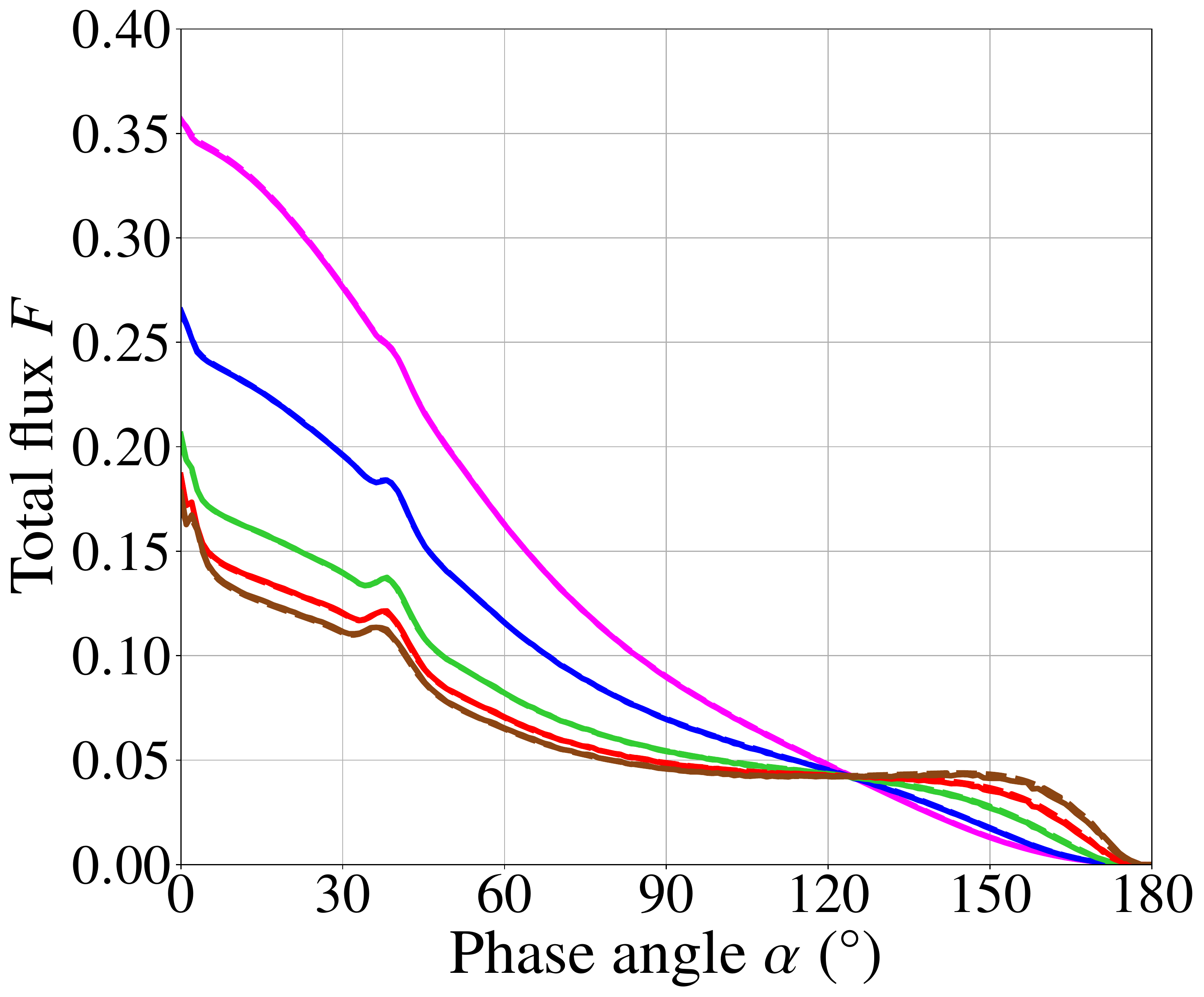}
\includegraphics[width=0.33\textwidth]{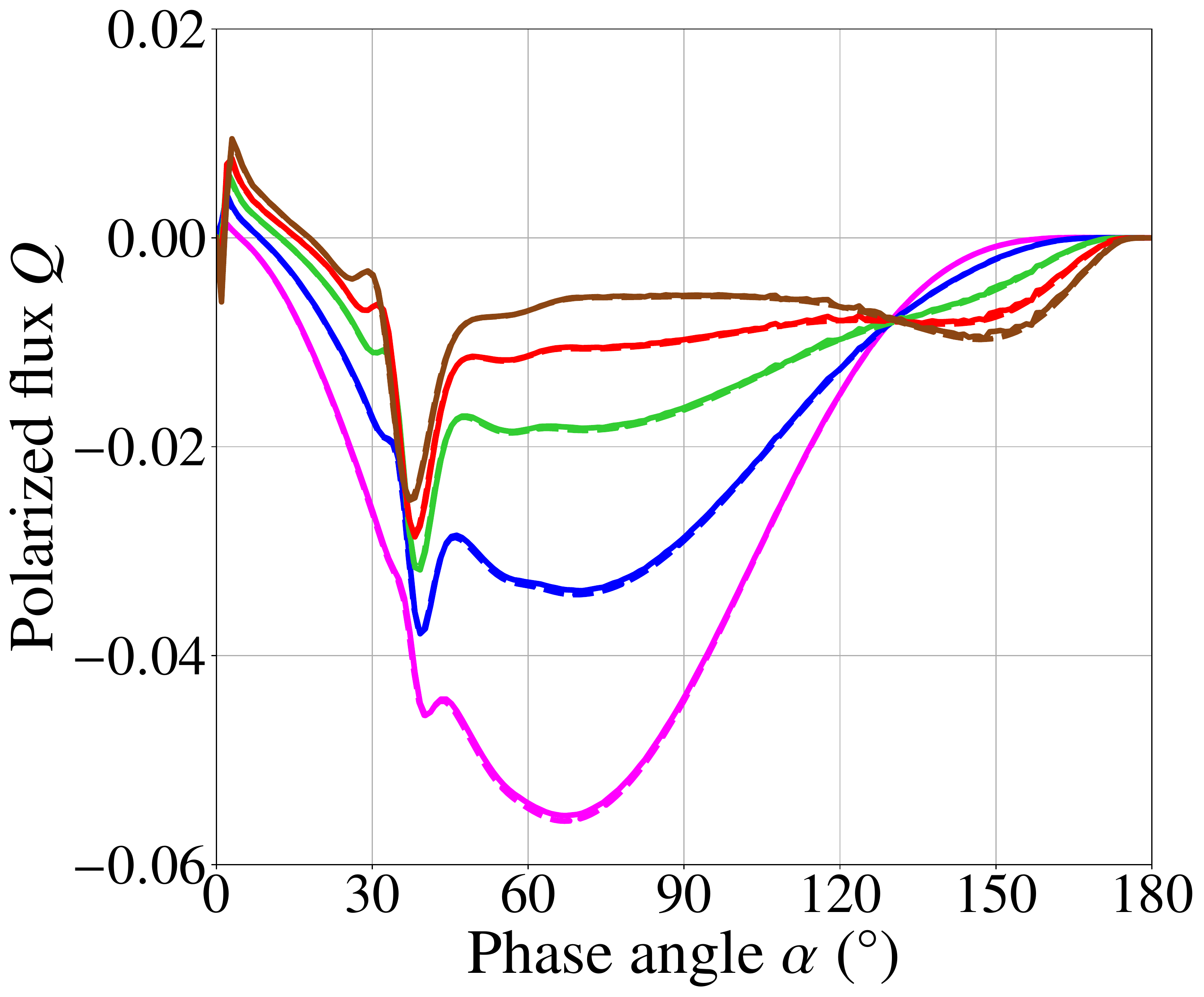}
\includegraphics[width=0.33\textwidth]{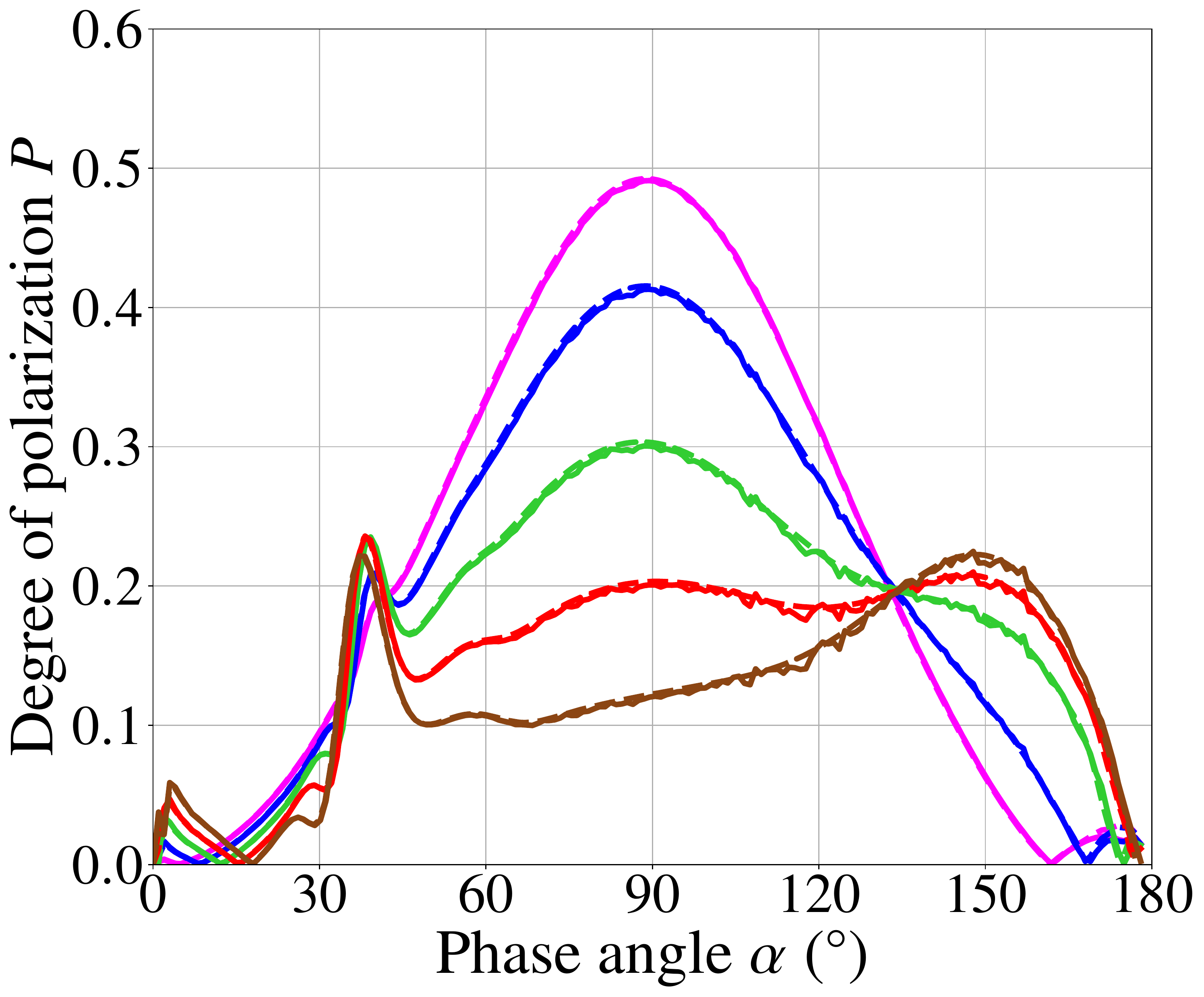}
\caption{Comparison between $F$ (left), $Q$ (middle), and $P$ (right) 
         of the reflected light as computed using the weighted sum 
         method (dashed lines) and as computed for pixelated, 
         horizontally inhomogeneous ocean planets with patchy clouds
         (solid lines). The solid lines are the same as those shown in 
         the second rows and second columns of 
         Figs.~\ref{fig_patchycloudsF}-\ref{fig_patchycloudsP}.
         The cloud coverage fraction $f_{\rm c}$, which is the 
         weighting factor in the weighted sum approach, 
         is 0.50 and the wind speed $v$ is 7~m/s. 
         The wavelengths are the same as those used
         in Figs.~\ref{fig_patchycloudsF}-\ref{fig_patchycloudsP}, i.e.\ 350~nm (pink), 443~nm (blue), 550~nm (green),
         670~nm (red), 865~nm (brown). 
}	
\label{fig_comparisonweightedsum}         
\end{figure*}

Figure~\ref{fig_comparisonweightedsum} shows that the average phase 
curves of $F$, $Q$, and $P$ for partly cloudy ocean planets with 
randomly distributed cloud patterns follow the phase curves computed 
using the weighted sum very accurately. For the curves in this figure,
the wind speed $v$ is 7~m/s and the cloud coverage fraction 
$f_{\rm c}$ 0.50. However, we find that for any combination of 
$v$ (1~m/s, 7~m/s, or 13~m/s) and $f_{\rm c}$ (0.25, 0.50, or 0.75) 
the agreement between the average phase curves and the weighted sums
is excellent. 

It should be emphasized that we use \textit{random} patchy cloud 
cover patterns. Hence, in the case of a planet with a 
non-random cloud pattern, e.g.\ a cloud pattern that is correlated to
underlying continents, horizontally inhomogeneous planet models 
as presented in this paper may be needed to accurately model 
measured phase curves.
In addition, using a weighted sum of horizontally homogeneous 
planets to compute phase curves of horizontally inhomogeneous planets
does not provide information about the variability in the phase curves
that is due to the variability in the cloud pattern, as shown by the 
shaded areas around the phase curves in  
Figs.~\ref{fig_patchycloudsF}-\ref{fig_patchycloudsP}.

\end{appendix}

\end{document}